%% file: main.tex
\documentclass{article}
\usepackage{arxiv}
\input{preamble_arxiv}
\usepackage{float}
\usepackage{needspace}
\input{meta}
\renewcommand{\shorttitle}{From topology-optimization brief to solver-checked release}
\hypersetup{pdftitle={\papertitle},pdfauthor={Shaoliang Yang and Jun Wang},pdfkeywords={\paperkeywords}}

\title{\papertitle}
\author{%
Shaoliang Yang\\
Department of Mechanical Engineering\\
Santa Clara University\\
500 El Camino Real, Santa Clara, CA 95053, USA
\And
Jun Wang\thanks{Corresponding author. Email: \href{mailto:jwang22@scu.edu}{jwang22@scu.edu}.}\\
Department of Mechanical Engineering\\
Santa Clara University\\
500 El Camino Real, Santa Clara, CA 95053, USA
}
\date{November 2026}

\begin{document}
\raggedbottom
\maketitle
\begin{abstract}
\input{abstract}
\end{abstract}
\keywords{\paperkeywords}

\input{body}

\begingroup\small
\setlength{\bibsep}{0pt}\setlength{\parskip}{0pt}\linespread{0.97}\selectfont
\bibliographystyle{unsrtnat}
\bibliography{references}
\endgroup

\clearpage
\appendix
\raggedbottom
\renewcommand{\thesection}{S\arabic{section}}
\renewcommand{\thetable}{S\arabic{table}}
\renewcommand{\thefigure}{S\arabic{figure}}
\setcounter{section}{0}\setcounter{table}{0}\setcounter{figure}{0}
\begin{center}{\Large\bfseries Supplementary material}\end{center}
\addcontentsline{toc}{section}{Supplementary material}
\input{supp_appendix}
\end{document}

%% file: preamble_arxiv.tex
\usepackage[T1]{fontenc}
\usepackage[utf8]{inputenc}
\usepackage{microtype}
\usepackage{amsmath,amssymb,booktabs,tabularx,longtable,array}
\usepackage{graphicx,adjustbox}
\usepackage[font=small,labelfont=bf]{caption}
\usepackage[numbers,sort&compress]{natbib}
\usepackage[dvipsnames]{xcolor}
\definecolor{navy}{HTML}{234866}
\usepackage{tikz}
\usetikzlibrary{positioning,arrows.meta}
\usepackage{listings}
\usepackage{placeins}
\usepackage[colorlinks=true,linkcolor=navy,citecolor=navy,urlcolor=navy]{hyperref}
\usepackage{xurl}
\hypersetup{pdftitle={From topology-optimization brief to solver-checked release: separating intent, admissibility and acceptance knowledge for language agents},pdfauthor={Shaoliang Yang and Jun Wang}}

%% file: meta.tex
\newcommand{\papertitle}{From topology-optimization brief to solver-checked release: separating intent, admissibility and acceptance knowledge for language agents}
\newcommand{\paperkeywords}{engineering knowledge representation; language agents; topology optimization; design intent; numerical verification; interactive engineering software}
\newcommand{\repourl}{https://github.com/nbbllxx0/From-brief-design-to-verified-release}

%% file: abstract.tex
An engineering solver can converge even when a language agent has misread the user's problem. AgenticTO makes this distinction inspectable by keeping three records separate: the proposed problem specification, deterministic formulation checks, and numerical acceptance of the solved design. We evaluated the workflow on 50 author-built topology-optimization briefs, injected faults, five numerical problems, and complete agent sessions. The tests used linear-elastic models with one load case. Geometry tools raised the primary provider's correct specifications and refusals from 37 to 44 of 50, while also increasing wrong specifications delivered as finished. Input-format feedback alone recovered most of the accuracy gain. The wrong specifications passed the formulation rules, which also missed most injected faults that changed the intended problem. Clarification corrected the remaining errors when an oracle supplied reference facts, but also prompted questions on fully specified briefs. Perfect accuracy under that oracle therefore does not establish performance with engineers. Numerical trials reached 73.7 million elements, although released designs used finer grids than requested. The records expose how a proposed problem becomes an accepted numerical result; agreement with the brief still needs a separate judgment.

%% file: body.tex
\section{Introduction}
\label{sec:intro}
Large-scale topology optimization can place material in a three-dimensional domain containing millions of finite elements. Using that capability still requires an engineer to state a precise problem. A support must constrain the intended displacement components on the intended region. A quoted load must become a force vector with consistent units and a defined spatial distribution. Mounting pads, access holes, and clearances must become retained-solid or retained-void regions. A numerically successful computation cannot establish that these choices match the brief. The briefs used in this study state such requirements in topology-optimization terms, with axes, coordinates, volume fraction, and element size given; requirements stated as performance targets or as computer-aided design (CAD) geometry are not tested.

Language agents offer a way to translate such requirements into computational actions, and language models are being applied across mechanics, product design, and manufacturing \citep{Mustapha2024Survey}. Recent systems coordinate topology-optimization formulation and solver execution \citep{Park2026TopOptAgents,Lin2026TOMaster}; other workflows also use multiview critique to pursue design preferences \citep{Stewart2026TOAgents}. However, a plausible explanation, a valid input file, and a converged result answer different questions. A support can select real nodes and remove every rigid-body mode while restraining a direction that the engineer meant to leave free. An agent can add a sensible-looking solid pad that was never requested and thereby change the material budget. Conversely, a correctly transcribed brief can describe a mechanism with no static equilibrium. These failures require evidence at different stages of the workflow.

\begin{figure}[p]
\centering\includegraphics[width=0.85\linewidth]{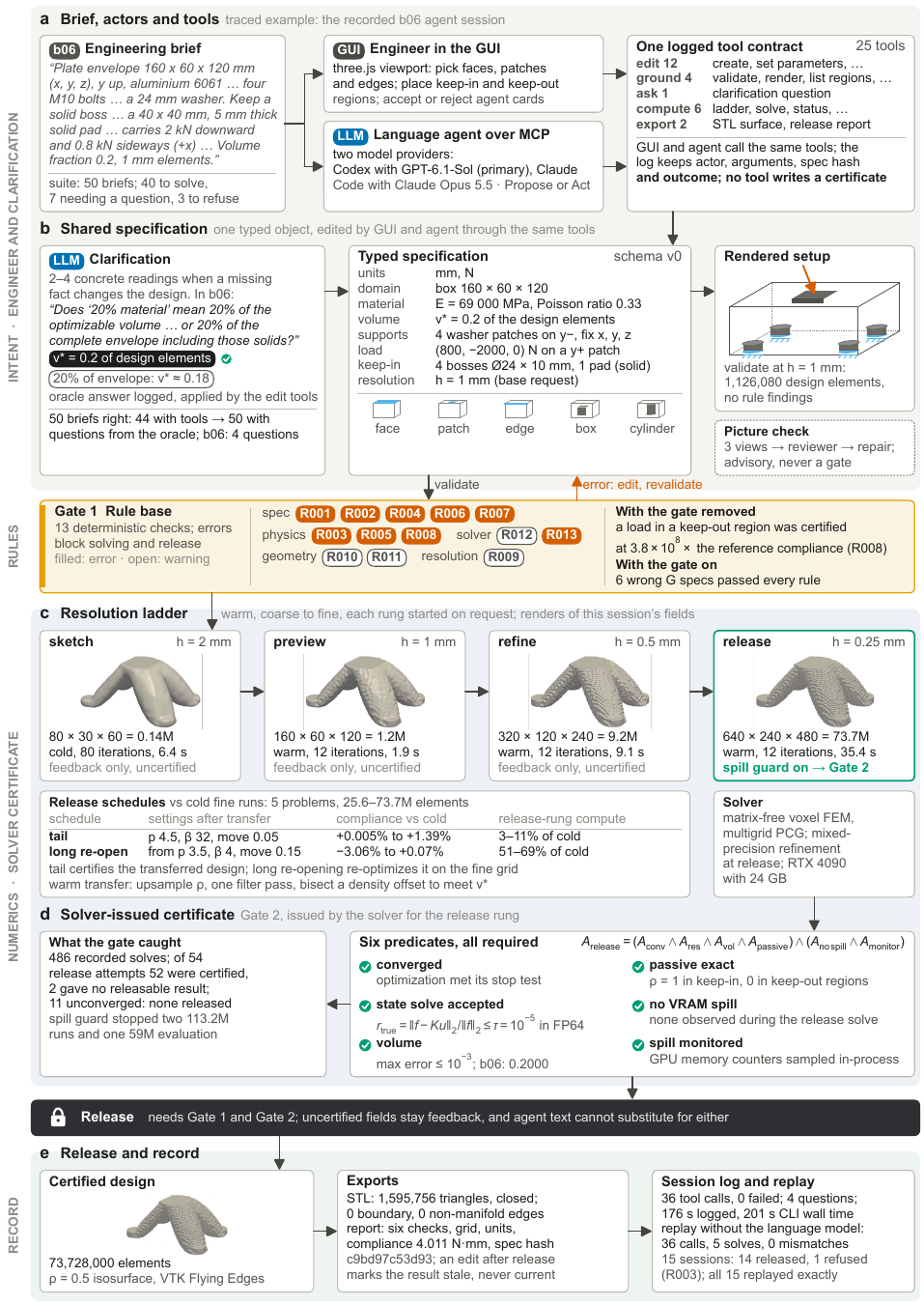}
\caption{AgenticTO workflow, traced through one recorded agent session (b06 motor mount). Renders are $\rho=0.5$ isosurfaces of the saved fields; under the final-phase-only tail schedule the fine rungs keep the transferred voxel steps, so the release certifies the transferred design on a grid finer than specified. The schema combines the b06 brief's two force components into one force (Section~\ref{sec:suite}). Benchmark scores use the adjudicated reference readings. G: grounding tools without questions; LLM: large language model; MCP: Model Context Protocol; $p$: penalization; $\beta$: projection sharpness; FEM: finite-element method; VRAM: video random-access memory; PCG: preconditioned conjugate gradient; FP64: 64-bit floating point; CLI: command-line interface; VTK: Visualization Toolkit; STL: stereolithography file.}
\label{fig:workflow}
\end{figure}

We investigate a knowledge layer between the engineer, the agent, and the numerical solver. Its content is problem-formulation knowledge: which regions are supported, loaded, retained, or kept empty and in which units, when such a formulation is valid, and what evidence accepts a numerical result (Table~\ref{tab:knowledge}). The layer follows the engineering-informatics practice of holding design knowledge in explicit, executable form \citep{LaRocca2012KBE} and checking design models against encoded rules before acceptance \citep{Eastman2009RuleChecking}. Its central object is a typed design specification whose regions are defined geometrically rather than on a mesh. Both the graphical user interface (GUI) and the agent edit that object through the same tools. A deterministic rule base checks selected geometric, physical, and resource conditions before a solve. The solver returns a certificate containing the acceptance evidence for its numerical result. Tool calls, specification hashes, and solve outcomes form a provenance log that can be replayed without asking the language model to make its decisions again. Figure~\ref{fig:workflow} traces this workflow through one recorded agent session. Read the main arrow path first: a specification is proposed, rules check its formulation, and the solver certifies a numerical result before release. The numeric annotations document one example; Section~\ref{sec:system} defines the individual checks.

The three records make authority explicit. The agent proposes the specification; deterministic rules check selected formulation conditions; the solver certifies its numerical result. The rules cannot determine whether an unmentioned washer should be retained solid, and a numerical certificate cannot recover an unstated load direction. We call the records inspectable because they can be read and replayed. Agreement with the brief is tested separately against a reference reading: the complete interpretation used for comparison.

The evaluation tests where these records help and where they fail. Grounding tools improved right outcomes but also increased wrong specifications delivered as finished. Clarification against stored reference facts corrected the remaining primary-provider errors, while asking many questions on fully stated briefs. Rules and numerical certificates caught different failures, and neither established agreement with the brief. These findings motivate separate evidence for problem formulation, numerical acceptance, and reference agreement.

Scale raises the cost of a wrong or uncertified solve: validating a specification takes milliseconds, whereas a release at tens of millions of elements takes minutes of graphics processing unit (GPU) computation. Large-scale topology optimization is already established \citep{Aage2017Giga,Wu2016HighRes,Traff2023GPU,Hou2025CuPy}; our solver is inherited from companion work. We evaluate whether the shared representation and release contract connect a written brief to an inspectable result, not a new kernel-speed or discretization-accuracy claim.

The numerical workflow separates quick feedback from final release. A short final continuation can certify a transferred design, while a longer re-opening can improve it at greater cost (Section~\ref{sec:ladder}). The system also selects a release grid finer than the specification's stated resolution, which changes physical regularization (Section~\ref{sec:spec}). Both the schedule and the actual grid are recorded, so a quick release must not be read as a complete fine-grid optimization or as acceptance of the engineer's stated resolution.

The paper makes four contributions. First, it specifies a shared design-intent representation, deterministic rule base, and logged tool contract for agent and GUI operations.

Second, it evaluates grounding and clarification on 50 three-dimensional briefs with two model providers, distinguishing right outcomes from wrong specifications delivered as finished.

Third, it measures the numerical cost and compliance consequences of coarse-to-fine release on five problems, together with a common-grid evaluation of two resolution paths.

Fourth, it tests the rule gate and numerical certificate through rule removal and injected faults, then evaluates release, replay, interface latency and modeled interaction cost in complete sessions.

J.W. adjudicated the benchmark's reference readings and contested outputs; S.Y. repeated the original adjudication blind to condition, provider and first verdict. J.W. alone adjudicated the later conditions. These author judgments are not independent ground truth (Section~\ref{sec:suite}). A certificate likewise covers stated computational checks, not global optimality or manufacturing approval.

\section{Related work and scope of the contribution}
\label{sec:related}
\subsection{Language agents for engineering computation}
TopOptAgents organizes six language-model roles around formulation, validation, code generation, execution, and design assessment \citep{Park2026TopOptAgents}. TO-Agents combines topology optimization with multiview critique and an independent judge to pursue subjective design preferences \citep{Stewart2026TOAgents}. TO-Master uses typed computational tools in a conversational workflow and covers a broader range of physics and geometry inputs than the present system \citep{Lin2026TOMaster}. Large-model-guided topology optimization (LMTO) uses a large visual--language model so that designs follow a stated concept, while their structural performance remains comparable to or better than that of designs from bi-directional evolutionary structural optimization \citep{Liang2025LMTO}. That work uses the large model to shape the design; ours uses it to state the mechanical problem, which is why we score specifications rather than shapes. These works establish that an agent can coordinate an optimization workflow. Their objectives differ from our focus on scored specification equivalence, explicit clarification, and the evidence required to release a numerical result.

The distinction is methodological rather than a claim that earlier tools perform no checking. Boundary-condition checks and visual review already appear in agent workflows. We separate the correctness of the translated problem from its admissibility as a discrete mechanical problem and from the acceptance of the state solve, then measure failures at those boundaries. The second-provider experiment also tests whether giving an agent the same clarification tool reproduces the benefit. We avoid ranking systems by success rates drawn from different task suites, since a subjective-preference task, a code-execution task, and a fully specified mechanics task have different failure criteria.

Beyond topology optimization, MechAgents demonstrates language-agent collaboration for mechanics, and FEABench evaluates tool-mediated multiphysics reasoning \citep{Ni2024MechAgents,Mudur2025FEABench}. A constrained natural-language interface for FEniCS uses a structured specification and deterministic dispatch to separate language interpretation from numerical execution \citep{Upadhyay2026FEniCSNL}. This is a close architectural precedent. Our contribution lies in the particular three-dimensional design representation, shared GUI tools, resolution ladder, release evidence, and associated evaluation, rather than in the general idea of placing a deterministic solver behind a language interface.

\subsection{Explicit engineering knowledge and rule checking}
Knowledge-based engineering captures design rules and procedures in executable form so that they can be applied consistently and reused \citep{LaRocca2012KBE}. Automated rule checking applies the same idea to acceptance: requirements are interpreted into executable rules, the design model is prepared for them, and the checking results are reported \citep{Eastman2009RuleChecking}. Recent work in building information modeling encodes requirements as semantic-web rules and reports which kinds of requirement, including geometric and vague ones, can be checked automatically \citep{Pauwels2024BIMSemWeb}. Language models have also been combined with explicit method knowledge, for example to automate the Theory of Inventive Problem Solving (TRIZ) procedure \citep{Jiang2025AutoTRIZ}. Our rule base follows the rule-checking tradition on a small scale, with 13 deterministic checks and stated limits. Most of these checks, such as empty selections, rigid-body adequacy, and memory estimates, are standard pre-solve diagnostics in finite-element practice. Here the agent proposes a specification, the rules decide whether it may be solved, and the solver decides whether its numerical result may be released. This arrangement separates those decisions from judging whether the specification matches the brief. Much of the specification holds what a topology-optimization input deck holds. Beyond it, the representation carries named geometric regions that survive a change of mesh, units tied to the modulus, an assignment of authority over each record, and a release policy; in the evaluation, building the typed regions through the tools removed the text-only condition's invalid geometry, such as zero-length cylinders (Section~\ref{sec:grounding}). We did not compare the representation itself with an alternative, such as generated solver code.

In the terms of the function--behaviour--structure framework~\cite{Gero2004FBS}, the specification records structure and the behavior asked of it, not the function behind them: it holds where a load acts but not why. Requirements-engineering work links ambiguity during elicitation with tacit knowledge~\cite{Ferrari2016Ambiguity}. Our clarification protocol requests additional facts when the brief leaves a design-changing choice open. Table~\ref{tab:knowledge} states this knowledge model: what each record holds, where that knowledge comes from, who has authority over it, and which failures it can and cannot expose.

This separation is related to the distinction in computational science between verification, which asks whether the model equations are solved correctly, and validation, which asks whether the model represents the intended problem for its intended use \citep{Oberkampf2002VV,Oberkampf2010VV}. Here the specification and clarification address a form of the second question for a language-agent workflow, and the solver certificate addresses the first only in part: it bounds the algebraic residual of the state solve and checks the volume and passive-region constraints on one grid, but it does not estimate discretization error, which solution verification in those frameworks also requires. Unlike validation in that sense, which compares a model with physical observations, comparing a specification with the engineer's stated brief is requirements verification; clarification, which elicits facts the brief omits, is closer to requirements validation.

Design intent has a broader meaning in computer-aided design, where it refers to how a model is meant to behave when it is modified, which modelers embed through features and constraints \citep{Otey2018DesignIntent}. Design-rationale systems record why a design takes its form \citep{Regli2000Rationale}, and requirements traceability is the ability to follow a requirement forward and backward through the artifacts derived from it \citep{Gotel1994Traceability}. We use design intent in a narrower, checkable sense: the problem the engineer means to have solved, recorded as a specification whose equivalence with a reference reading can be computed. The provenance log traces a released field back to the specification and the tool calls that produced it, but the specification does not record why each field holds its value.

\subsection{Interactive and multiresolution topology optimization}
Interactive design systems let engineers express preferences through actions on geometry or intermediate structures. Forte connects user input to generative structural design \citep{Chen2018Forte}, while HiTop exposes feature-size controls during topology optimization \citep{Ha2023HiTop}. Such work shows that the interaction mechanism is part of the design method. A language panel alone does not establish that an interface is understandable or efficient. We therefore report backend response times and modeled actions separately, and leave human usability to a future participant study.

Multiresolution topology optimization also has an established numerical meaning. The multiresolution paradigm of \citet{Nguyen2010MTOP} and later adaptive approaches \citep{Gupta2020MultiRes} can separate design and analysis discretizations. Our ladder instead advances the working voxel problem through several resolutions and exposes these stages as feedback and release choices. We use ordinary warm transfer followed by specified continuation schedules. This is a system policy built from numerical methods, rather than a new multiresolution finite-element formulation.

Large-scale implementations provide the computational basis. High-resolution systems and GPU codes already address large three-dimensional problems \citep{Wu2016HighRes,Hou2025CuPy}; giga-voxel methods exceed these counts \citep{Aage2017Giga}, and GPU methods reach the same tens-of-millions range \citep{Traff2023GPU}. Element count alone therefore does not identify a new contribution. The present numerical experiments compare release policies within the same solver family. They are not cross-paper speed comparisons, and their warm-start compute ratios exclude earlier rungs unless explicitly stated.

\subsection{Relationship to companion work}
AutoSiMP already maps language to structured topology-optimization inputs and chooses numerical settings \citep{Yang2026AutoSiMP}; later companion work changes continuation parameters from solver state \citep{Yang2026Controllers}. Consequence-guided clarification instead chooses questions from the mechanical consequences of alternative readings \citep{Yang2026Clarify}. Earlier question-ranking research provides a related precedent \citep{Rao2018Clarify}. Here the agent chooses the alternatives and whether to ask; an oracle answers from reference facts. We evaluate the resulting specifications and question burden, not the earlier question-choice method.

The numerical backend is inherited from fused operators, Galerkin multigrid and guarded state-solve work \citep{Yang2026Fused,Yang2026Galerkin,Yang2026Floor}. Our contribution is the shared specification, rule and release contract, and their evaluation as a complete system. The clarification study~\citep{Yang2026Clarify} has 36 tasks without a language model, and our study of impossible problems~\citep{Yang2026Impossible} uses 30 primal/adjoint pairs; neither supplies tasks, references or language runs here. AutoSiMP was not run as a baseline because it chiefly addresses two-dimensional problems and has no released implementation that can run through this tool contract.

\section{System and release contract}
\label{sec:system}
\subsection{Design problem and explicit representation}
\label{sec:spec}
We consider one-load-case, isotropic linear-elastic minimum-compliance problems on box-shaped voxel domains. The finite-element state equation and objective are
\begin{equation}
 K(\bar\rho)u=f,\qquad C(\bar\rho)=f^{\mathsf T}u,
 \label{eq:state}
\end{equation}
where $u$ is displacement, $f$ is the assembled load vector, and $K$ is the constrained stiffness operator. The physical density $\bar\rho$ is obtained from design variables by filtering and projection \citep{Wang2011Projection} in a density-based material-distribution method \citep{Bendsoe1989}, using solid isotropic material with penalization (SIMP) as reviewed by \citet{Sigmund2013Approaches}. With design elements $\mathcal D$, retained-solid elements $\mathcal S$, and retained-void elements $\mathcal V$, the material constraint is
\begin{equation}
 \frac{1}{|\mathcal D|}\sum_{e\in\mathcal D}\bar\rho_e\leq v^*,
 \qquad \bar\rho_e=1\ (e\in\mathcal S),\quad
 \bar\rho_e=0\ (e\in\mathcal V).
 \label{eq:volume}
\end{equation}
Equal-size voxels make the element average a volume fraction. In particular, $v^*$ applies to designable material, rather than to the entire envelope. Adding a retained-solid pad changes both the feasible set and the total permitted material. That choice must remain visible in the specification.

The specification schema (version 0, provided as a machine-readable file with the supplement) records the domain dimensions, length and force units, Young's modulus, Poisson's ratio, volume fraction, supports, loads, passive regions, requested resolution, and an optional minimum-feature requirement. Its region types are faces, surface patches, edges, boxes, and cylinders. Coordinates are absolute in a frame whose origin is the minimum corner of the domain. A surface patch stores its full widths in the face's two coordinate directions. A support names the displacement components it fixes; a load stores a total force that is shared equally among the selected nodes. This convention preserves total force across meshes but is not a general pressure-integration rule.

Equal nodal sharing is not consistent pressure integration, and the comparator cannot detect that difference. Edge loads and supports also produce singular elastic fields, so their compliance is not mesh-converged as the grid is refined. The supplement identifies the affected reference readings and comparison boundaries.

The representation is geometric before it is discrete. A mounting face remains a named face when the mesh changes. The problem builder then selects nodes and element centers, creates fixed-degree-of-freedom masks, distributes forces, and constructs passive masks. Any snapping or disappearing region can be reported during validation. Material modulus is expressed in the force-per-length-squared unit implied by the specification. Thus changing from newtons to kilonewtons while retaining millimeters requires changing the numerical modulus as well as the forces.

Resolution is stated as an element edge length or a cell count. The system may select a finer release grid, and did so in every released agent session. The engineer accepts the stated resolution, not this finer grid. Because the filter radius is specified in element units, refinement shortens its physical length: the release answers a more finely regularized problem. Filtering regularizes topology optimization \citep{Bourdin2001Filters,Bruns2001Nonlinear,Lazarov2011Helmholtz}; a radius in physical length would keep that length fixed, while robust projection formulations address minimum member size \citep{Wang2011Projection}. The optional minimum-feature field is a requested length, not a measurement of the resulting design. Its ratio to the element spacing of the grid being checked, including release, triggers a warning below 2.4. It does not certify a physical member size or manufacturing constraint.

Keeping the accepted specification's regularization on the release grid would need a filter of constant physical length, and a pilot with such a filter did not meet the stagnation test within 300 iterations on any of the three problems tried, so the released designs carry the finer regularization.

\subsection{Deterministic rule base}
\label{sec:rules}
The rule base contains 13 checks (Table~\ref{tab:t2_rules}). Errors block solving and release, and the rules run again on the grid of every solve, the release rung included, before it starts. Warnings and informational findings remain visible, with a suggested corrective action. The checks range from empty node selections to rigid-body adequacy, incompatible passive regions, and estimated GPU memory demand. They are deterministic functions of the specification and the grid, rather than judgments generated by a language model.

We call a specification admissible when it defines a well-posed and feasible discrete linear-elastic optimization problem on the requested grid: a nonzero load is applied, the supports remove all rigid-body modes, design elements remain after the passive regions, and no element is both retained and removed. R003, R006, R007, and the zero-force test of R004 check these necessary conditions. 

Some rules are stricter than admissibility. They reject empty selections, loads lost on fixed components, and supports or loads entirely inside keep-out regions (Table~\ref{tab:t2_rules}). The last case can remain numerically solvable because void elements retain a small artificial stiffness. Thus admissibility, passing the rule set, and agreement with the brief remain different tests. Grid construction and GPU-memory checks govern execution; warnings are advisory. On coarse multigrid levels, a lost support produces a warning if a rigid mode is left free, and an informational note if the coarse body remains restrained.

Defects outside the rule set include a load only partly inside a keep-out region, retained solids disconnected from the load path, and a modulus implausible for the stated units, such as the two unit slips in Section~\ref{sec:providers}.
\begin{table}[t]
\centering\small
\caption{Knowledge model. Each record primarily holds one kind of problem-formulation knowledge under one authority, with the exceptions listed in its cells, and exposes some failures but not others.}
\label{tab:knowledge}
\begin{tabularx}{\linewidth}{p{21mm}*{3}{X}}
\toprule
Record & Knowledge and its source & Representation and authority & Exposes; does not expose \\
\midrule
Specification (intent) & Domain, units, material, supports, loads, retained regions and stated resolution; from the brief and clarification & Typed geometric record; the engineer accepts it, but the system sets the release grid and physical filter length & Makes the proposed reading comparable with a reference. The record alone, its default free-text notes and the tested per-value source tags did not reliably distinguish well-formed wrong readings \\
Rule base (admissibility) & Necessary formulation conditions, stricter faithfulness/load-path checks and execution advice; from finite-element practice & Deterministic functions of specification and grid; an error blocks solving and release & Exposes selected empty regions, free rigid modes and passive conflicts. Misses partial keep-out loads, disconnected solids, implausible moduli and other intent errors \\
Certificate (numerical acceptance) & Design stagnation, state residual, volume and passive consistency, plus execution conditions; from the solver & Solver-issued record required for release; actual grid, filter and schedule are recorded, not accepted by the engineer & Exposes numerical/resource failures. Does not establish intent, physical consistency beyond its predicates, or stress and manufacturing performance \\
\bottomrule
\end{tabularx}
\end{table}

Rigid-body checking tests the three translations and three rotations of a body, including missing rotational restraints on supports with many nodes. Loads on fixed displacement components are dropped by the problem builder: more than half lost blocks the solve; any positive share up to and including half raises a warning. The remaining vector can still be solved correctly, so the rule makes that change visible.

\input{tables/t2_rules}

\subsection{Shared tools, provenance, and replay}
\label{sec:tools}
The agent accesses the system through the Model Context Protocol (MCP)~\citep{MCPSpec}, while the GUI calls the same session operations through the backend. Tools create or inspect a specification, edit parameters and regions, validate and render the setup, inspect the ladder, start or await a solve, examine field statistics, undo or redo edits, request clarification, and export results. The agent receives structured outcomes rather than a private route to modify a certificate field.

A session log stores each tool call, its actor and arguments, the resulting specification hash, and the outcome. This log describes the actual sequence, including failed calls and waits. It follows the provenance practice of scientific workflow systems, which record how each result was derived so that it can be inspected and reproduced \citep{Davidson2008Provenance}. Replay executes recorded actions without re-running the language model. It therefore checks execution reproducibility conditional on the recorded decisions. It does not imply that a fresh language-model run will make the same decisions, nor does it independently validate the engineer's intended task.

Session replay uses deterministic coarsest-grid assembly because the upstream solver's atomic additions can run in different orders. Solver-identity tests use a separate atomic-free Jacobi configuration. Changing the coarse-grid policy changes the numerical execution; the later padding-aware replay reports compliance differences and is not called bit-exact.

Long solves return a job identifier that the agent can inspect or await repeatedly. Independent worker events keep status current. The log retains failed calls, waits and cancellations, distinguishing orchestration from solver time.

\subsection{Grounding, clarification, and picture checking}
\label{sec:clarify}
The grounding protocol asks the agent to construct the complete specification through tools, inspect selected regions, validate the setup, and render the support and load placement. Tool responses provide direct geometric facts: selected regions, displacement directions, force components, and rule findings. This feedback can repair schema or geometric-construction errors without requiring the agent to generate numerical solver code.

Clarification is a separate capability. When a missing fact changes the design, the agent can submit two to four concrete readings. Each reading supplies parameter values or a proposed item, rather than only a verbal label. The interface can compute consequence previews and presents the choices as selectable chips. The answer is recorded and then applied through the normal edit tools. This makes the question, its alternatives, and the chosen interpretation part of the computational record.

In the experiments, a scripted oracle compares the proposed choices with the stored intended specification. It selects a matching option, or states that none matches and supplies the stored clarification facts. Because the oracle has direct access to those facts, it can supply more precise feedback than an unprepared user. Its answers also do not depend on the consequence images. The benchmark therefore tests an agent's questioning and use of supplied information under a cooperative oracle; it does not isolate the effect of consequence visualization or estimate human answer quality.

The additional picture-check condition starts from the same grounding-and-clarification result. A reviewer model receives three rendered views and the brief, then returns a structured mismatch decision and a list of concrete issues. A repair run receives any reported issues, checks them against the brief and tool responses, and may revise the specification or ask another question. This check targets region placement and directions visible in a picture. It cannot verify a hidden unit conversion solely from the rendered geometry.

The picture check is advisory. Its parser treats an unparseable reply as no actionable mismatch; none of the 139 reported replies was unparseable. The formulation and numerical gates remain in force.

\subsection{Progressive computation and release policies}
\label{sec:ladder}
The ladder organizes computation by purpose. A sketch produces an early density field; preview and refine improve the working resolution; release supplies a result on the reported final grid with spill monitoring enabled. The GUI starts each rung on request. It does not automatically solve after every edit. When a suitable previous field exists, the next computation begins from that field.

Warm transfer upsamples the coarse density, filters it once and adjusts a constant offset by bisection to meet the target projected volume on the new mask. The default tail continues only the final sharpened phase. A short re-opening permits limited change; a long re-opening starts less sharply and allows more boundary motion. The supplement specifies all three schedules.

A saturated transferred field can pass the tail's stopping test without changing solid/void states. Such a release certifies the transferred design. Long re-opening can improve it, but neither policy guarantees a cold-run optimum or new members.

Multigrid uses coarser grids to accelerate each state solve. Padding adds zero-stiffness cells around the physical grid, and ordinary coarsening can make that padding act like a clamp on a free physical surface. The padding-aware policy prevents this artificial coarse-grid clamp. On padded grids it also dilates the coarse support masks, retaining small physical supports that sampling could otherwise lose. These changes affect the preconditioner; fine-level acceptance still uses the recomputed true residual. Each solve records its policy, distinguishing original and padding-aware timings.

\subsection{Solver-issued certificate and export}
\label{sec:certificate}
The certificate combines four numerical predicates with two execution predicates:
\begin{equation}
 A_{\mathrm{release}}=\bigl(A_{\mathrm{conv}}\land A_{\mathrm{res}}
 \land A_{\mathrm{vol}}\land A_{\mathrm{passive}}\bigr)
 \land\bigl(A_{\mathrm{no\ spill}}\land A_{\mathrm{monitor}}\bigr).
 \label{eq:certificate}
\end{equation}
The numerical predicates require optimization convergence, a design-stagnation test defined below; an accepted final state solve; absolute error in the final design-volume fraction no greater than $10^{-3}$; and exact retained-solid and retained-void values. The execution predicates require no observed video random-access memory (VRAM) spill and active spill monitoring. Final state acceptance uses the true relative residual, recomputed in 64-bit floating point (FP64), at the solve's configured tolerance:
\begin{equation}
 r_{\mathrm{true}}=\frac{\|f-Ku\|_2}{\|f\|_2}\leq\tau.
 \label{eq:residual}
\end{equation}
The experiments normally use $\tau=10^{-5}$ for optimization; the independent small-problem cross-check uses a tighter acceptance tolerance. Spill monitoring samples the Windows GPU-memory counters through an in-process implementation. The two execution predicates are therefore defined only on Windows hosts; another platform would need its own monitor. A passing no-spill check means no spill was observed under that monitoring policy. The two spill predicates record execution conditions on a shared desktop GPU rather than numerical accuracy: they state whether the run was monitored and free of memory oversubscription. Only the four numerical predicates carry over unchanged to another host; the two execution predicates are host guards.

Optimization convergence is the solver's stop test. In the final continuation phase, the spread (maximum minus minimum) of the compliance over the last ten iterations must not exceed $10^{-4}$ of its mean, and either the largest change of any physical density must be at most 0.01 or, in the reported runs, at most 100 parts per million of the design elements may change by more than 0.01. The reported runs use the 100-parts-per-million alternative: the allowed moving-element count is the integer part of that fraction of the design elements, with a minimum allowance of one. A run terminated by the iteration cap fails the predicate. This is a design-stagnation test, not a first-order optimality measure. A saturated projection can meet it, which is why a tail release certifies a transferred design rather than re-optimizing it.

The solver records the grid, units, compliance, termination, residual and stopping settings. It also records stated and release element sizes, their physical filter lengths, the warm-start grid and the continuation schedule. These fields distinguish a transferred design from a re-optimized one and accepted resolution from released resolution; for the reported sessions these grid, filter and schedule fields were reconstructed from logs rather than read from the original certificates (Section~\ref{sec:sessions}). Export as a released result requires both gates. Intermediate fields remain available as feedback.

The GUI presents a certified density as an isosurface at $\rho=0.5$, extracted with the Visualization Toolkit's Flying Edges implementation~\cite{Schroeder2015FlyingEdges}. The same surface is used for stereolithography (STL) export. The numerical certificate belongs to the density problem that was solved; extracting a surface does not transfer that certificate to a new mechanical analysis of the surface or a manufactured part. Where we compare binarized parts, we perform an additional common-grid state evaluation explicitly.

\subsection{Graphical interface and interaction scope}
\label{sec:gui}
\begin{figure}[t]
\centering\includegraphics{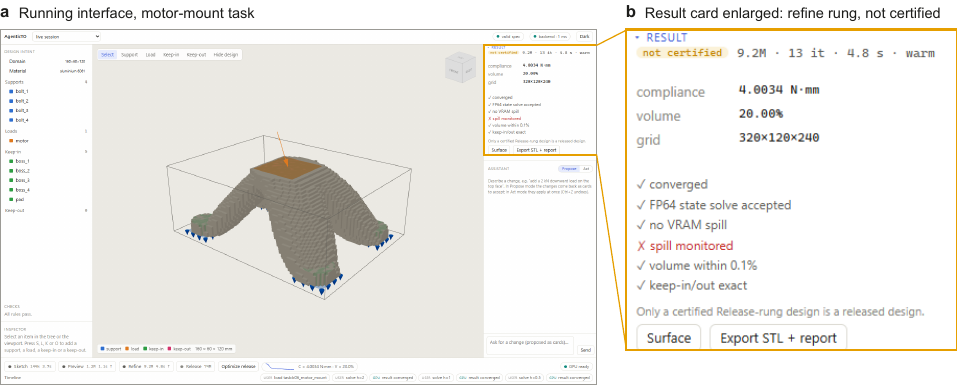}
\caption{Running graphical user interface (GUI) on the motor-mount task. (a) The viewport, specification inspector, resolution ladder, and result card expose the current problem and numerical status. (b) The result card for the 9.2-million-element refine rung, enlarged, from a separate GUI run of the task whose timing differs from the agent session in Figure~\ref{fig:workflow}: five predicates pass, but spill monitoring runs only on release rungs, so this intermediate field is shown as feedback and is not certified. A released result requires all six predicates of Eq.~\eqref{eq:certificate}. FP64 denotes 64-bit floating point; VRAM is video random-access memory; STL is the stereolithography surface format. The screenshot documents implemented software and provides no evidence of usability with human participants.}
\label{fig:gui}
\end{figure}

The implementation uses React~\citep{ReactSoftware} for the interface and three.js~\citep{ThreejsSoftware} for three-dimensional rendering. A FastAPI~\citep{FastAPISoftware} web backend connects it to session operations, and WebSocket connections deliver events as jobs progress. The backend uses NumPy and CuPy for array computation on the CPU and GPU, SciPy for connectivity and shape checks, and Pydantic to validate specifications against the schema \citep{Harris2020NumPy,Okuta2017CuPy,Virtanen2020SciPy,PydanticSoftware}. Matplotlib draws the setup pictures and the charts, and PyVista the design renders in the figures \citep{Hunter2007Matplotlib,Sullivan2019PyVista}. The viewport supports face, patch, and edge selection and placement of retained-solid and retained-void regions. An inspector exposes numerical values, while the ladder shows job progress and intermediate fields. A result card displays certificate checks and the actual release grid. Figure~\ref{fig:gui} shows the running interface.

The assistant has Propose and Act modes. Propose presents changes as action cards with a geometric overlay so that the engineer can inspect and accept them. Act applies edits and may start computations through the same tools. Clarification choices remain explicit in both workflows. Propose mode follows published guidelines for human--artificial intelligence (AI) interaction, which ask that a system's actions be easy to dismiss and easy to correct when it is wrong \citep{Amershi2019Guidelines}. Undo operations restore the session's earlier specification; a changed specification makes an earlier result stale rather than silently relabeling it as current.

The evaluated implementation is limited to this interaction set. It does not include imported STL design domains, brush editing, branching alternatives, comparison workspaces, parameter sweeps, or an automatically triggered sketch on every edit. These distinctions matter when interpreting the latency and action counts: they refer to implemented operations on the supported box-domain workflow.

\input{sections_evaluation_results}
\input{sections_discussion}

%% file: tables/t2_rules.tex
\begin{table}[t]
\centering\small
\caption{Verification rule base (deterministic; any error blocks solving and release)}
\label{tab:t2_rules}
\begin{adjustbox}{max width=\linewidth}
\begin{tabular}{lrrr}
\toprule
Code & Severity & Condition & Kind \\
\midrule
R001 & error & resolution cannot be met (grid too fine or too coarse) & spec \\
R002 & error & a support selects no nodes & spec \\
R003 & error & supports leave a rigid-body mode free & physics \\
R004 & error & a load selects no nodes, or every load is zero & spec \\
R005 & error / warning & load applied on fixed DOFs (dropped and recorded) & physics \\
R006 & error & no design elements remain after passive regions & spec \\
R007 & error & elements both keep-in and keep-out & spec \\
R008 & error & a support or load lies entirely inside a keep-out & physics \\
R009 & warning & smallest required feature not resolvable at this h & resolution \\
R010 & warning & a region was snapped to the grid & geometry \\
R011 & warning & a passive region contains no element center & geometry \\
R012 & warning / info & supports lost on coarse multigrid levels & solver \\
R013 & error & memory need exceeds the GPU & solver \\
\bottomrule
\end{tabular}
\end{adjustbox}
\par\smallskip\footnotesize DOFs: degrees of freedom; GPU: graphics processing unit; $h$: element edge length.
\end{table}

%% file: sections_evaluation_results.tex
\section{Evaluation design}
\label{sec:eval}
\subsection{Questions and study structure}
We evaluate whether the explicit records preserve numerical behavior, support agreement with a brief, and block inappropriate release. The evidence is organized by the object being tested (Table~\ref{tab:studies}). The studies have different units, and counts from different studies are not pooled.
\begin{table}[t]
\caption{Evaluation studies, their units and the question each answers}
\label{tab:studies}
\centering\small
\begin{tabularx}{\linewidth}{p{24mm}p{34mm}X}
\toprule
Study & Unit and scale & Main question \\
\midrule
Solver verification & Three identity problems; two independent cross-checks & Does the specification layer preserve numerical behavior? \\
Brief interpretation & 50 briefs per primary condition & Which readings are right, wrong but delivered, or refused? \\
Numerical continuation & Five completed problems; two common-grid paths & What changes with warm starts and finer grids? \\
Acceptance checks & 486 recorded solves; separate fault injections on 47 references & Which failures do rules and certificates expose? \\
Complete workflow & 15 sessions: 14 releases and one correct refusal; latency on the 14 releases & Do release, replay and warm interaction work together? \\
\bottomrule
\end{tabularx}
\end{table}
The supplement maps these studies to records and scripts. It separately retains development pilots and runs superseded after an oracle correction.

\subsection{Brief suite and reference readings}
\label{sec:suite}
The suite contains 50 briefs: 40 fully stated solve tasks, seven requiring clarification and three requiring refusal (Table~\ref{tab:t1_suite}). Half describe engineering-style parts and half derive from benchmark families. Only six benchmark attributions were confirmed; the supplement distinguishes adaptations, untraced attributions and generic examples, and supplies every brief. Briefs are identified as b01 to b50 in the figures and the supplement. Each task has a reference specification, expected response class and, where needed, clarification facts and accepted alternative readings.

The schema supports one load case. Nine references combine force components on one region as simultaneous loading, which the brief need not establish. The supplement reports the sensitivity excluding those nine tasks. After unit normalization, the comparator checks the two specifications on a common grid, including supports, nodal loads and passive regions. It accepts equivalent geometric syntax, not arbitrary geometric changes. Every credited record whose reference states an element size was compared at that size; this discrete equivalence is not continuous geometric identity.

The references were drafted with Claude Opus 5.5 during system development, and the 50-brief suite was fixed before the benchmark. J.W. adjudicated all references and 41 contested outputs. S.Y. repeated that original adjudication blind to condition, provider, messages and first verdict, agreeing on 33/41 contested records (Cohen's $\kappa=0.60$ \citep{Cohen1960}). Scores use J.W.'s verdicts. Later conditions were adjudicated by J.W. alone with condition visible. These author judgments are not independent ground truth. The supplement preserves the changes, alternative scores, pilot/oracle history and complete comparison tolerances.
\input{tables/t1_suite}

\subsection{Conditions, providers, and protocol}
\label{sec:conditions}
Every language condition emits the same typed specification. The primary comparison uses:
\begin{description}
\item[R: rule parser.] Deterministic extraction without a language model.
\item[T: text only.] A model emits JavaScript Object Notation (JSON) from the brief and schema; refusal is allowed.
\item[G: grounding.] The agent builds, validates and renders through tools, assuming missing facts rather than asking.
\item[G+C: clarification.] G may ask the scripted reference-aware oracle.
\item[G+C+V: picture check.] The same G+C output receives three-view review and possible repair.
\end{description}
The four follow-ups isolate selected changes: T+S adds schema-feedback retries; GNR removes rule feedback; G+P records each value's source; G+C+P permits questions only about items tagged assumed. Full prompts, tool availability and repair rules are in the supplement.

G+C+V is an incremental check, not an independent fresh run. Comparing G+C with G measures permission and instructions to ask against required guessing; it does not isolate an interface effect. The oracle selects a reference-matching option or supplies stored facts if none matches (Section~\ref{sec:clarify}). It does not depend on the consequence images.

The primary provider is OpenAI's Codex command-line interface (CLI) with GPT-6.1-Sol, and the second is Anthropic's Claude Code CLI with Claude Opus 5.5, both at nominal high reasoning effort. These client settings are not assumed equivalent. The clients used subscriptions, not separately billed application programming interface (API) calls; tokens and wall times are recorded without an inferred API price. Dates, versions and the smaller default-effort diagnostic are in the supplement. Primary paired tests use the first repeat, one observation per task and condition. Unequal, overlapping repeat subsets are reported separately as run variation, not additional independent tasks.

\subsection{Outcomes and statistical comparisons}
\label{sec:metrics}
Right behavior means a reference-equivalent specification on an answerable task, or refusal on a task designated unsolvable. A schema-invalid output, tool failure or refusal on an answerable task is a failure. Primary first-repeat rates use all 50 briefs, including the three refusal tasks; subset repeats use their stated denominator. The secondary matching-specification count credits an equivalent specification even if the system later refuses. Delivered wrong counts schema-valid specifications returned as finished on non-refusal tasks that fail the reference comparison; an assumption note does not make such a result right.

Rates receive 95\% Wilson score intervals \citep{Wilson1927}. Paired comparisons use the same task in each condition, following McNemar's paired design \citep{McNemar1947}. If $b$ tasks are corrected and $c$ are broken by the second condition, we calculate a two-sided exact binomial-tail probability:
\begin{equation}
 p=\min\left\{1,\;2\sum_{j=0}^{\min(b,c)}\binom{b+c}{j}2^{-(b+c)}\right\},
 \label{eq:mcnemar}
\end{equation}
with $p=1$ when $b+c=0$. Under the null of equal marginal outcome probabilities, each discordant task has probability one half of favoring either condition. Reported $p$ values are unadjusted across the displayed comparisons, which should be read as component analyses of this suite. The intervals do not quantify uncertainty in the reference readings or generalization to an independently sampled industrial population.

Consequence-weighted error evaluates the predicted design under the intended problem, divides compliance by the reference intended-design compliance and subtracts one. We report the large consequence values and material-budget changes separately: an invented solid can improve stiffness by using more material. This is distinct from specification agreement.

\subsection{Numerical settings and timing boundaries}
\label{sec:hardware}
The workstation uses an RTX 4090 with 24 GB VRAM, an i9-13900KF and 96 GB host memory. Shared desktop GPU use caused two guarded 113.2-million-element attempts to stop on spill; they are excluded from released scale. The inherited solver uses structured hexahedra, matrix-free operators, multigrid state solves and density filtering/projection. Release runs check spill every ten iterations. The five completed release-policy comparisons use a 300-iteration cap. The supplement gives schedules, memory context, failed duct-bridge attempts and failed constant-physical-filter pilots.

Compute is recorded solver time. Original session time spans the first logged call through final export and excludes CLI startup; CLI wall time includes it. Padding-aware totals combine measured numerical replays with original agent and other session time, rather than new live agent runs. Backend interaction latency excludes browser frame scheduling. These clocks answer different questions and are kept separate.

\section{Results}
\label{sec:results}
\subsection{The session extensions preserve the inherited solver and replay exactly}
\label{sec:verification}
Without passive regions, the session solver reproduced the inherited SIMP implementation on three verification problems with the same iteration count, compliance agreement within $10^{-9}$, and maximum density difference zero. This identity check uses the atomic-free Jacobi configuration. It establishes that the tested session extensions preserve that path; it does not establish bit-exact identity for arbitrary upstream multigrid reductions.

Two specification-built problems, including passive regions, were independently assembled with scikit-fem~\cite{Gustafsson2020skfem} and solved with SciPy's sparse direct solver~\cite{Virtanen2020SciPy}. Relative compliance differences were $8.64\times10^{-11}$ on a 16,000-element passive cantilever and $5.73\times10^{-10}$ on a 50,000-element L-bracket. The inherited solve was accepted at true-residual tolerance $10^{-8}$ in this cross-check. These fixed-problem comparisons test assembly, boundary conditions, and physical scaling. They are not validation against a physical experiment or a large-scale mesh-convergence study.

Two initial scripted sessions replayed their recorded specification, render, and solve outcomes. In the complete-session study, all 15 original sessions also replayed with no reported mismatch under their recorded deterministic configuration. Replaying the original log demonstrates a stronger execution property than saving an agent's final text, while leaving the quality of the recorded decisions to the other evaluations.

\subsection{Right outcomes rose under grounding; oracle-answered clarification set an upper bound of 50}
\label{sec:grounding}
\begin{figure}[t]
\centering\includegraphics{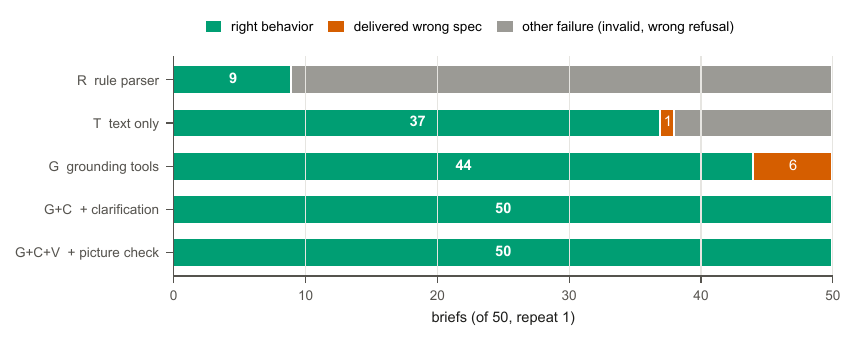}
\caption{Outcomes on 50 briefs for the primary provider, first repeat. R: rule parser; T: text-only specification; G: grounding tools with no questions; G+C: grounding and clarification; G+C+V: the same G+C result followed by a three-view picture check. Right behavior includes correct refusal; delivered wrong denotes a valid specification returned as finished that fails the adjudicated reference comparison.}
\label{fig:x2}
\end{figure}

Table~\ref{tab:t4_x2_conditions} and Figure~\ref{fig:x2} summarize the first repeat. The rule parser constructed six matching specifications and achieved nine right-behavior outcomes out of 50. T constructed 34 matching specifications and achieved 37 right outcomes, while G achieved 44 on both counts. G+C and G+C+V each achieved 50 right outcomes under the adjudicated reference readings. Their Wilson interval is approximately 93--100\%, which is compatible with performance below 100\% even before considering reference uncertainty.

The T-to-G comparison corrected eight tasks and broke one ($p\approx0.039$); G-to-G+C corrected six and broke none ($p\approx0.031$). These descriptive paired tests depend on few discordant tasks and on the author-adjudicated references. Before adjudication, T-to-G corrected seven and broke two ($p\approx0.18$), and every changed verdict moved from wrong to right. Under S.Y.'s verdicts, the G-to-G+C test is $p\approx0.063$, although G+C and G+C+V still reach 50. A single reversed verdict can move either primary test above 0.05. Full paired counts, secondary-specification tests and adjudication sensitivities are in the supplement.

Grounding changed the failure profile. Of the eight T failures it corrected, six were schema-invalid outputs and two were refusals of clarification briefs. Yet G delivered six wrong specifications as finished, compared with one for T. A higher right-outcome count therefore coexisted with more wrong delivery. G differs from T in tool-based construction, rule feedback, rendering and instructions to assume rather than refuse; only rule feedback was ablated. It also used about four times the median input tokens, without a budget-matched T condition (Table~\ref{tab:t4_x2_conditions}). These comparisons do not isolate a single grounding component. The supplement gives the task-class breakdown and the assumption-note diagnostic.

Schema feedback alone recovered most of the grounding gain: T+S was right on 42/50, compared with 37/50 for T and 44/50 for G. The G comparison therefore cannot isolate geometric inspection, rendering or rule feedback. The supplement reports the paired tests and output diagnostics.

All six wrong G specifications passed every rule. Five guessed a missing fact because G could not ask; one added unrequested support bosses to a fully stated brief. The paired G+C condition was right on those same tasks against oracle answers, reaching 50/50, but asked 59 questions: 14 on clarification tasks and 45 on 26 fully stated tasks. Five corrections therefore compare required guessing with permission to ask; only one correction concerned a fully stated task. This is a cooperative-oracle upper bound, not demonstrated performance with engineers.

The 45 exchanges on fully stated briefs were not classified by whether the brief alone supplied the answer. The oracle could supply stored facts missing from the offered choices. Thus the comparison measures permitting and using questions, not the causal benefit of clarification visualization or any one grounding component.

The follow-up controls did not provide a reliable error detector. Removing rule feedback left the three refusals intact and did not improve right outcomes. Value-source tags marked all eight G+P errors but also most right specifications. Restricting questions to tagged assumptions reduced the burden to six questions, none on fully stated briefs, but lost three G+C outcomes. The picture check changed no primary-provider outcome after G+C was already right on every brief. Table~\ref{tab:t4_x2_conditions} and the supplement give all counts and costs. These controls bound the primary result; they do not identify a general accuracy gain from rules, tags or images.

Repeats retained the protocol differences but used unequal, overlapping task subsets. The supplement reports each repeat separately; pooling them into a larger independent test set would overstate the evidence.
\input{tables/main_t4_x2_conditions}

Wrong readings also had unequal mechanical consequences. One G error increased compliance by about 164\%; another left effectively no load path under the intended loading, with a consequence ratio of order $10^8$. Three G outputs used about 3--10\% more material than intended and appeared 0.7--2.5\% stiffer, so these negative compliance changes were not improvements under the original budget. A single mean consequence score would conceal intent errors and resource violations. The supplement gives all primary and follow-up cases.

\subsection{The second provider asked fewer questions and retained four errors}
\label{sec:providers}
The second provider reached 42/50 right outcomes for T, 46/50 for G+C and 47/50 for G+C+V (Table~\ref{tab:t7_providers}). G+C asked seven questions rather than 59 and delivered four wrong specifications. Two guessed a load on a clarification task; two mixed kilonewtons and millimeters with a modulus left in megapascals. Only the first two could have been resolved by asking about the missing fact. The four discordant G+C outcomes favor the primary provider, but the paired comparison is not significant ($p\approx0.12$).

Adjudication matters most for second-provider text-only outputs: their right count is 39 under S.Y.'s verdicts, 37 when both authors must accept a reading, and 44 when either may. The supplement gives the paired tests, modulus-range diagnostic and default-effort subset. One model per provider and one complete second-provider repeat cannot establish equivalent effort settings, provider-independent accuracy or a general capability ranking.

\input{tables/main_t7_providers}

\subsection{Rules and certificates caught different failures}
\label{sec:gates}
Table~\ref{tab:gates} separates failure counts by their own denominators. Of 54 recorded release attempts, 52 carried passing certificates; two did not supply releasable results. The full census contains 486 solves and 11 non-converged computations. Three separately documented spill stops enforce resources, including two at 113.2 million elements; they are not successful releases. These counts demonstrate gate operation, not an estimated false-acceptance rate.

Removing formulation validation exposed the complementary failure. Two rigid-body-deficient briefs failed in the solver. A load inside keep-out elements instead converged and passed the numerical certificate, at about $3.8\times10^8$ times the corrected reference compliance: the void's small artificial stiffness made the contradictory problem solvable. The formulation rule blocks it before solving. This deliberately constructed case shows a mechanism, not its frequency. On the brief benchmark, rule feedback caused no measured repair; removing it left all three refusals unchanged (Section~\ref{sec:grounding}).

Injected faults show the same boundary. We changed one feature at a time in the 47 references that passed every rule. The rules caught 145 of 148 faults constructed to match their own rule classes, but only 21 of 226 intent-changing faults. They also missed the tested partial keep-out loads and modulus unit slips outside those classes. These are conformance checks on the suite's own references, not a population false-acceptance estimate. The supplement gives every fault class and denominator. It also reports an exploratory strain-energy check on the solved keep-out field; that post hoc predicate is not part of the reported certificate, and one contradictory case cannot define its threshold.

The two gates are therefore complementary but insufficient for intent. A certificate tests the defined computation; the rules test selected conditions on its formulation. All six wrong G specifications passed the rules, and their converged consequence evaluations show why numerical acceptance cannot substitute for comparing a specification with the brief. In these runs only oracle-backed clarification prevented those errors. The hypothetical acceptance combinations and their distinct denominators are detailed in the supplement.

\begin{table}[t]
\centering\small
\caption{Observed failures at different acceptance boundaries. Counts have distinct denominators and are not combined into a false-acceptance rate.}
\label{tab:gates}
\begin{tabularx}{\linewidth}{p{22mm}p{30mm}Xp{30mm}}
\toprule
Boundary & Evidence & Observed outcome & Stopped by \\
\midrule
Intent & G on 50 briefs & Six wrong specifications passed the deterministic rules. & No in-system check; oracle-answered clarification (G+C) \\
Formulation & Three rule-bypassed refusal tasks & Two rigid-body cases failed numerically; the load-in-void case converged and was numerically certified. & Rules (all three); without them, the certificate stopped two \\
Numerical release & 54 recorded release attempts & 52 certified; two supplied no releasable result. & Certificate \\
Convergence & Census of recorded solves & Eleven finished without converging and remained unreleasable. & Certificate; the rules had passed them \\
Resource use & Three separately documented stops & Spill guard stopped two 113.2M runs and one 59M evaluation. & Execution predicates \\
\bottomrule
\end{tabularx}
\end{table}

\subsection{Fourteen of 15 sessions reached certified releases}
\label{sec:sessions}
Fourteen complete agent sessions exported an STL and report with a passing certificate and a specification matching the reference. The remaining hinged-arm task was correctly refused after a rigid-body rule error. The 15 briefs were author-selected, and sessions were single primary-provider runs except for a cantilever rerun after four exposed software faults were fixed. The supplement reports the selection and development history.

\begin{figure}[t]
\centering\includegraphics{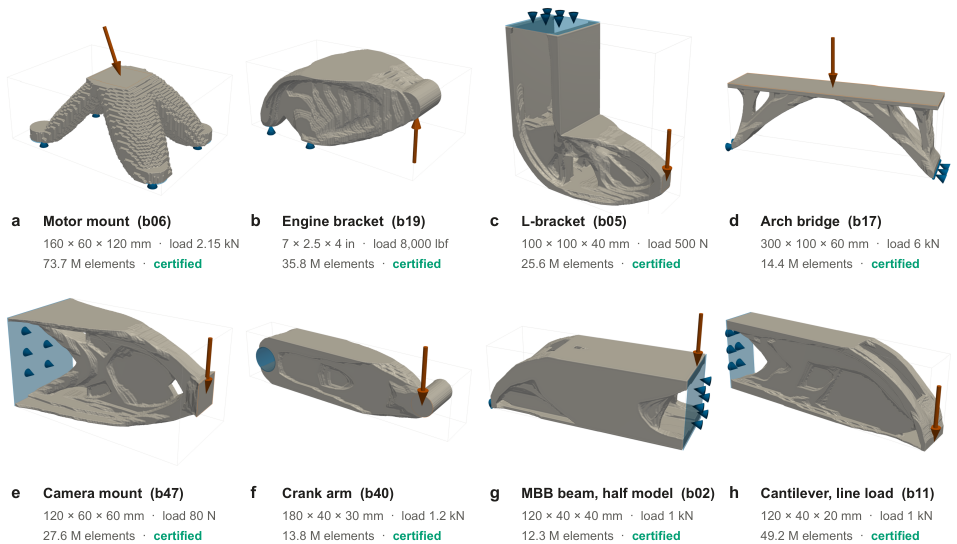}
\caption{Released designs from eight complete agent sessions. Each part is the $\rho=0.5$ surface exported at the session's release rung, which passed the numerical acceptance checks (certified), lightly smoothed for display. Blue marks restrained regions: cones touch the restrained surface but do not show which displacement components are fixed, and the translucent cylinder in (f) is a restrained bore. Vermillion arrows show the load resultant on its region; the grey box is the design domain. Labels give the domain, the load resultant, and the release grid. The motor mount (a) was released with the tail schedule (12 iterations per warm rung) and the other sessions with the long re-opening (61 iterations on the release rung). The camera mount (e) is a clarification task; MBB denotes the Messerschmitt--B\"olkow--Blohm benchmark beam. Certificates apply to the solved density fields, not to stress, buckling, or manufacturing performance.}
\label{fig:designs}
\end{figure}

Thirteen releases used long re-opening; the motor mount used the tail schedule and certified a transferred design. All release grids were two to eight times finer in element edge length than the matching specification stated, shortening the physical filter length by the same factor. Session success thus means a certified release derived from a matching specification, not a release validated at the stated resolution. The reconstructed certificate fields expose this difference. All 15 original logs replayed exactly under the recorded configuration. Release sizes were 12.3--73.7 million elements; Figure~\ref{fig:designs} shows eight parts and the supplement all 14.

The original 14 released sessions took 77--1877 s from first tool call to final export, with median 179 s. CLI wall time was 93--1898 s, with median 199 s. The difference includes startup and execution outside the logged session. Keeping both figures prevents changing the timing boundary to fit a minutes-scale description.

\begin{figure}[t]
\centering\includegraphics{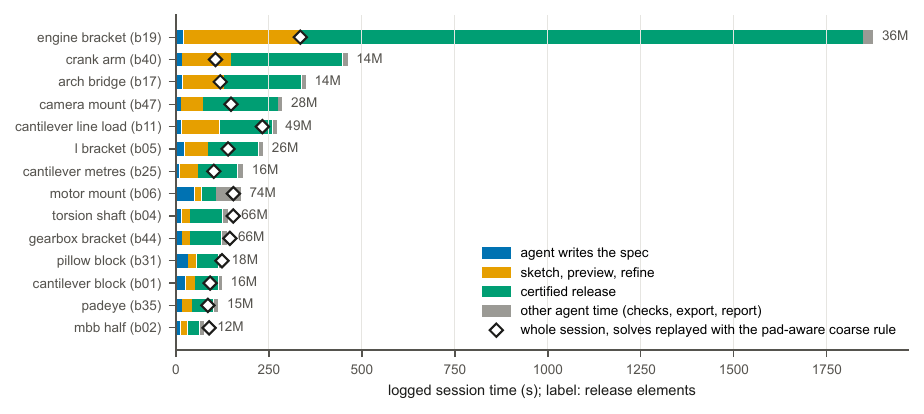}
\caption{Original logged session durations for the 14 released tasks, partitioned into specification, intermediate solves, release, and other agent work. Diamonds replace original solve durations with padding-aware replay durations while retaining original agent time; they are reconstructed estimates, not new live sessions. Element labels are rounded. Command-line interface startup is excluded from this plot and reported separately in the text.}
\label{fig:time}
\end{figure}

Padding caused the largest timing outlier. A matched warm-start investigation reduced the engine-bracket release compute from about 1508 s to about 212 s. Replaying every recorded solve with the padding-aware policy, while retaining original agent and other time, gave reconstructed sessions of 86--334 s, median 132 s (Figure~\ref{fig:time}). All 14 remained certified, with maximum absolute relative compliance change about $1.01\times10^{-5}$. These are numerical replays with original language-model time, not new live agent sessions. The supplement gives solver counters and per-task timing boundaries.

The longest agent session also used about 2.96 million input tokens, compared with roughly 0.42--1.18 million on the other released tasks. Repeated status and solve-related interactions can increase context accounting along with wall time. This is one reason to report orchestration, solver time, and token use separately. The per-task session table, policy-replay distinction, and release gallery are provided in the supplement.

\subsection{Long re-opening approached cold compliance at lower final-rung cost}
\label{sec:escalation}
\begin{figure}[t]
\centering\includegraphics{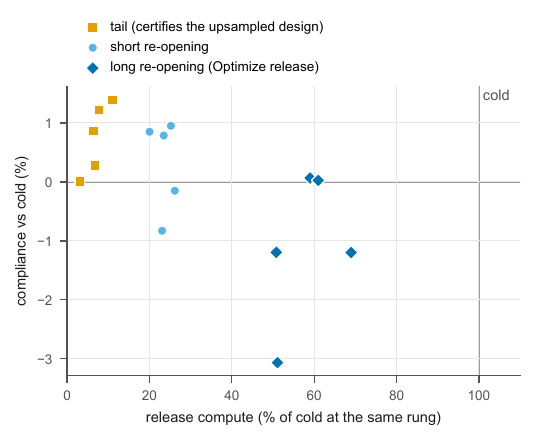}
\caption{Warm release policies relative to cold fine-grid runs on five completed problems. Compliance differences and compute fractions refer to the final release rung; previous rungs are excluded. Tail continuation mainly certifies the transferred design, whereas long re-opening re-optimizes it. The bridge cold reference required a different successful solver configuration. All plotted final results passed their numerical certificates.}
\label{fig:arms}
\end{figure}

On five completed problems, all final arms passed their certificates at 25.6--73.7 million elements (Table~\ref{tab:t3_x1_escalation}). Long re-opening differed from a single cold run by $-3.06\%$ to $+0.07\%$ compliance at 51--69\% of cold final-rung compute; negative values mean lower compliance. The tail used 3--11\% at $+0.005\%$ to $+1.39\%$ compliance but changed essentially no binary design, chiefly certifying transferred topology (Figure~\ref{fig:arms}).

Including prior warm rungs raises the long re-opening share to 59--80\%, or 61--84\% with preview. These are compute comparisons, not empty-session speedups. Each arm has one run, warm starts and shorter schedules are confounded, and regularization histories differ. The bridge cold reference required a different successful solver. Failed duct-bridge and physical-filter pilots are reported in the supplement. The cold runs are local references; hundredths-of-a-percent differences are best read as ties.

\input{tables/main_t3_x1_escalation}

\subsection{Warm and cold paths differ when judged on a common part model}
\label{sec:resolution}
For a span bridge and a motor mount, every candidate was transferred to the finest grid, binarized to match the design-material fraction and evaluated under the same finest-grid state model. The bridge warm path ended at 2382 versus 3333 N\,mm for cold, 28.5\% lower, at about 413 versus 466 s. The motor mount ended at 3.915 versus 3.912 N\,mm, about $+0.09\%$, at 315 versus 466 s (Figure~\ref{fig:resolution}; all rungs in the supplement).

The coarse optimizer under-predicted its resulting part's finest-grid compliance by 0.25--12.3\%. This combines analysis resolution, projection, volume-matched binarization and, for the bridge, edge-support singularity. Element-based filters also give warm and cold paths different physical regularization histories. These two examples support checking the actual reported part model; they are not mesh-convergence estimates, a general coarse-start advantage or manufacturing validation.

\begin{figure}[t]
\centering\includegraphics{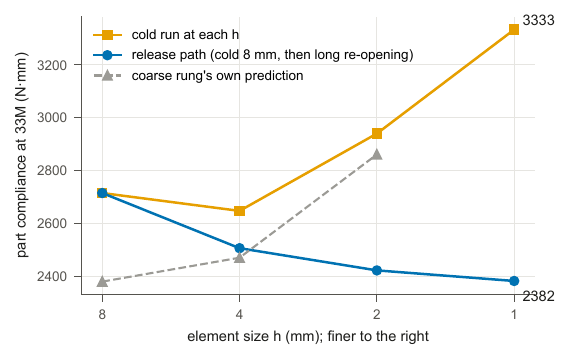}
\caption{Span-bridge resolution paths judged by a common 32.77-million-element part evaluation. Each density is upsampled and binarized at the prescribed design-material fraction before its compliance is evaluated. The warm path improves monotonically on this example; the cold fine run is a local reference rather than a certified optimum. The motor-mount results are tabulated in the supplement.}
\label{fig:resolution}
\end{figure}

\subsection{Warm interactions met the latency target; cold sketches often missed it}
\label{sec:interface_results}
The backend latency study uses the 14 released tasks. Validation, edit commit, and edit-to-warm-sketch interactions were repeated 20 times per task; the complete ladder was run once per task. Per-task 95th-percentile validation latency ranged from about 1.2 to 10.5 ms. Per-task 95th-percentile edit-to-warm-sketch latency ranged from 0.59 to 2.40 s, meeting on all tasks our own three-second upper target for warm feedback, which is not taken from a usability standard. These measurements exclude frontend frame scheduling and are not a guarantee about every future desktop workload.

Cold sketches missed the target more often: only four of 14 finished within three seconds, and durations ranged from approximately 2.5 to 6.7 s. A first density frame generally arrived much earlier, while completion required the cold continuation schedule. Preview and refine also had outliers that reached their iteration caps. In the separate latency study, default warm releases took approximately 3.1--56.4 s of compute, and all 14 were certified; these runs used tail continuation and must not be described as cold high-resolution optimizations.

The Keystroke-Level Model (KLM) estimates error-free expert interaction time by summing keystroke and button-press, pointing, homing, and mental-preparation operators \citep{Card1980KLM}. The modeled median setup-and-release interaction time was 480 s for a solver script, 208 s for a typed structured specification, 104 s for GUI-only setup, and 25 s for the GUI with an existing brief pasted into the assistant. If the entire brief had to be typed, the assistant condition increased to 141 s. Solver and agent waiting are excluded from all those model totals. The agent path includes a mental operator for each proposed change card and a click to accept them all, but not the time an engineer needs to check the specification against the brief, which the agent path requires.

The pasted-brief result therefore concerns removing manual translation of an already written requirement. It is not a measured reduction in elapsed task time, and typing a new brief can cost more than direct GUI entry on simple tasks. Every generated scripting setup was executed and checked against the problem builder, providing a functional basis for its operator count. Nevertheless, errors, debugging, learning, trust calibration, and perceived workload are outside this model.

Replaying the G+C specifications as proposal cards yielded 423 cards, all accepted unchanged by the scripted oracle against the reference readings. G yielded cards requiring edits or rejection, consistent with its wrong session outcomes. Oracle acceptance is a deterministic agreement result, not a human approval rate. An automated GUI test passed all 90 recorded checks on small meshes. That verifies exercised software interactions; it does not substitute for a participant study.

%% file: tables/t1_suite.tex
\begin{table}[t]
\centering\small
\caption{Task suite (50 briefs)}
\label{tab:t1_suite}
\begin{adjustbox}{max width=\linewidth}
\begin{tabular}{lr}
\toprule
Property & Value \\
\midrule
Briefs & 50 \\
Expected behavior & clarify 7, refuse 3, solve 40 \\
Unit systems (length, force) & mm, N (46); in, lbf (1); m, kN (1); m, N (1); mm, kN (1) \\
Support and load region kinds & patch 80, face 41, edge 15, cylinder 8 \\
Briefs with passive regions & 19 (solid 17, void 9 regions) \\
Clarify tasks: stored oracle facts per brief & 2, 1, 2, 1, 1, 1, 1 \\
Briefs with credited alternative readings & 14 \\
Source of the problem & engineering brief 25; benchmark-modeled 25 (family confirmed 6, source untraced 6, generic 13) \\
\bottomrule
\end{tabular}
\end{adjustbox}
\end{table}

%% file: tables/main_t4_x2_conditions.tex
\begin{table}[t]
\centering\small
\setlength{\tabcolsep}{4pt}
\caption{Specification grounding on 50 briefs: primary provider, first repeat. Before adjudication, the T-to-G comparison corrected seven tasks and broke two ($p\approx0.18$).}
\label{tab:t4_x2_conditions}
\begin{tabular}{lrrrrrr}
\toprule
Condition & \shortstack{Matching\\spec.} & \shortstack{Right behavior\\(95\% interval)} & \shortstack{Delivered\\wrong} & Questions & \shortstack{Input\\tokens} & \shortstack{Wall\\(s)} \\
\midrule
R & 6 & 9 [10--31\%] & 0 & -- & -- & -- \\
T & 34 & 37 [60--84\%] & 1 & -- & 20k & 10 \\
G & 44 & 44 [76--94\%] & 6 & -- & 86k & 28 \\
G+C & 50 & 50 [93--100\%] & 0 & 59 & 104k & 37 \\
G+C+V & 50 & 50 [93--100\%] & 0 & 59 & 135k & 49 \\
\midrule
T+S & 39 & 42 [71--92\%] & 4 & -- & 20k & 9 \\
GNR & 42 & 42 [71--92\%] & 8 & -- & 87k & 20 \\
G+P & 42 & 42 [71--92\%] & 8 & -- & 106k & 33 \\
G+C+P & 47 & 47 [84--98\%] & 3 & 6 & 107k & 33 \\
\bottomrule
\end{tabular}
\par\smallskip\footnotesize R: rule parser; T: text only; G: grounding; C: clarification; V: added picture check. Below the rule, follow-up conditions run after the suite: S: schema feedback; GNR: G without rule feedback; P: value-source tags (with C, questions only about items tagged assumed). Their contested outputs were adjudicated afterwards by the same rule. Intervals are Wilson confidence intervals \citep{Wilson1927}. Tokens and wall time are medians. Scores use the adjudicated reference readings.
\end{table}

%% file: tables/main_t7_providers.tex
\begin{table}[t]
\centering\small
\setlength{\tabcolsep}{4pt}
\caption{Two providers at nominal high effort: 50 briefs, first repeat.}
\label{tab:t7_providers}
\begin{tabular}{llrrrrr}
\toprule
Condition & Provider & Right & \shortstack{Delivered\\wrong} & Questions & \shortstack{Input\\tokens} & \shortstack{Paired $p$\\vs Codex} \\
\midrule
T & Codex & 37 & 1 & -- & 20k &  \\
T & Claude Code & 42 & 8 & -- & 9k & 0.18 \\
G+C & Codex & 50 & 0 & 59 & 104k &  \\
G+C & Claude Code & 46 & 4 & 7 & 44k & 0.12 \\
G+C+V & Codex & 50 & 0 & 59 & 135k &  \\
G+C+V & Claude Code & 47 & 3 & 8 & 59k & 0.25 \\
G+C+P & Codex & 47 & 3 & 6 & 107k &  \\
G+C+P & Claude Code & 47 & 3 & 5 & 95k & 1 \\
\bottomrule
\end{tabular}
\par\smallskip\footnotesize T: text only; G+C: grounding and clarification; G+C+V: added picture check; G+C+P: questions only about items tagged assumed (a follow-up condition). Model identifiers are specified in Methods. Token counts are medians. Scores use the adjudicated reference readings.
\end{table}

%% file: tables/main_t3_x1_escalation.tex
\begin{table}[t]
\centering\small
\setlength{\tabcolsep}{4pt}
\caption{Warm release policies relative to cold fine-grid runs. Every listed final result was certified.}
\label{tab:t3_x1_escalation}
\begin{tabular}{lrrrrrrr}
\toprule
Problem & \shortstack{Elements\\(million)} & \shortstack{Cold\\(s)} & \shortstack{Tail\\ $\Delta C$} & \shortstack{Tail\\compute} & \shortstack{Long\\ $\Delta C$} & \shortstack{Long\\compute} & \shortstack{Long\\IoU} \\
\midrule
Cantilever & 65.5 & 118 & +0.28\% & 7\% & -1.20\% & 69\% & 0.96 \\
Bridge & 65.5 & 249 & +1.39\% & 11\% & -1.19\% & 51\% & 0.88 \\
Torsion & 65.5 & 121 & +0.86\% & 6\% & +0.07\% & 59\% & 0.96 \\
L-bracket & 25.6 & 251 & +0.00\% & 3\% & -3.06\% & 51\% & 0.86 \\
Motor mount & 73.7 & 423 & +1.22\% & 8\% & +0.03\% & 61\% & 0.96 \\
\bottomrule
\end{tabular}
\par\smallskip\footnotesize $\Delta C$: compliance relative to cold at the same grid. Compute is the final-rung share of cold; previous rungs are excluded. Long: long re-opening. IoU: intersection over union with the cold binary mask, pooled for large grids. The bridge cold run required a different successful solver configuration.
\end{table}

%% file: sections_discussion.tex
\section{Discussion}
\label{sec:discussion}
\subsection{Inspectable records as a design position}
The main system result is a division of responsibility that can be inspected in the artifacts. The specification records what problem is being solved. Rules decide whether selected prerequisites hold. The solver certifies its numerical outcome. The provenance log connects these objects to the actions that produced them. This division lets the engineer ask a concrete question about a released result: which specification, grid, solver settings, and acceptance checks support it?

The load-in-void case shows why all those objects matter. A finite stiffness floor can make a contradictory problem numerically solvable. An accurate residual then indicates that the solver has addressed that numerical problem, even when the resulting structure has no useful interpretation under the brief. Similarly, the grounding benchmark shows that a rule-clean specification can contain an invented retained region. Improving one layer does not remove the need to inspect the others.

The release label should therefore be read as a conjunction of explicit checks within a defined software contract. It is not an assurance that the entire engineering problem has been exhaustively checked. The rule set is finite, intent references are incomplete, and the physical model covers only one part of structural performance. Keeping the certificate's predicates visible supports informed use of a result and avoids delegating final authority to a fluent agent explanation.

\subsection{Clarification benefit and question burden}
Within the current suite, permitting clarification changed the primary provider's behavior in a useful way: all six remaining G misreadings, five of them on briefs with a design-changing gap, were corrected without breaking a credited outcome. However, it also produced 45 questions on fully stated tasks. A deployed system would need a policy for distinguishing a material uncertainty from a detail that the brief already resolves. Limiting questions to the items that the agent tagged assumed was one such policy: it removed the questions on fully stated briefs but lost three right outcomes, because the agent did not ask about items it had itself tagged assumed (Section~\ref{sec:grounding}). Otherwise an engineer can exchange the burden of numerical setup for the burden of repeatedly confirming explicit information.

The second-provider result suggests that this policy is partly behavioral. The same tool and similar instructions did not induce the same frequency of questioning. A stronger application could require explicit resolution of specified missing fields or track unanswered assumptions as state, rather than depending entirely on an agent's willingness to ask. Tracking assumed values as state is the subject of assumption-based truth maintenance \citep{deKleer1986ATMS}, and stating what each record assumes and guarantees would bring the knowledge model of Table~\ref{tab:knowledge} closer to contract-based design \citep{Benveniste2018Contracts}; the present records formalize neither.

Inspecting the specification did not reliably reveal an invented reading. Free-text notes mentioned an assumption in five of the six G errors but also in 19 of the 41 right G specifications, and typed source tags marked every wrong G+P specification but also most right ones (Section~\ref{sec:grounding}).

Permission to ask about tagged items did not lead the agent to ask about them; a requirement to resolve every design-changing item tagged assumed before release remains untested, and it would itself need evaluation, because indiscriminate mandatory questioning could increase friction without improving the specification.

The oracle also limits what is established. It compares options with a stored reference, normalizes units, and can supply facts when none matches. Real engineers may be uncertain, inconsistent, or unable to map a verbal requirement to one of the proposed choices. Consequence previews might help, but the current oracle ignores them when choosing. Testing that interface mechanism requires participants or a different controlled experiment. The present results support the asking-and-answering protocol under the stated oracle, rather than a causal benefit from the consequence images.

\subsection{Scale, optimization history, and response time}
Fine resolution has two roles in this workflow. It permits smaller geometric features, and it evaluates the transferred design under the reported discretization. These roles should not be conflated. A tail release can provide the latter while preserving the coarse topology. A long re-opening changes more of the design and approaches or improves the cold reference on the five completed examples, yet remains dependent on the transferred basin of attraction.

The bridge resolution path emphasizes that dependence. Its finest cold run had worse part compliance than several coarse-derived candidates evaluated on the same fine model. More elements did not cure an unfavorable optimization trajectory. A useful next step is to select among several starts or paths with a fixed total budget, including failed attempts, rather than treating one cold fine result as an optimal reference. The present comparisons retain cold runs as reproducible local baselines.

Interactivity also has several time scales. Validation operates in milliseconds; warm sketches usually return in seconds; high-resolution re-optimization and complete agent workflows can require minutes. The interface supports progressive feedback across those scales. Calling the entire release process real time would hide cold-sketch misses, padded-grid outliers, language latency, and the difference between tail certification and re-optimization.

Shared desktop memory is part of that operating context. The spill guard prevented a nominal 113.2-million-element target from becoming a valid result, and the largest supported release remained 73.7 million. This is evidence that the resource gate acted, not that 100 million elements were achieved under the release contract. A dedicated GPU environment might change the feasible scale, but it was not tested here.

\subsection{Implications for engineering informatics}
For engineering informatics, the practical result is a way to inspect how an agent's proposed problem became an accepted numerical result. The specification records the proposed reading, the rules state which formulation checks passed, and the certificate records acceptance of the solved field. Their authority is explicit, so an agent's explanation cannot substitute for a missing check.

This division is a design convention for linear-elastic compliance, not a derived necessity. Its boundaries overlap: the rules include faithfulness and execution checks, and the certificate includes execution predicates. The evaluation tests those boundaries rather than proving the architecture superior. Rule removal supplied one constructed contradiction that passed numerical acceptance; fault injection showed that the rules caught faults of their own classes but few intent-changing faults. Removing rule feedback changed no benchmark outcome through a finding. These are existence examples and conformance checks, not estimates of general false-acceptance rates.

Most importantly, the separate records made the grounding protocol's six wrong readings comparable with a reference but did not single them out. Only clarification against the reference-holding oracle prevented them in the primary-provider runs. Inspectability therefore supports review of intent without establishing that intent has been verified against the brief. A solved-field strain-energy predicate separated one constructed contradiction from valid problems (Section~\ref{sec:gates}), but is not part of the reported certificate.

As an untested hypothesis, the same division could suit other agent-driven analyses in which a converged computation can answer the wrong question, such as finite-element setup, flow simulation, or building-model checking. The present evidence covers linear-elastic compliance on voxel domains with one solver family. Transfer to another domain would require its own formulation rules, acceptance predicates, and evaluation.

\subsection{Threats to validity and remaining scope}
Reference quality is the main threat. The author-built suite and accepted alternatives were adjudicated by the authors, not independent engineers. The blinded repeat exposes disagreement but does not remove that dependence; later conditions had only one adjudicator. The sensitivity results in Section~\ref{sec:suite} and the supplement should accompany the accuracy counts. An independent held-out suite would test transfer beyond tasks developed alongside this system. The current briefs also supply topology-optimization parameters rather than general performance requirements or CAD geometry.

The release-grid choice limits the engineering interpretation. Every released session used a finer grid and a shorter physical filter length than the accepted specification states. The certificate exposes that choice but does not ask the engineer to accept it. A deployed workflow should resolve and accept the actual grid and regularization before release. Similarly, the numerical certificate covers the stated voxel-density problem, not stress, fatigue, contact, nonlinear response or manufacturing performance. A closed surface still needs analysis before fabrication or use.

The numerical comparisons use single runs, local cold references and different regularization histories. Failed constant-physical-filter pilots and the unusable duct-bridge case limit generality; the bridge cold reference also required a different solver configuration. These comparisons establish measured tradeoffs on completed cases rather than a global optimum or a universally better continuation policy.

Only two provider models, one complete second-provider repeat and unequal, overlapping repeat subsets were studied. The suite is small, deliberately heterogeneous and restricted to one load case. Numerical and latency measurements share one desktop workstation. Padding-aware session times combine numerical replays with original agent time, and KLM omits waiting and errors. No participants were recruited, so satisfaction, learning, mental effort and trust remain unmeasured. Exact numerical replay further depends on the recorded solver, deterministic reduction, hardware behavior and full environment; an action log alone does not guarantee it.

\section{Conclusion}
AgenticTO connects a written design brief to a geometric specification, deterministic formulation checks, and a solver-issued release certificate through one logged tool interface. The distinction matters because a converged, rule-clean computation can still answer a problem the engineer did not request.

On the author-adjudicated 50-brief suite, grounding improved correct outcomes over text-only specification, although schema feedback recovered most of that gain. Grounding also delivered more wrong specifications as finished. Clarification prevented the remaining primary-provider errors when answered from reference facts, but generated many questions on fully stated briefs, and the second provider asked much less often. Assumption labels did not reliably distinguish wrong from right readings. The blinded second adjudication and single-author follow-up decisions bound how these accuracy results should be interpreted.

Five completed numerical comparisons reached solver-certified releases at 25.6--73.7 million elements. Long re-opening came within $-3.06\%$ to $+0.07\%$ of the cold local references at 51--69\% of their final-rung compute; tail continuation mainly certified a transferred topology. Fourteen selected agent sessions exported certified designs and one task was correctly refused, with exact replay of the original action logs. The system-selected release grids were finer than the stated resolution, so release success does not establish fidelity to that resolution or its physical regularization.

External adjudication, held-out briefs, and a participant study remain the next validation steps. Within the tested scope, the system makes formulation and numerical acceptance inspectable while preserving the need for a separate judgment of agreement with the brief.

\section*{CRediT (Contributor Roles Taxonomy) authorship contribution statement}
Shaoliang Yang: methodology, software, formal analysis, writing -- original draft.\\
Jun Wang: conceptualization, funding acquisition, supervision, writing -- review and editing.

\section*{Funding}
This work is supported by Santa Clara University.

\section*{Declaration of competing interest}
The authors declare that they have no known competing financial interests or personal relationships that could have appeared to influence the work reported in this paper.

\section*{Data and software availability}
The application code, the 50 briefs with their adjudicated reference readings and the adjudication record, the saved agent outputs and tool logs, the numerical result records, the figure and table generators, and the replay implementation are available at \url{\repourl}. The code is released under the MIT License, and the briefs, records and documentation under the Creative Commons Attribution 4.0 International (CC BY 4.0) license.Running new solves also requires the authors' voxfem solver package, which is distributed separately from the repository; the records, tables and most figures can be inspected and regenerated without it. The supplementary material reproduces the brief texts, prompts, and detailed result tables.

\section*{Declaration of generative AI and AI-assisted technologies}
The authors designed the study, conducted the experiments and analysis, and wrote the manuscript, using artificial intelligence (AI) assistance at their discretion. OpenAI Codex assisted manuscript editing. Codex and Claude-based tools supported software development, experiments, and analysis. After using these tools, the authors reviewed and edited the content as needed. The authors retain full responsibility for the final content.

%% file: supp_appendix.tex
\noindent Specification-accuracy scores use the reference readings as adjudicated by one author (Section~\ref{S:sec:adjudication}); the other author's blinded adjudication is compared in Section~\ref{S:sec:adjudication2}. The complete brief texts are reproduced as experimental inputs, including their original wording and recorded source attributions. Source attributions are not a claim that their engineering restatements have already been independently verified.

\section{Notation, records, and evidence boundaries}
A specification is the structured statement of the design problem; a reading is one complete interpretation of a brief. Right behavior credits an equivalent intended or accepted reading, or a correct refusal on a designated refusal task. R denotes a rule parser; T a text-only large language model; G grounding tools without questions; G+C grounding and clarification; G+C+V the same result followed by a visual check. GUI means graphical user interface, GPU graphics processing unit, MCP Model Context Protocol, and CLI command-line interface. KLM means Keystroke-Level Model, a model of expert interaction time (\href{https://doi.org/10.1145/358886.358895}{Card et al., 1980}). SIMP means solid isotropic material with penalization. FP64 denotes 64-bit floating point. STL denotes the stereolithography surface-file format. VRAM means video random-access memory, CAD computer-aided design, and API application programming interface. IoU means intersection over union of binary material masks. PCG means preconditioned conjugate gradient. A certificate reports numerical acceptance predicates for a stated computation, rather than structural safety. T+S adds schema-feedback retries to T; GNR removes rule feedback from G; G+P adds value-source labels to G; G+C+P permits clarification only about values labeled assumed.

The study identifiers below match the archived records and generator scripts. Each table caption identifies its denominator. Different studies use different meshes, starts, schedules, and timing boundaries. In particular, release-policy compute shares omit earlier rungs, complete-session timing includes agent work but excludes CLI startup, and padding-aware totals replace solve durations within the original session. KLM totals omit solver and agent waiting.

\begin{table}[H]
\centering\small
\caption{Studies and source records. Paths are relative to the root of the public repository (main-text Data and software availability), except the GNR summary, which is among the arXiv ancillary files.}
\begin{tabularx}{\linewidth}{>{\raggedright\arraybackslash}p{10mm}>{\raggedright\arraybackslash}p{42mm}>{\raggedright\arraybackslash}X}
\toprule
Study & Purpose & Primary records and generator \\
\midrule
X0 & Extensions and replay & \path{results/X0/}; \path{scripts/x0_replay.py} \\
X1 & Warm release policies & \path{results/X1/escalation.json}; \path{scripts/x1_escalation.py} \\
X2 & Specification benchmark, rule-feedback ablation & \path{results/X2/}; \path{anc/supplement/gnr_summary.json}; \path{results/X2/failure_categories.json}; \path{scripts/x2_run.py}; \path{scripts/x2_rescore.py}; \path{scripts/x2_gnr_ablation.py}; \path{scripts/x2_failure_categories.py}; follow-up conditions: \path{results/X2/provenance_summary.json}; \path{scripts/x2_provenance_batch.sh}; \path{scripts/x2_provenance.py} \\
X3 & Complete agent sessions & \path{results/X3/}; \path{scripts/x3_run.py}; \path{scripts/x3_regularization.py} \\
X3 replay & Padding-aware timing & \path{results/X3_aware/}; original agent times retained \\
X5 & Common-grid part evaluation and connectivity & \path{results/X5/resolution.json}; \path{results/X5/connectivity_fullres.json}; \path{scripts/x5_resolution.py}; \path{scripts/x5_chain.py}; \path{scripts/x5_connectivity_fullres.py} \\
X6 & Latency, actions, cards & \path{results/X6/}; \path{scripts/x6a_latency.py}; \path{scripts/x6b_klm.py}; \path{scripts/x6c_cards.py} \\
Gates & Census, rule removal, fault injection, keep-out energy, post hoc flags on delivered specifications & \path{results/C4/census.json}; \path{results/C4/rules_off.json}; \path{results/C4/fault_matrix.json}; \path{results/C4/void_energy.json}; \path{results/C4/assumption_flags.json}; \path{results/C4/modulus_plausibility.json}; \path{results/C4/value_traceability.json}; \path{scripts/c4_gate_ablation.py}; \path{scripts/c4b_void_energy.py}; \path{scripts/c4c_fault_matrix.py}; \path{scripts/c4d_assumption_flags.py}; \path{scripts/c4e_modulus_plausibility.py}; \path{scripts/c4f_value_traceability.py} \\
\bottomrule
\end{tabularx}
\end{table}

The suite reference files and scoring implementation are included as machine-readable supporting material. The reference file includes the accepted alternatives from the follow-up adjudication as well as the original benchmark readings; the original benchmark's count of credited alternative readings and the follow-up decisions are reported separately below. The comparator permits a 3\% relative difference in Young's modulus and an absolute Poisson-ratio difference below 0.02, to admit nearby handbook values when a brief names a material without fixing its constants. Load values use relative tolerance $10^{-6}$ with a small scale-dependent absolute tolerance. Geometric masks must agree on the comparison grid. The comparator tries the element size of the intended specification and then two, four, eight and sixteen times that size, and uses the first that both specifications can build with at most four million elements; if none qualifies it keeps the coarsest grid it could build. No saved comparison exceeded that limit, and each saved scoring record states the grid it used. Every credited record whose reference states an element size was compared at that size. Equivalence here is therefore a reproducible discrete criterion, not a proof of equality of continuous geometric descriptions.

\section{Detailed specification-benchmark results}
\label{S:sec:suppbench}
\input{tables/t5_x2_mcnemar}
\input{tables/t6_x2_repeats}
The two-sided exact paired tests use discordant tasks only. Fixed means wrong in the first condition and right in the second; broken means the reverse. Reported tests are unadjusted. An unchanged outcome under the picture check is not an equivalence demonstration on new tasks. Unequal repeat sizes reflect the saved completed runs and should not be pooled as independent task samples. The second G+C repeat covered b01--b22 and ten further tasks (b25, b31, b39, b44--b50); the third G+C repeat and both G+C+V repeats covered the same 20 tasks (b01, b03, b04, b06--b10, b19, b22, b25, b31, b39, b44--b50). The saved records do not state a rule for choosing these subsets.

For the smaller second-provider default-effort diagnostic, the grounding-and-clarification condition achieved 16/20 right outcomes, with four delivered-wrong results and six questions. The primary provider achieved 20/20 on the corresponding tasks, with 34 questions; the paired $p$ value was 0.125. The second provider's text-only and visually checked conditions achieved 12/20 and 16/20. These default-effort runs were rescored against the adjudicated readings, including the readings accepted in the follow-up adjudication (which credited their b04 lugs in all three conditions and the text-only b47 pad), but were not themselves in the contested set. This diagnostic is distinct from the full 50-task high-effort comparison.

The reference-based consequence scoring is heavy-tailed. Text-only output includes twelve non-refusal tasks without a valid specification and one guessed volume fraction that uses approximately 33\% more material. Grounding-only output includes an intended-loading penalty of approximately 164\%, a missing-load-path case of order $10^8$, and three outputs with increased mass, two from added retained solids and one from a smaller clearance. These observations motivate reporting compliance consequences with material use and outcome class instead of a single mean.

The rule-feedback ablation (GNR) ran on 31 October 2026 with the same provider, model, reasoning effort and command-line client version as the first G repeat; the tool server's validation returned only schema errors and an empty findings list, without the grid summary that G's validation returns. After the follow-up adjudication, its eight delivered-wrong specifications were b07, b10, b31, b39, b44, b45, b46 and b47; G was right on b10 and b31. In G's first repeat, validation returned informational coarse-multigrid support notes on 11 tasks, one warning about a load on fixed displacement components, and blocking errors only on the three refusal briefs; none of these reached b10, and b31 received one informational coarse-multigrid support note. Median input tokens were about 87,000 for GNR and 86,000 for G. Two of the 15 run-derived accepted readings of the first adjudication came from G runs (b25, b48); the follow-up adjudication accepted GNR's drive lugs on b04 and its 4 mm boss on b48.

The follow-up conditions T+S, G+P and G+C+P, the second G+P repeat on the 20-brief subset, and the second provider's G+C+P ran on 1 November 2026 with the same providers, models, reasoning effort and command-line client versions as the corresponding first repeats (Section~\ref{S:sec:suppprompts} reproduces the added prompt text). The source tags are kept in the session record, not in the specification, so the comparator, the rules and all earlier runs are unchanged; validation adds an informational source-label finding that lists items tagged assumed or not yet tagged and never blocks. Every delivered G+P and G+C+P specification tagged every item.

T+S retried once on b06, b19, b31, b33, b39 and b40 and never needed a second retry; on b41, whose T output had been schema-invalid, its first reply was valid. T+S delivered wrong specifications on b07, b10, b44 and b47, refused the clarification briefs b45, b46 and b48, and broke b43 (refused, a fully stated brief) and b44.

G+P delivered wrong specifications on b07, b10, b33, b39, b44, b45, b46 and b47; G was right on b10 and b33. Its second repeat delivered seven wrong specifications on the 20-brief subset (b07, b10, b39, b44, b45, b46, b47), all tagged on a failed field, and tagged nine of its ten right ones (material tags set aside).

G+C+P asked on b07 (twice), b10, b45, b46 and b48 and was wrong on b39, b44 and b47; on b44 and b47 it asked nothing, and on b39 it added retained washer pads and a plate. On b07 and b48 it asked about the support or load direction and set a retained plate or boss 1 mm deep itself, readings that were accepted in adjudication. With material tags counted, it tagged 42 of its 44 right delivered specifications, and 21 without them. The second provider's G+C+P asked once each on b07, b45, b46, b47 and b48 and was wrong on b10, b12 and b44; it corrected the two G+C unit slips (b13, b37). Its b12 error is a unit slip in a modulus that it tagged stated (210,000 with forces in kilonewtons), and its 3 mm camera pad on b47 and 4 mm boss on b48 were accepted in adjudication; it tagged 18 of its 44 right specifications with material tags counted and 7 without. Median input tokens were about 20,000 for T+S, 106,000 for G+P, 107,000 for G+C+P, and 95,000 for the second provider's G+C+P against 44,000 for its G+C.

J.W. adjudicated their contested records, together with GNR's, on 1 November 2026 from a sheet that listed each record with its brief, reference readings, the fields that differ, its consequence, and, where the comparator found an equivalent specification, the earlier row and any 31 October verdict on the same reading. The 42 records held 27 distinct readings, eight of them judged wrong on 31 October; every such verdict was confirmed. Nine records were accepted (b04 from GNR and both G+P repeats, b07 from G+C+P, b16 from G+P, b47 from the second provider's G+C+P, and b48 from GNR, G+C+P and the second provider's G+C+P) and 33 were confirmed as wrong (\path{paper/adjudication/followup/}). The accepted b16 reading had a failed solve in the consequence scoring, so its consequence value is not used.

Their consequences under the intended loading were scored as for the other conditions (\path{results/X2/provenance_summary.json}, mechanical-consequence results): the second G+P repeat also lost the gearbox-bracket load path and overspent material on the camera mount (43\%) and the pump bracket (23\%), and the second provider's G+C+P lost the gearbox-bracket load path and otherwise changed compliance by at most 6\%.

Three post hoc checks of the saved first-repeat specifications ask whether a wrong reading could have been flagged without the reference; none is part of the evaluated system. The first flags a specification whose description or region notes contain the stem `assum': it flagged five of the six wrong and 19 of the 41 right G specifications, whereas the closing messages mentioned an assumption in all 47.

The second converts Young's modulus to megapascals and flags values outside 0.1--1000 GPa: under the second provider it flagged the two G+C unit slips (b13, b37) and no right specification, and on the references it would catch 14 of the 47 injected modulus slips.

The third tags each number that places a support, load, or passive region, and each force component, as traceable when it equals a number written in the brief (also multiplied or divided by 1000), zero, a domain size, half of one of these, or a sum or difference of two of these, optionally halved; a specification with any untraced value is flagged. It flagged three of the six wrong G specifications (b44, b46, b47; b44 only for a torque-derived force that its reference shares, not for its wrong axis and added lugs) and one of the 41 right ones (b04, whose force is derived from a stated torque), and one of the 40 reference readings of fully stated briefs. The three unflagged errors (b07, b39, b45) chose a load direction or magnitude, or added retained regions, from numbers in the brief. Over all eight first-repeat condition and provider cells (364 schema-valid specifications) it flagged 12 of 32 wrong and 11 of 332 right ones; all 11 flags on right specifications, and 5 of the 12 on wrong ones, fell on values that the reference reading also contains.

\section{Adjudication of the reference readings}
\label{S:sec:adjudication}
The briefs, intended readings, and accepted alternative readings were drafted with Claude Opus 5.5, the second provider's model, during system development. A pilot on the first ten briefs of an earlier version of the suite preceded the benchmark and changed the suite and the scorer: an ambiguous brief became the fully stated torsion shaft and its original wording the clarification task b44, volume fractions were stated over the design volume, and accepted alternative readings were encoded as specifications. The 50-brief suite was then fixed before the benchmark runs. During them, an oracle error on numbers written as strings was fixed and both clarification conditions were rerun; the invalid runs are kept.

One author (J.W.) adjudicated the reference set on 31 October 2026. The intended readings had been drafted with language-model assistance; the adjudicator read each brief as an engineer would and checked its intended reading, its oracle facts for clarification tasks, its refusal designation, and its cited benchmark source. The adjudicator also judged the contested set: every first-repeat run, across the five primary-provider conditions and the three high-effort second-provider conditions, whose specification was valid, not refused, and matched neither the intended nor an accepted reading. Each of the 41 contested runs was judged against the brief text, the intended specification, the fields that differed, and the mechanical consequence under the intended loading. A run was marked wrong when its specification departed from the brief or the clarification facts in a way that changes the stated problem, including loads and material constants even when the optimal topology would be unchanged, and acceptable when the brief leaves the difference open. Each acceptable specification was added to its brief as an accepted reading, and all 672 saved runs were rescored without a language model or solver. The 27 runs whose verdict changed all moved from wrong to right.

Nineteen contested runs were accepted: thin retained plates and bosses at named interfaces whose thickness the brief leaves open; square support patches standing for round washers; roller supports that restrain only the vertical direction where the brief frees the span direction without fixing the width; and, for the press riser, support restricted to the bottom surface beside the coolant channel. Twenty-two runs were confirmed as wrong: guessed volume fractions, load magnitudes, up axes, clearances, lug footprints and an unstated minimum feature size on clarification tasks, measured against the stated clarification facts; retained support bosses and a retained shaft region that the brief does not request and that change the material budget; a load applied through a volume instead of the stated face; and four runs, on two briefs, that worked in kilonewtons and millimeters but left Young's modulus in megapascals, which leaves the optimal topology unchanged but states a material 1000 times too stiff. Runs on clarification tasks under the grounding-only condition, which could not ask, were judged against the clarification facts; such a guess is a plausible completion of the brief, but not the stated problem. Two second-provider text-only runs ended by asking for the missing load or clearance while still returning a completed specification; they count as wrong specifications delivered as finished.

Forty-nine intended readings were confirmed. The press-riser support was corrected from the full bottom face to the two bottom strips beside the open coolant channel. Because the brief states that the block is bolted over its full area, and the extra fixed nodes over the open channel touch only void elements, the full-face reading is kept as admissible, so no run that fixed the full face changed; the one run that supported only the two strips, previously scored wrong, now counts as right. One accepted reading of the half MBB beam (the Messerschmitt--B\"olkow--Blohm beam, a standard topology-optimization benchmark), which also fixed the end support along the span, was withdrawn because the brief says the supports carry vertical and sideways load only; no run depended on it. All three refusal designations were confirmed. Of the 25 benchmark-based briefs, the cited family was confirmed for six (the half MBB beam, the lower-edge line-load cantilever, the wheel, the engine bracket, the z-up MBB beam and the Michell-type cantilever), each rescaled to physical units and, for the engine bracket and the Michell cantilever, adapted rather than restated. The cited source could not be traced to a specific published problem for six, and the other 13 generic family descriptions are treated as engineering examples. The per-brief verdicts appear with the brief texts in Section~\ref{S:sec:briefs}.

\begin{table}[H]
\centering\small
\caption{Benchmark results before and after adjudication (first repeat).}
\label{S:tab:adjudication}
\begin{tabular}{lrr}
\toprule
Result & Before & After \\
\midrule
Briefs with credited alternative readings & 3 & 14 \\
Primary provider right outcomes, T / G / G+C / G+C+V & 37 / 42 / 50 / 50 & 37 / 44 / 50 / 50 \\
Primary provider delivered wrong, T / G / G+C & 1 / 8 / 0 & 1 / 6 / 0 \\
T $\rightarrow$ G, fixed / broken ($p$) & 7 / 2 (0.18) & 8 / 1 (0.039) \\
G $\rightarrow$ G+C, fixed / broken ($p$) & 8 / 0 (0.0078) & 6 / 0 (0.031) \\
Second provider right outcomes, T / G+C / G+C+V & 33 / 42 / 43 & 42 / 46 / 47 \\
Second provider delivered wrong, G+C & 8 & 4 \\
Provider comparison under G+C ($p$) & 0.0078 & 0.12 \\
\bottomrule
\end{tabular}
\end{table}
\FloatBarrier

\section{Second adjudication}
\label{S:sec:adjudication2}
The other author (S.Y.) repeated the adjudication on 31 October 2026 from a separate package. Part A listed the 50 briefs with their reference readings as drafted, before the first adjudication. Part B listed the 41 contested records under neutral identifiers in an order shuffled with a fixed seed, each with its brief, the drafted reference reading, the fields that differ, the candidate specification in a normalized form, and its consequence under the intended loading. The package hid the condition, provider, agent messages, and first-round verdicts and comments, and it removed free-text descriptions and number formats that could identify the provider.

S.Y. accepted all 50 drafted readings. The one difference from the first round is the press riser: S.Y. read the support as the bottom material beside the open coolant channel, as J.W. did, but judged the drafted full-face reading equivalent, which is the accepted status the first adjudication gave it, so no reference changes. With one negative rating among 100, a chance-corrected agreement coefficient is not informative for Part A. On the contested records the two adjudicators agreed on 33 of 41 (Cohen's $\kappa=0.60$). The eight disagreements form two groups. S.Y. rejected five square support patches standing for round washers, all from second-provider text-only runs, because a square whose side equals the washer diameter has about 27\% more area than the washer and so changes the stated support region. S.Y. accepted three retained interfaces that J.W. had rejected: the keyed shaft cylinder of the pedal arm and a one-element boss layer of the cable anchor, both from second-provider text-only runs, and the mounting plate and screw interfaces of the television bracket from a grounding-only run. Table~\ref{S:tab:adjudication2} gives the right outcomes and paired tests when the contested records are scored by J.W. (as in the paper), by S.Y., by crediting only readings both accept, and by crediting readings either accepts.

\begin{table}[H]
\centering\footnotesize
\caption{Right outcomes (first repeat, 50 briefs) and exact paired tests under each combination of the two adjudications of the 41 contested records. Paired entries are fixed / broken ($p$); for the provider pair, tasks on which the second provider was ahead / behind under text only.}
\label{S:tab:adjudication2}
\begin{tabular}{lrrrrrlll}
\toprule
Contested records scored by & T & G & G+C & G+C+V & 2nd T & T $\rightarrow$ G & G $\rightarrow$ G+C & T, provider pair \\
\midrule
J.W. (used in the paper) & 37 & 44 & 50 & 50 & 42 & 8 / 1 (0.039) & 6 / 0 (0.031) & 7 / 2 (0.18) \\
S.Y. & 37 & 45 & 50 & 50 & 39 & 9 / 1 (0.021) & 5 / 0 (0.063) & 3 / 1 (0.62) \\
Both must accept & 37 & 44 & 50 & 50 & 37 & 8 / 1 (0.039) & 6 / 0 (0.031) & 2 / 2 (1.00) \\
Either accepts & 37 & 45 & 50 & 50 & 44 & 9 / 1 (0.021) & 5 / 0 (0.063) & 8 / 1 (0.039) \\
\bottomrule
\end{tabular}
\par\smallskip\footnotesize 2nd T: second provider, text only. The rule parser's 9/50 right outcomes, T $\rightarrow$ G+C (13 / 0), and the second provider's G+C (46) and G+C+V (47) results are the same under every combination.
\end{table}
\FloatBarrier

\section*{Protocol and diagnostic detail}
\addcontentsline{toc}{section}{Protocol and diagnostic detail}
The following records give the detailed comparisons summarized in the main text. They retain the original conditions, denominators and caveats; the primary outcome and the protocol key are defined in main-text Sections 4.3--4.4.

\subsection*{Provider configuration and repeat records}
The primary provider is OpenAI's Codex command-line interface (CLI), model GPT-6.1-Sol, with reasoning effort recorded as high. The saved records identify Codex CLI version 0.165.0. A second-provider experiment uses Anthropic's Claude Code CLI version 2.1.312, model Claude Opus 5.5, at effort high on T, G+C, and G+C+V for all 50 tasks, and on G+C+P as a follow-up. The Claude runs disable user settings and built-in tools, leaving the specified MCP interface; the picture check of G+C+V instead enables only the built-in file-reading tool, which opens the rendered images. The runner passes the model explicitly to Codex but not to Claude Code, which therefore used its default model: the generation records identify it as Claude Opus 5.5, whereas the picture-check records, written seconds later by the same client, do not store a model identifier or client version. These are recorded experimental configurations, not statements about current product defaults.

The primary benchmark run records are dated 30 October 2026; Section~\ref{S:sec:suppbench} gives the later ablation and follow-up dates. Both providers were accessed through subscription CLIs, rather than billed application programming interface calls. We report observed tokens and latency, without inventing a monetary cost from subscription usage. Input-token totals include cached input when supplied by the CLI. A nominal effort label is not a calibrated amount of reasoning shared across providers, and token accounting differs between them.

Additional primary-provider runs assess variation on subsets. The saved G+C repeats contain 50, 32, and 20 tasks; G+C+V contains 50, 20, and 20. Text-only and grounding-only repeats use their separately reported denominators, and G+P was repeated on the 20-brief subset. These are repeated observations on overlapping tasks, not independent samples from an engineering-task population. A smaller second-provider default-effort diagnostic is reported in Section~\ref{S:sec:suppbench} and is not pooled with the matched-label comparison. Prompts, coordinate conventions, review schema, and repair instructions are reproduced from the runner in Section~\ref{S:sec:suppprompts}.

\subsection*{Paired-test sensitivity and task-class results}
The T-to-G primary comparison corrected eight tasks and broke one ($p\approx0.039$); the secondary specification comparison corrected 11 and broke one ($p\approx0.0063$). G-to-G+C corrected six outcomes and broke none ($p\approx0.031$), and T-to-G+C corrected 13 and broke none ($p\approx0.00024$). With six to 13 discordant tasks, these $p$ values rest on few outcomes. The full paired table is Table~\ref{S:tab:t5_x2_mcnemar}. They also depend on the adjudicated references. Before adjudication, the same T-to-G comparison corrected seven tasks and broke two ($p\approx0.18$; Table~\ref{S:tab:adjudication}), and the adjudicator saw the condition and provider of every contested record. Every verdict that adjudication changed moved from wrong to right. A single reversed verdict would move either the T-to-G or the G-to-G+C test above 0.05, so we read these tests as descriptive. The second adjudication shows this: under S.Y.'s verdicts, T-to-G corrected nine tasks and broke one ($p\approx0.021$), while G-to-G+C corrected five and broke none ($p\approx0.063$), because S.Y. accepted one grounding-only output that J.W. had rejected. G+C and G+C+V reach 50 under either adjudicator.

Tools changed the failure profile. Text-only output often failed to encode valid geometric objects, including zero-length cylinders used to represent washers or a surface load. Of the eight text-only failures that grounding corrected, six were schema-invalid outputs and two were refusals of clarification briefs. The G protocol differs from T in several ways at once, namely tool-based construction, validation feedback from the rule base, rendering, and an instruction to assume rather than refuse, and only the rule feedback was ablated (below). G also used more inference, with median input tokens about four times those of T (main-text specification-outcome table), and no text-only condition was given a matched budget. Split by task class, six of the eight T-to-G corrections were on fully stated briefs, where T was right on 33 of 40 and G on 39 with none broken; on the seven clarification tasks, T was right on one and G on two, with two corrected and one broken. However, G returned six delivered-wrong specifications, compared with one for T. Several T failures were explicit refusals on tasks requiring clarification, whereas G was instructed to choose assumptions and continue. A higher count of valid specifications can therefore coexist with more delivered but wrong specifications. All six G outputs listed their guessed values in the closing message, yet each specification was delivered as finished rather than withheld. Five of the six specifications also mentioned an assumption in their own free-text notes, as did 19 of the 41 right G specifications, so the notes did not separate wrong from right readings.

\subsection*{Feedback, source tags and picture-check diagnostics}
An ablation separated the rule feedback. The grounding-without-rule-feedback condition (GNR) repeated G with the same prompt, tools, provider and settings, but validation checked only the schema and returned no rule findings or grid summary. GNR was right on 42 of 50 briefs (95\% interval 71--92\%) and delivered eight wrong specifications, all of which passed every rule. It corrected no G outcome and broke two ($p=0.5$), one by adding retained bosses to the pillow block, as G had on other briefs, and one by narrowing the hook plate's attachment. On these two tasks, G's own validation had returned no blocking or warning finding, only one informational multigrid note, so apart from that note and the grid summary the agent saw the same feedback in both conditions, and the differences are consistent with run-to-run variation; G's second repeat likewise broke two first-repeat outcomes and corrected none. GNR also refused all three refusal briefs without a rule finding, giving the equilibrium or clearance reason in its own words. Two of these briefs state the free motion in words, whereas in the valve bracket the conflict follows only from coordinates; three briefs cannot show how often an agent misses a defect that a rule would catch. In these runs, removing rule feedback changed no outcome through a finding and left the refusals in place; the evidence for the rule gate is the rule-removal case and the fault injection (Section~\ref{S:sec:rulecensus}).

Recording each value's source did not raise right outcomes. G+P was right on 42 of 50 briefs (71--92\%) and delivered eight wrong specifications; it corrected no G outcome and broke two ($p=0.5$), one by adding a retained skin and screw bosses to the fully stated shelf bracket, whose reference has none, and one by replacing the hook plate's attachment footprint with an edge load. GNR, run with the same G instruction, also broke two, and on the 20-brief subset G+P and the second G repeat had the same outcome on every brief, so the difference lies within the run-to-run variation of these conditions.

The tags marked every error: each of the eight wrong specifications tagged as assumed a design-changing item in a field that the comparator found wrong. They also marked 25 of the 39 right delivered specifications, and 38 of the 39 if a material tagged assumed is counted, because agents tagged a modulus looked up from an alloy name as assumed. Reviewing every tagged specification would have caught all eight errors at the cost of checking 33 of the 47 delivered ones, with material tags set aside.

Limiting questions to tagged items removed questions on fully stated briefs but lost right outcomes. G+C+P asked six questions, none on a fully stated brief, against 45 on 26 of the 40 for G+C. It was right on 47 of 50 (84--98\%), correcting no G+C outcome and breaking three ($p=0.25$). In each of the three, the agent had tagged the failing item assumed and did not ask about it: on two clarification briefs it asked nothing, and on the fully stated television (TV) bracket it added retained regions that the brief did not request, as G had. On two further clarification briefs it asked about a support or the load direction and then set the depth of a retained plate or boss itself, readings that adjudication accepted.

The second provider asked on no fully stated brief even under G+C. Under G+C+P it was right on 47 of 50 against 46 (two corrected, one broken, $p=1$; main-text provider-comparison table), with five questions against seven; two of its three errors fell on items that it had tagged assumed and did not ask about, and the third was a unit slip in a modulus that it had tagged stated. Its tags were more selective, marking seven of its 44 right delivered specifications when material tags are set aside. In these runs the agents' own tags located most of their errors, all of the primary provider's, but did not separate them from right readings, and tagging an item assumed did not make the agent ask about it.

The extra picture check left primary-provider outcomes unchanged. It flagged an issue on one task, but the repair retained the already matching specification. Median input tokens rose from about 104,000 for G+C to 135,000 for G+C+V, and median wall time rose from 37 to 49 s. Because G+C was already right on every brief, the check had no error to correct here, so the primary-provider runs cannot test its benefit. Under the second provider, the same check repaired one wrong G+C specification (the gearbox bracket) and broke none, raising right outcomes from 46 to 47 (main-text provider-comparison table). It does not establish that images are useless for different geometries or faults deliberately designed to challenge visual placement.

\subsection*{Mechanical consequences and fault-class detail}
Mechanical consequences were uneven. On non-refusal tasks, one G misreading produced a compliance penalty of about 164\%, and another left effectively no load path under the intended loading, with a consequence ratio of order $10^8$. Three G outputs used about 3--10\% more material than intended, two by adding retained solids and one by shrinking a required clearance, and appeared 0.7--2.5\% stiffer. The negative compliance changes are not improvements under the original material budget. The remaining load-magnitude error had little topology consequence for the linear single-load minimum-compliance objective, while still representing an incorrectly stated problem. These cases show why a single mean consequence score would obscure both intent errors and resource violations. The follow-up conditions repeated the pattern. The gearbox-bracket error recurred in T+S, G+P and G+C+P with a consequence ratio of order $10^8$, and guessed volume fractions on the camera mount used 33--38\% more material than intended in the same three conditions. The sensor-arm reading cost 99\% more compliance in T+S and 85\% in G+P. The remaining G+C+P error, the regions added to the TV bracket, used 6.4\% more material.

A fault-injection check measured the same division on the suite's reference specifications. Into each of the 47 that pass every rule (the other three are the refusal briefs), we injected one fault at a time and counted it caught when a new blocking error appeared (Table~\ref{S:tab:faultmatrix}). Four fault classes mirror rules R004, R003, R008, and R005, so they test whether the rules work as defined, not how much of admissibility they cover. The rules caught every zero force (47 of 47), every support set reduced to one restrained direction (47 of 47), every load moved inside a keep-out region (8 of 8), and 43 of 46 loads moved onto fully fixed supports, 145 of 148 faults of their own classes; the three misses moved one of several loads and drew a warning. Two defects that main-text Section~3.2 names as outside the rule set passed every rule: a load straddling a keep-out face (5 references) and a modulus divided by 1000 (47). Of 226 intent-changing faults, the rules caught 21: none of the flipped force signs, forces multiplied by 1000 (a unit slip), or changed volume fractions (47 each); an invented keep-in pad only where it overlapped a keep-out region (R007, 6 of 47); and a support moved to the opposite face only where more than half of the applied load then fell on fixed components (R005, 15 of 38). These are injected faults on the suite's own references, not a sample of agent errors. A plausibility predicate on the solved field could move part of the admissibility check into the certificate. Measured after the fact, the share of the compliance carried by strain energy in keep-out elements was 0.99999997 for the constructed load-in-keep-out problem and at most $2.4\times10^{-8}$ for its corrected version and for eight valid reference problems with keep-out regions (Section~\ref{S:sec:rulecensus}). One contradictory case cannot set a threshold, and the predicate is not part of the certificate reported here.

\subsection*{Representation and supporting-result details}
\addcontentsline{toc}{subsection}{Representation and supporting-result details}
Applying the follow-up adjudication verdicts re-scored all 942 saved outputs; it changed no suite condition and credited four runs of the default-effort diagnostic.

\subsubsection*{Geometry, loading and physical regularization}
For a brief that describes a distributed pressure, equal sharing places more force on the boundary nodes of the region than a consistent pressure load would, and the comparator, which scores the stored nodal loads, cannot see that difference. Ten of the 50 reference readings spread a load over a whole face and three along an edge line. Edge supports and loads, used in eight of the 50 reference readings, give an elastic field that is singular along the edge, so compliance grows as the grid is refined; certified compliances of such a problem on different grids are not mesh-converged values. In the main-text release-policy table and the main-text common-grid figure only the bridges use edge supports, and each comparison there is made on one grid, except the under-prediction of main-text Section 5.7, which compares a coarse optimizer's own compliance with that of its part on the finest grid.

Resolution is stated as an element edge length or a number of cells along the shortest edge. The ladder may choose a release grid finer than that base request, but it reports the actual grid and rejects conflicting explicit solve arguments. In the agent sessions, the release grid was the ladder's release rung, chosen by the system rather than requested by the engineer, and in every released session it was finer than the resolution stated in the specification. The engineer's acceptance of the specification therefore covers the stated resolution, not the release grid. Because the filter radius is set in element units, a finer release grid also shortens the filter's physical length, so the released part answers a more finely regularized problem than a run at the stated resolution would. The optional minimum-feature field requests a feature length; the rule warns when it spans fewer than 2.4 elements at the chosen grid spacing. The experiments use continuation filters specified in element units; they do not establish a certified physical minimum-member size or a mesh-independent manufacturing constraint. Filtering regularizes topology optimization \citep{Bourdin2001Filters,Bruns2001Nonlinear,Lazarov2011Helmholtz}. Defining the filter radius in physical length keeps that length fixed as the mesh changes; robust projection formulations address minimum member size \citep{Wang2011Projection}. The release grid is therefore a numerical resolution, not a manufacturable feature size.

\subsubsection*{Rule predicates and support checks}
The other blocking rules are stricter than admissibility. R002, the empty-region test of R004, and R005 (more than half of the applied load on fixed displacement components) reject a specification whose problem may still be well posed, because the stored problem no longer matches what was specified. R008 rejects a support or load entirely inside a keep-out region (the main-text rule table); this is a limited load-path check. Here the ersatz stiffness, the artificial stiffness retained in void elements, keeps the discrete problem solvable when it fails (main-text Section 5.4), and it sits before the solve because it needs no solved field, not because it differs in kind from a check on the solved field. R001 and R013 are execution conditions, stating whether the requested grid can be built and fits in GPU memory, and the warnings R009--R012 are advisory. An admissible specification can therefore fail a stricter rule, and a specification that passes every rule can still be wrong.

Rigid-body checking tests whether the supports restrain the six rigid modes (translations and rotations) of a three-dimensional body. It can detect a missing rotational restraint even when the selected support contains many nodes. A separate rule detects a load or support entirely contained in a retained-void region. This is relevant because a positive ersatz stiffness can make the algebraic problem solvable even when the specified physical load has no meaningful material path. The rule does not rely on the optimizer becoming unstable before reporting the contradiction.

Loads on fixed displacement components require additional care. The problem builder drops those components and records the change. The rule blocks the solve if more than half of the applied load is lost; any positive share up to and including half raises a warning. A solver can correctly solve the altered vector, so silently dropping such components would otherwise hide a change in the task. Similarly, a selected geometric region that contains no nodes is different from a nonempty region restrained in the wrong direction; the rule base catches the first, while intent comparison is needed for the second.

\subsubsection*{Replay, asynchronous tools and certificate fields}
Bit-exact numerical replay required a change at the coarsest multigrid assembly. The upstream solver uses atomic scatter-add operations whose reduction order can vary. AgenticTO supplies a deterministic coarsest assembly for sessions. The extension-verification study separates solver identity from this session contract: it tests exact identity with an atomic-free Jacobi-preconditioned configuration, then tests logged sessions with deterministic multigrid. A different coarse-grid policy is a different numerical execution; its later timing replay is reported with compliance differences rather than called bit-exact.

Long-running tools also require an explicit completion protocol. A start operation returns a job identifier, status queries inspect that job, and a wait operation can be repeated after a timeout. Worker events are pumped independently of an outstanding wait. Otherwise an agent can observe an obsolete status and cancel or duplicate a job that has already finished. The log retains these orchestration decisions, making such failures inspectable instead of merging them into solver time.

The verification parser currently treats an unparseable reviewer response as containing no actionable mismatch. None of the 139 reviewer responses in the reported runs was unparseable (90 for the primary provider and 49 for the second). We report this implementation limit because the image check is an advisory stage, rather than an acceptance authority. The deterministic formulation and numerical gates remain in force regardless of the reviewer's output.

The solver creates this record, including grid, element count, units, compliance, termination, residual, tolerance, and stopping settings. The certificate also records the release grid's element size beside the specification's stated one, the filter radius in elements and as a length at both sizes, the continuation schedule by name, the grid of the warm start, and the iterations on the release grid, so a tail release can be told from a long re-opening and the released regularization from the accepted one. For the reported sessions, these certificate fields were reconstructed from the session logs (main-text Section 5.5). The tool interface exposes it for inspection. Export as a released result requires the rule gate and the solver certificate, rather than a final sentence written by the agent. Intermediate fields can be displayed for feedback without being labeled releasable.

\subsubsection*{Reference comparison and adjudication}
Each task stores an intended specification, expected response class, clarification facts where relevant, and any accepted alternative readings, that is, other readings accepted as faithful to the brief. Acceptance here concerns fidelity to the brief; mechanical admissibility is defined separately in main-text Section 3.2. Where a brief gives a downward and a sideways force component on one region, as the motor-mount brief does, the reference reads them as one simultaneous force, the only reading the single-load-case schema can hold; whether they act together is not a clarification fact. An engineer may read such components as separate load cases, which the schema cannot represent, so a deployed system should flag such briefs as outside its scope rather than combine the components. Nine of the 50 reference readings combine force components on one region; without them, right outcomes were 32 of 41 for T, 36 of 41 for G, and 41 of 41 for G+C. Correctness is evaluated after normalizing units and building the two specifications on a common grid. The comparison checks domain, volume fraction, resolution, minimum-feature request, fixed-displacement masks, nodal loads, and passive masks. Names and the choice of equivalent geometric syntax do not determine correctness. A face support and a box selecting exactly that face can describe the same problem.

S.Y. repeated the adjudication with condition, provider, agent messages and first verdict hidden, agreeing on 33 of 41 contested records (Cohen's $\kappa=0.60$ \citep{Cohen1960}). Scores use J.W.'s verdicts; requiring both adjudicators to accept a reading leaves every primary-provider count unchanged. Follow-up conditions were adjudicated once by J.W., with condition visible. All affected outputs were re-scored. These are author-adjudicated references, not independent ground truth; the supplement reports alternative scores and every changed reading.

\subsubsection*{Questions and provider differences}
Five G errors were guessed readings on clarification tasks, where the brief leaves a design-changing fact open and G was not allowed to ask; the sixth added retained support bosses that a fully stated brief did not request. A guess on an ambiguous brief differs from a wrong answer given after clarification, but each still delivered a problem the engineer had not stated. All six passed the deterministic rules. A valid support mask, plausible load vector, or consistent material model does not indicate that the engineer requested it. G+C resolved the remaining reference mismatches by asking, but asked 59 questions in total: 14 on the seven clarification tasks and 45 on 26 of the 40 fully stated tasks. Five of the six corrections were on clarification tasks, where G had been told not to ask and the oracle answers from the stored reference facts; on fully stated briefs, G-to-G+C corrected one task and broke none. The G+C result is therefore an upper bound under an oracle that holds the reference; it does not show how the same questions would fare with an engineer who answers partly or wrongly. The words corrected and resolved here describe differences between protocols, not effects of a single component: most G errors were guesses that the G protocol required, and of the G components only rule feedback was ablated.

The 45 exchanges on fully stated briefs were not classified by whether the answer follows from the brief alone, so some stated-brief corrections may rest on reference facts that the brief does not determine. G+C asked on all seven clarification tasks. Counting a question on a fully stated brief as a cost, G+C was right without asking on 14 of the 40 fully stated tasks, against 39 of 40 for G. The result trades delivered-wrong specifications for a question burden that remains substantial.

At the same nominal high effort, the second provider achieved 42/50 right outcomes for T, 46/50 for G+C, and 47/50 for G+C+V (the main-text provider table). Its T count depends most on adjudication: it is 39 under S.Y.'s verdicts, 37 when both adjudicators must accept a reading, and 44 when either may, mainly through the square-washer patches. Its G+C run asked seven questions, compared with 59 for the primary provider, and delivered four wrong specifications. The paired G+C provider comparison has four discordant outcomes, all favoring the primary provider ($p\approx0.12$), so this sample shows no significant provider difference in right outcomes. Under text only, the second provider was ahead on seven tasks and behind on two ($p\approx0.18$); under S.Y.'s verdicts it was ahead on three and behind on one ($p\approx0.62$).

The wrong outcomes were assumed load readings on two clarification tasks and two unit slips, in which the model worked in kilonewtons and millimeters but left Young's modulus in megapascals. A modulus range check, which the rule set lacks, would have flagged both after the fact: converted to megapascals, their moduli lie far outside 0.1 to 1000 GPa, as no right first-repeat specification of either provider does. On the suite's references, the same range would catch only 14 of the 47 injected modulus slips (division by 1000), because a slipped steel modulus still lies inside it. Roller-support readings that left the transverse direction free were accepted in adjudication. A model can explain its assumption while still returning a specification outside the current credited set. A separate default-effort diagnostic on 20 tasks showed a similar questioning pattern, but it is not a controlled effort-response curve.

Thus the tools and workflow executed under both providers. The second provider asked far less often and kept four errors on briefs that the primary provider's G+C run got right. Two of those four errors were assumed readings on clarification tasks that a question could have resolved; the other two were unit slips, on which the model asked no question and which no rule checks. Only the first two are consistent with the benefit of clarification depending on the agent choosing to ask, and the difference in right outcomes is not significant on this sample. The comparison covers one model per provider and one complete second-provider repeat. It cannot support provider-independent accuracy, a claim that the nominal effort settings are equivalent, or a general ranking of provider capabilities.

\subsubsection*{Release-policy and common-grid comparisons}
On the five completed problems, final grids ranged from 25.6 to 73.7 million elements. Every reported final arm in the main-text release-policy table converged and received a passing certificate. Each problem has one run per arm, so no seed spread bounds the warm--cold differences, and the cold references use filter radii in element units, so they are local references for a mesh-dependent problem rather than the best attainable fine-grid designs. The fast tail's compliance difference relative to the cold run ranged from approximately $+0.005\%$ to $+1.39\%$. Its final-rung compute was only 3--11\% of cold, but on the release rung the binary design did not change after transfer on four problems and changed in 0.003\% of the pooled elements on the motor mount. This is fine-grid certification of a coarse-derived topology.

The long re-opening reached $-3.06\%$ to $+0.07\%$ of cold compliance at 51--69\% of cold final-rung compute (the main-text release-policy figure). Including the preceding refine-rung long re-opening raised the corresponding compute share to 59--80\%, and adding the shared preview rung as well gave 61--84\% of the cold run's compute. The ratios therefore describe the specified comparison boundary, rather than an end-to-end speedup from an empty session. The saving also combines two effects that were not separated, the warm start and the shorter schedule, because no run used the re-opening schedule from a uniform start; with filter radii in element units, the warm and cold arms also followed different regularization histories (main-text Section 5.7). Because the residual tolerance bounds the algebraic error of each state solve rather than the compliance error, compliance differences of a few hundredths of a percent are best read as ties.

The warm and cold designs differed. Binary intersection over union, measured after pooling large masks to limit comparison cost, was about 0.86--0.96 for the long re-opening. Cold solutions could contain more tunnels. Lower compliance from a warm path is evidence of a different local optimization trajectory, not evidence that the warm method reaches the cold design faster or finds the global optimum.

The bridge cold reference at 65.5 million elements failed twice with the mixed-precision iterative-refinement solver and was completed with the FP64 multigrid configuration and dilated coarse supports. The warm bridge arms succeeded with iterative refinement. This makes the bridge compute comparison partly dependent on the successful solver configurations, and prevents treating it as a pure initial-density ablation. We retain it with that qualification rather than omit the failed cold solves.

\begin{figure}[!htb]
\centering\includegraphics[width=\linewidth]{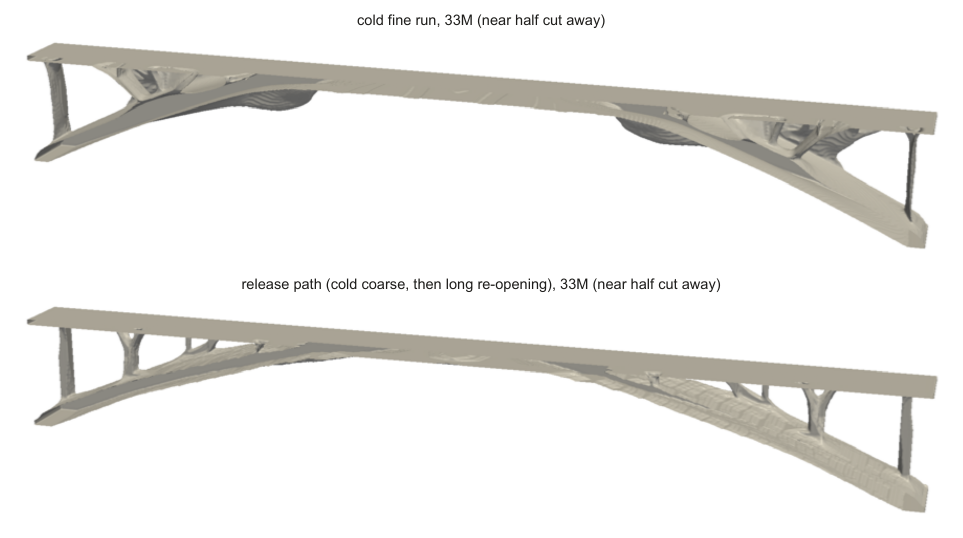}
\caption{\label{S:fig:suppbridge}The span bridge at approximately 33 million elements: a cold fine run and the release path. Both are local numerical outcomes. The path has lower compliance in the common-grid volume-matched part evaluation; finer resolution alone does not establish a better optimum. Closed $\rho=0.5$ isosurfaces at full resolution, near half cut away; both designs are single connected solids, and the cold fine design encloses 92 interior cavities (6-connected) against 12 for the release path.}
\end{figure}

Because the filter radius is set in element units, the coarse start of the bridge's warm path regularized at a longer physical length than the cold fine run, so the two arms followed differently regularized continuation paths to the same final problem. Along that warm release path, finest-grid part compliance decreased from 2715 to 2506, 2422, and 2382 N\,mm as the grid advanced from approximately 0.064 to 0.512, 4.096, and 32.768 million elements. The final path result was 28.5\% below the cold finest-grid result of 3333 N\,mm. Total compute for the path was about 413 s, against 466 s for the cold finest run alone. The cold fine design was also more intricate, with 92 enclosed cavities against 12 in the final path design (Figure~\ref{S:fig:suppbridge}): a finer grid alone did not produce a better design.

On the motor mount, the path reduced part compliance from approximately 4.012 to 3.970, 3.933, and 3.915 N\,mm. The cold finest result was 3.912 N\,mm, a difference of about 0.09\% from the path. Total path compute was about 315 s, against 466 s for the cold fine run. The compact mount thus showed a much smaller initialization effect than the slender bridge.

A coarse optimizer's own compliance under-predicted the finest-grid compliance of its resulting part by approximately 0.25--12.3\% in these experiments (bridge 2.7--12.3\%, motor mount 0.25--2.7\%); the bridge values include compliance growth at its edge supports (main-text Section 3.1). This comparison combines analysis resolution, projection, and conversion to a volume-matched binary part. It is not a mesh-convergence estimate holding all those choices fixed. Likewise, the element-based filter permits a shrinking physical length scale as the grid is refined. The study supports checking the actual reported fine-grid part model and following a useful continuation path on these two examples; it does not show that finer optima are always better or that the extracted structure satisfies manufacturing requirements.

\subsubsection*{Gate census and the validation-removal case}
The solve census contains 486 recorded solves, including intermediate computations and development diagnostics. It identifies 54 completed or recorded release attempts under its defined release/spill-monitoring criterion, of which 52 carried passing certificates. Two did not provide releasable results. Three additional spill-guard stops are documented separately: two at 113.2 million elements and one 59-million-element evaluation. The census also contains 11 computations that finished without optimization convergence. Intermediate converged solves without release monitoring are intentionally uncertified and must not be counted as erroneous release failures.

The certificate kept those 11 non-converged outcomes from being shown as finished designs. Any convergence flag would do the same, so the census documents the gate's operation, not an advantage over other acceptance checks or an estimated false-acceptance rate. The three spill stops demonstrate resource enforcement. They do not count as successful timings or increase the largest released mesh.

Disabling formulation validation provides a more direct test of the complementary rule gate. Two rigid-body-deficient briefs failed in the solver: one at its first state solve and the other after 20,000 preconditioned conjugate-gradient (PCG) iterations. The load-in-keep-out task behaved differently. It converged and received a passing numerical certificate, yet its compliance was approximately $3.8\times10^8$ times that of the corrected reference problem. The algebraic acceptance checks were satisfied although the load acted on keep-out elements held at the minimum stiffness; the rule that blocks this case needs no solved field, so it runs before the solve. This is one constructed case, chosen to show the mechanism rather than to estimate how often it occurs. In the benchmark runs the rules did no measurable work. Across the first-repeat G and G+C runs of the primary provider and the G+C runs of the second provider, agents called validation about once per run, blocking errors fired only on the designated refusal briefs (R003 on the slider bracket and hinged arm, R008 on the valve bracket), every such run ended in a refusal, and no run repaired a specification after a rule error (Table~\ref{S:tab:rulecensus}). With the rule findings removed, the same agent still refused all three (main-text Section 5.2), so in these runs the rules duplicated the agent's refusals rather than causing them, and they were not a measured source of corrected specifications.

This result establishes the narrower role of the numerical certificate. It checks whether the defined numerical computation met its requirements; it does not determine whether the defined load is supported by the intended material. The R008 formulation rule blocks that task before the expensive solve. Conversely, all six G intent errors passed the formulation rules, and no numerical predicate compares a specification with its brief. In this study only clarification against the oracle prevented them. Read as acceptance designs (the main-text acceptance-boundary table), the rules alone would have shown as finished designs the six intent errors, the eleven non-converged computations, and the release attempt that converged without a certificate (the other failed attempt produced no design); the certificate alone would have released the load-in-keep-out problem and the six intent errors, whose designs converged in the consequence evaluation although no certificate was issued for them; rules and certificate together could have released the six intent errors if their solves passed the certificate.

\subsubsection*{Complete sessions and padding-aware timing}
The complete-session study started from 15 briefs. Fourteen reached an exported STL and report with a passing release certificate and a specification matching the intended reading. The hinged-arm task, one of the three designated refusal tasks, was refused as intended after the rigid-body rule (R003) reported that its support left an unrestrained rotational mode, so all 15 sessions ended in the right behavior. The 15 briefs comprise eight benchmark-modeled problems, four fully stated engineering parts, two clarification tasks, and the refusal task. The authors chose them rather than sampling them, and each reported session is a single run with the primary provider, except that the cantilever block was run three times: its first session exposed three system faults and its second a fourth (event delivery), and all were fixed before the reported third run (Section~\ref{S:sec:suppsessions}).

\begin{figure}[!htb]
\centering\includegraphics[width=0.85\linewidth]{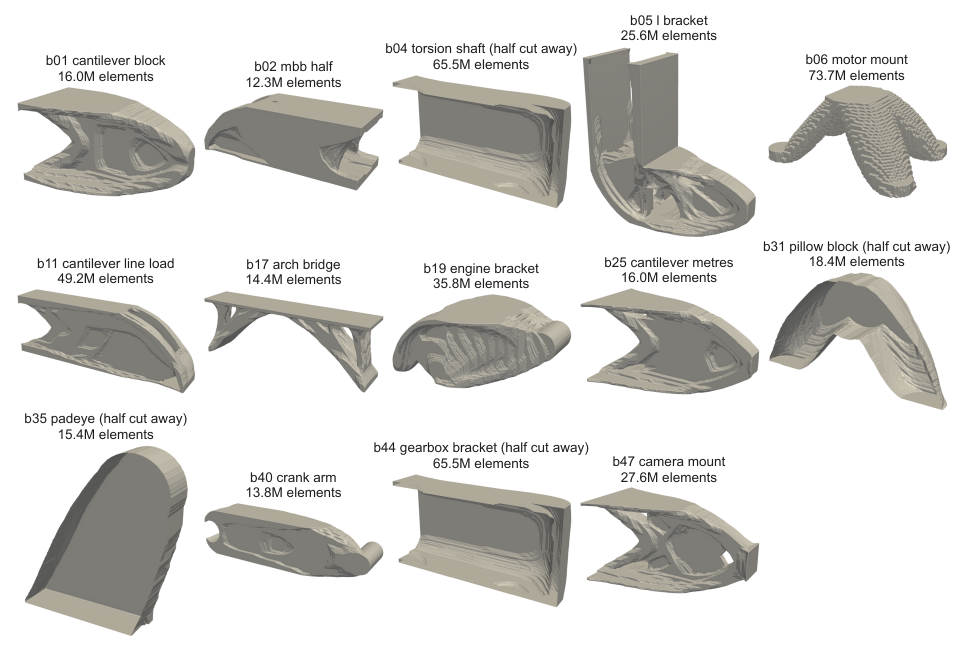}
\caption{\label{S:fig:suppgallery}The 14 released designs in the complete-session study. Labels identify the source task and rounded element count. Release certificates apply to the solved density fields. The gallery does not establish stress, buckling, manufacturing, or physical-test performance.}
\end{figure}

Thirteen of the 14 releases re-optimized on the release grid with the long re-opening schedule (61 iterations); the motor-mount session used the tail schedule (12 iterations) and therefore certified a transferred design. Every release grid was finer than the resolution in the session's specification, by a factor of two to eight in element edge length. The released part therefore answers a more finely regularized problem than the specification scored as matching the brief, and the success count does not register that difference. The certificate's regularization fields, reconstructed from the logs, make it visible: the release grid was two times finer than stated in one session, four times in ten and eight times in three, so the filter's physical length was a half, a quarter or an eighth of its length at the stated resolution (Section~\ref{S:sec:suppnumerical}). A session success here means a certified release derived from a matching specification, not a release scored against the brief at the stated resolution. All 15 original logs replayed exactly under the recorded configuration, including the refusal path. The successful release sizes in this study ranged from 12.3 to 73.7 million elements; the 25.6--73.7-million range in main-text Section 5.6 belongs specifically to the release-policy study. The main-text release gallery shows eight of the released parts with the supports and loads that their specifications state; Figure~\ref{S:fig:suppgallery} shows all 14.

Padding caused the largest outlier. The original engine-bracket session spent about 1508 s of compute in its release solve. A dedicated matched warm-start investigation reduced this to approximately 212 s with the padding-aware coarse policy at a relative compliance change of about $-6.5\times10^{-8}$. Across the padding diagnostics, the highest mean PCG iteration count per state solve across the tested configurations fell from 429 to 39. These observations concern the preconditioner and grid padding, not a changed acceptance residual or a language-model improvement.

\subsection*{Companion-work boundaries and extended validity account}
\addcontentsline{toc}{subsection}{Companion-work boundaries and extended validity account}
AutoSiMP already translates natural-language requirements into structured topology-optimization inputs and combines configuration with evaluation and retries \citep{Yang2026AutoSiMP}. Earlier optimization-controller work changes continuation parameters from observed solver state \citep{Yang2026Controllers}. Here, the agent primarily assembles and checks the design specification and chooses among fixed numerical schedules. We do not attribute the underlying density method or adaptive-continuation idea to the present study.

Consequence-guided clarification evaluates alternative readings of incomplete specifications using their mechanical consequences \citep{Yang2026Clarify}. The present interface incorporates alternative readings and consequence previews, but the benchmark oracle answers from stored facts. It therefore measures whether a model requests and uses clarification, not whether previewed consequences improve a person's choice. In natural-language processing, clarification questions have been ranked by the expected value of the information their answers supply \citep{Rao2018Clarify}; our benchmark instead scores whether the specification produced after asking matches the intended reading. We also do not claim that this interface implements the full minimax question-selection or cross-reading acceptance guarantee of that companion method.

The numerical backend builds on work on fused matrix-free operators, Galerkin multigrid, and guarded state-solve acceptance \citep{Yang2026Fused,Yang2026Galerkin,Yang2026Floor}. The new code adds session behavior, passive-region handling, deterministic replay support, and a padding-aware coarse support policy. Small verification experiments test those extensions without replacing the companion solver's evidence. Finally, observations that a language model may acknowledge an impossible problem while still labeling it solved motivate an acceptance status determined by executable checks \citep{Yang2026Impossible}. The gate-removal experiment evaluates that principle at the system level. Relative to those papers, the increment here is the shared typed specification with its rule gate and release contract, the logged agent--GUI tool interface, and their evaluation on the 50-brief suite.

No material is shared with those two studies: the clarification study selects questions without a language model on 36 tasks derived from published problems, the impossible-problem study scores standalone language models on 30 pairs of mechanics problems, and the briefs, refusal briefs, oracle and runs reported here were created for this study.

No other agent system was run on the suite as a baseline: AutoSiMP addresses mainly two-dimensional problems and has no released implementation that can run through the present tool contract.

The most immediate threat is reference quality. All 50 intended readings and 41 contested responses were adjudicated twice, but by two authors rather than by independent engineers, and six cited benchmark sources could not be traced.

The second pass, blinded by its package, agreed on 49 of 50 readings and 33 of 41 contested records; it left the grounding-and-clarification counts unchanged, credited a grounding-only output that J.W. had rejected, and so moved the G-to-G+C test to $p\approx0.063$; it placed the second provider's text-only count between 37, when both must accept a reading, and 44, when either may. Unit equivalence and mask comparison make the scoring reproducible, but they cannot establish that the stored reading is the only reasonable engineering interpretation. The contested outputs of the follow-up conditions below the rule in the main-text specification-outcome table, including their second repeat and second-provider run, were adjudicated once, by J.W., who saw each record's condition. Until an adjudication outside the author team is available, exact accuracy counts, card-acceptance totals, and paired $p$ values remain provisional.

An independent held-out suite would further test transfer beyond tasks developed alongside the system. The briefs also state topology-optimization parameters (axes, coordinates, volume fraction, element size) on box envelopes; requirements stated as performance targets or as CAD geometry are not tested.

The release grid is chosen by the system and is finer than the stated resolution, so a released design answers a more finely regularized problem than the specification that the benchmark scores. The evaluated version thus gives the system authority over that regularization: the engineer accepts the stated resolution, and the certificate records the release grid's physical filter length beside the accepted one, but no step asks the engineer to accept it. A deployed version should show both for acceptance and record them in the certificate.

The suite is small and heterogeneous, and some tasks are deliberately paired to isolate ambiguity. The single-load-case schema also shaped nine reference readings, which combine force components on one region (main-text Section 4.2). Model knowledge of common benchmark forms may help performance. Claude Opus 5.5 drafted the briefs and reference readings and also ran as the second provider, so provider comparisons may favor it. Only two providers and one model from each were studied, the second provider has one complete repeat, and the additional primary repeats cover unequal overlapping subsets. Tool prompts, inference settings, CLI behavior, and model updates may affect future results. Logged replay preserves an executed trajectory without establishing repeatability of fresh language decisions.

The numerical evaluation is limited to isotropic, small-displacement, linear-elastic compliance minimization on voxel boxes with one load case. It does not evaluate stress limits, buckling, fatigue, contact, nonlinear materials, manufacturing constraints, or structural tests. The reported residual, volume, passive-region, and spill predicates do not certify those properties. Even a closed and manifold exported mesh requires further analysis before fabrication or use.

The release-policy experiments have acknowledged failures and exclusions. The constant-physical-filter arms did not converge, the duct bridge did not provide the intended comparison, and one cold bridge required a solver change. Filter radii in element units also alter physical regularization with resolution. The two part-evaluation examples therefore support the stated release paths rather than a general convergence theorem or broad optimizer superiority.

Timing evidence is similarly bounded. Numerical studies are single-workstation runs with shared desktop resources, and most high-resolution comparisons are not randomized repeated timing campaigns. Padding-aware session totals are reconstructed from measured solve replays plus original agent time; they are not new agent runs. Backend latency does not include browser frame timing, and KLM estimates omit waiting and errors. No participants were recruited, so satisfaction, workload, learning, and trust remain unmeasured.

Finally, reproducibility depends on both artifacts and dependencies. The session log fixes the decisions needed to replay, but exact replay also requires the recorded solver, deterministic reduction policy, hardware behavior, and numerical environment. Reproducibility requires the solver package together with the application code; on other hardware, exact replay also needs matching reduction behavior and a fully specified numerical environment.

\subsection*{Detailed evaluation protocol}
\addcontentsline{toc}{subsection}{Detailed evaluation protocol}
The evaluation asks four questions. Does the explicit representation preserve the numerical behavior of the inherited solver? How often do different language protocols recover the intended specification, and how often do they deliver a wrong one as finished? What computation and compliance tradeoffs follow from a warm release? Finally, which failures are blocked by formulation rules, numerical certificates, and the complete session workflow?

The study/unit map is given in main-text Section 4.1.

The suite contains 50 briefs: 40 designated for solving, seven requiring clarification, and three requiring refusal (the main-text brief-suite table). Twenty-five are engineering-style requirements for brackets, mounts, bridges, housings, and related parts; the other 25 are modeled on published benchmark families in physical units. Adjudication confirmed the cited family for six of these, all rescaled or adapted, could not trace the cited source for six, and treats the remaining 13 generic family descriptions as engineering examples. The suite includes five length--force unit pairs, several support and load region types, and 19 tasks with passive regions. Fourteen tasks have credited alternative readings after adjudication. The full brief texts and recorded source attributions appear in Section~\ref{S:sec:briefs}.

Each task stores a reference specification, response class, clarification facts and accepted alternative readings. These alternatives concern fidelity to the brief, not mechanical admissibility. After unit normalization, a common-grid comparator checks the domain, material, resolution, volume fraction, supports, nodal loads and passive regions. Names and equivalent geometric syntax do not affect the score. The single-load-case schema combines force components on one region; it cannot represent separate load cases. Nine references use such combinations. Excluding them leaves right outcomes of 32/41 for T, 36/41 for G and 41/41 for G+C. The supplement gives the coordinate conventions, scope qualification and comparison tolerances.

Specifications are compared after unit normalization on a common grid. The supplement gives the material/load tolerances, geometric-mask comparison and grid-selection rules. Every credited record whose reference states an element size was compared at that size. Equivalence is a reproducible discrete criterion, not a proof of equality of continuous geometric descriptions.

The briefs and reference readings were drafted with Claude Opus 5.5, the second provider's model, during system development, and the 50-brief suite was fixed before the benchmark. J.W. later adjudicated all intended readings and the 41 contested delivered-wrong records across conditions. Forty-nine intended readings were confirmed and one support footprint was corrected. Contested outputs were credited when the brief left their differing geometry open. The supplement records the pilot changes, oracle correction, accepted readings and complete adjudication history.

S.Y. repeated the original adjudication blind to condition, provider, agent messages and first verdict. Agreement was 33/41 contested records (Cohen's $\kappa=0.60$ \citep{Cohen1960}). Scores use J.W.'s verdicts, with sensitivities reported under S.Y.'s and joint decisions. Later conditions were adjudicated once by J.W., with condition visible. These are author-adjudicated references, not independent ground truth. The supplement preserves the full chronology and re-scoring counts.
Every language-model condition emits the same typed specification and is scored by the same comparator. The comparisons therefore test protocols on that representation, not the representation itself. The five primary protocols are:
\begin{description}
\item[R: rule parser.] A deterministic parser with no language model; R is unrelated to rule codes R001--R013.
\item[T: text only.] The model emits JavaScript Object Notation (JSON) from the brief and schema, with refusal allowed.
\item[G: grounding.] The model uses region inspection, validation and rendering, but must choose assumptions rather than ask clarification questions.
\item[G+C: grounding and clarification.] G gains access to the scripted clarification oracle.
\item[G+C+V: picture check.] The corresponding G+C result undergoes three-view review and possible repair (main-text Section 3.4).
\end{description}
Later comparisons test the role of feedback and source information:
\begin{description}
\item[T+S: schema feedback.] T receives validator errors for at most two retries, without grounding tools.
\item[GNR: no rule feedback.] G's validator returns schema errors only, with no rule findings or grid summary.
\item[G+P: value-source tags.] G records whether each specification item was stated, derived, assumed or answered by clarification. Validation lists assumed items as an advisory finding; the schema and comparator are unchanged.
\item[G+C+P: restricted questions.] The same tags are recorded, and a question is accepted only about items the agent tagged assumed.
\end{description}

G+C+V reuses the corresponding G+C output before visual review. It is thus an incremental check on that result, not an independent fresh agent run. The G condition is also an intentionally constrained guessing protocol. Its comparison with G+C estimates the effect of permitting and instructing questioning in this harness; it does not estimate the effect of merely installing a question tool in an otherwise unchanged conversation.

The primary provider is Codex command-line interface (CLI), model GPT-6.1-Sol at recorded high reasoning effort. The second is Claude Code CLI, model Claude Opus 5.5 at high effort, tested on T, G+C and G+C+V for all 50 tasks and on G+C+P as a follow-up. Client versions, tool isolation and run dates are given above under Provider configuration and repeat records. These are historical run configurations, not current product defaults.

Primary paired tests use the first repeat, with each task contributing once per condition; repeat results are reported separately. Repeats use overlapping task subsets with separately reported denominators; they are not independent samples from an engineering-task population. The smaller second-provider default-effort diagnostic is not pooled with the matched-label comparison. Table~\ref{S:tab:t6_x2_repeats} gives every repeat, and Section~\ref{S:sec:suppprompts} the prompts, coordinate conventions, review schema and repair instructions.

The primary outcome, right behavior, means a specification equivalent to the intended or an accepted reading on a non-refusal task, or refusal on a task designated unsolvable as stated. The secondary correct-specification count records equivalence of any constructed specification, including one that correctly represents a problem subsequently refused. A refusal on an incomplete but answerable task is a failure on the primary outcome, but differs from returning a wrong specification as finished.

A delivered-wrong outcome is a schema-valid, non-refused, non-refusal-task result that fails the reference comparison. Right-behavior rates use all briefs in the specified repeat, including correct refusals; delivered-wrong values are counts, not a rate over delivered outputs. The system delivers such a specification as finished rather than withholding it as unresolved, even when the closing message states an assumption. We distinguish such outcomes from invalid JSON, explicit refusal, and tool failures.

The paired-test definition is given in main-text Section 4.4.

Consequence-weighted error supplements specification comparison. For a density designed under the predicted specification, we evaluate compliance under the intended problem and divide by the reference intended-design compliance, then subtract one. We report severe failures and material-budget changes rather than an arithmetic mean: a missing load path can produce an enormous ratio, while an invented solid region can improve stiffness by spending more material. Matching specification fields, mechanical consequence, and material use are therefore separate observations.

The workstation uses one NVIDIA GeForce RTX 4090 with 24 GB VRAM, an Intel Core i9-13900KF, and 96 GB host memory. Desktop processes share the GPU and held roughly 2--5 GB in the relevant memory investigations. This shared setting explains why nominal device capacity was insufficient to validate the planned 113.2-million-element release. Two guarded attempts stopped on detected spill, and that scale is excluded from the reported release results.

The solver uses structured hexahedral elements, matrix-free fine-level operators, multigrid-preconditioned state solves, density filtering and projection, and optimality-criteria updates. Release runs enable spill checks every ten optimization iterations. Cold, tail, short re-opening, and long re-opening schedules share the terminal parameters, but differ in their initial density and continuation path. The optimization cap is 300 iterations in the release-policy experiments. The supplement records the schedule rows and implementation sources.

The release-policy study compares five completed problems at their own final grids: cantilever, bridge, torsion, an L-bracket, and a motor mount. A sixth duct-bridge case failed to provide a usable escalation comparison and is reported as an exclusion with failure details. A constant-physical-filter pilot also did not meet the stagnation test within 300 iterations on any of the three problems tried (about one million elements each) and was not promoted into the large-scale comparison. The successful sample therefore establishes feasibility and tradeoffs on those five problems, not universal escalation reliability.

Compute time refers to the solver's recorded computation. Original end-to-end session time runs from the first logged tool call to the final export or report and excludes CLI startup. We also retain CLI wall time, which includes startup and surrounding provider execution. The padding-aware session number is reconstructed by replacing original solve durations with measured replay durations while retaining original agent and other session time. It is not a new live agent session or an end-user measurement. Interface latency is measured through the backend; browser frame rendering is outside that timing boundary.

\subsection*{Continuation and knowledge-record details}
\addcontentsline{toc}{subsection}{Continuation and knowledge-record details}
Warm transfer upsamples the coarse density, applies one fine-grid filter pass, and adjusts a constant density offset by bisection so that the projected design volume matches the target at the new schedule's initial projection sharpness. The volume correction is needed because passive regions can voxelize differently at successive resolutions. An apparently feasible coarse field can otherwise become infeasible on the fine design mask.

We distinguish three continuation schedules after transfer. The default fast tail schedule uses only the final phase, with penalization $p=4.5$, projection sharpness $\beta=32$, and move limit 0.05. A short re-opening lowers the initial sharpness but permits limited change. The long re-opening starts at $p=3.5$, $\beta=4$, and move limit 0.15, then continues to the final sharpened phase. Table~\ref{S:tab:schedules} gives the full schedules.

A sharply projected upsampled field can be saturated away from the projection threshold. Continuing only the final phase then changes essentially no solid/void states. In that case, fine-grid release is an evaluation and acceptance of the transferred design. The longer re-opening allows boundary movement and measured compliance improvement, but it can retain much of the coarse topology. Neither warm policy guarantees the same local optimum as a cold fine-grid run or the creation of new members.

These checks are deliberately bounded. Passing them does not prove that a specification follows from its natural-language brief, that every possible physical defect is excluded, or that the retained design is connected after surface extraction. We test selected rules by disabling validation, rather than treating a clean rule report as a general correctness proof.

The three records of the main-text knowledge-model table are summarized below: what each holds and where it comes from, how it is represented and who has authority over it, and what it exposes.

\paragraph{Specification (intent)}
Domain, units, material, supports, loads, retained regions and stated resolution, from the brief and the engineer's answers to clarification. A typed schema with geometric regions; the engineer accepts it, except the release grid and the filter length it implies, which the system sets. It makes the reading inspectable and comparable with a reference. By default it records no typed field provenance, only free-text notes; an optional per-item source tag marked every wrong reading of the primary provider in tests but most right ones too, so neither reliably reveals a well-formed wrong reading by itself.

\paragraph{Rule base (admissibility)}
Necessary conditions for a well-posed, feasible problem, stricter faithfulness checks, a load-path plausibility check, plus execution and advisory checks, from finite-element practice. Deterministic functions of specification and grid; an error blocks the solve. It exposes empty selections, free rigid-body modes, contradictory passive regions and a load inside a keep-out region, but not a partial keep-out load, disconnected solids, an implausible modulus, or a wrong reading unless it also violates a rule.

\paragraph{Certificate (numerical acceptance)}
Design stagnation (the solver's `converged' test), residual, volume and passive-region predicates, plus execution predicates for spill monitoring, from the solver. A record issued by the solver and required for release. It exposes non-converged runs, inaccurate state solves, volume and passive violations and memory spill, but not a contradictory but solvable problem, an intent error, or stress and manufacturing performance. It records the release grid's physical filter length beside the one at the stated resolution, and the continuation schedule, but the engineer does not accept them.

All final arms on five completed problems passed their certificates at 25.6--73.7 million elements (the main-text release-policy table). Long re-opening differed from a single cold run by $-3.06\%$ to $+0.07\%$ compliance at 51--69\% of the cold run's final-rung compute (the main-text release-policy figure); negative values mean lower compliance. The fast tail used 3--11\% of cold compute at $+0.005\%$ to $+1.39\%$ compliance, but changed essentially no binary design after transfer. It chiefly certifies the coarse-derived topology.

Including earlier warm rungs raises the long re-opening's compute share to 59--80\%, or 61--84\% with preview. These are specified compute boundaries, not end-to-end speedups from an empty session. There is one run per arm, no uniform-start control with the shorter schedule, and different physical regularization histories. The cold runs are therefore local references; differences of a few hundredths of a percent are best read as ties.

The warm designs also retained different topology: pooled binary intersection over union was about 0.86--0.96 against cold designs. A passing certificate and similar compliance do not imply the same optimum. The bridge cold run required a different successful solver configuration after two mixed-precision failures; its warm arms used the default solver. The supplement documents that asymmetry.

The duct-bridge exclusion also bounds the result. Its preview failed to converge, finer fields developed a lattice-like structure, and warm or cold release attempts did not supply a stable comparison. Physical-filter-radius pilots failed within their iteration cap as well. These failures show that successful numerical acceptance and a favorable warm/cold tradeoff remain problem-dependent.

The resolution study evaluates two problems: a 1280 by 160 by 160 mm span bridge with an 8 mm retained deck, and the motor mount with retained mounting regions. Each candidate is transferred to the finest grid and binarized to match the design-material fraction, then evaluated under the same finest-grid state model. This comparison removes the advantage of judging a coarse design solely by its coarse analysis, while keeping the limitations of voxelization and the common analysis model.

On the span bridge, the warm path's finest-grid part compliance fell from 2715 to 2382 N\,mm, 28.5\% below the cold finest result of 3333 N\,mm (the main-text common-grid figure). Total path compute was about 413 s versus 466 s for the cold finest run. On the motor mount, the final path result was 3.915 versus 3.912 N\,mm, about $+0.09\%$, at 315 versus 466 s. The supplement lists every rung and the cavity counts. The bridge's stronger initialization effect is a result on these two problems, not a general advantage of coarse starts.

The coarse optimizer under-predicted its resulting part's finest-grid compliance by about 0.25--12.3\%. This combines analysis resolution, projection and volume-matched binarization; edge-supported bridge values also include singularity effects (main-text Section 3.1). The warm and cold paths use element-based filters, hence different physical regularization histories. These are common-grid part comparisons, not mesh-convergence estimates or manufacturing validation.

\section{Numerical continuation and release interpretation}
\label{S:sec:suppnumerical}
\begin{table}[H]
\centering\small
\caption{Continuation rows in the release-policy implementation. The second column is a cumulative schedule boundary; the final row extends to the stopping criterion or run cap. Filter radius is in element units; $p$ is the SIMP penalty exponent and $\beta$ the projection sharpness. The optimization run cap is 300 iterations.}
\label{S:tab:schedules}
\begin{tabular}{llrrrr}
\toprule
Policy & Boundary & $p$ & $\beta$ & Move limit & Filter radius \\
\midrule
Cold & 15 & 1.5 & 1 & 0.20 & 1.50 \\
Cold & 40 & 3.5 & 4 & 0.15 & 1.35 \\
Cold & 65 & 4.5 & 16 & 0.08 & 1.25 \\
Cold & 120 & 4.5 & 32 & 0.05 & 1.20 \\
Short re-opening & 4 & 4.5 & 8 & 0.10 & 1.20 \\
Short re-opening & 10 & 4.5 & 16 & 0.08 & 1.20 \\
Short re-opening & 1000 & 4.5 & 32 & 0.05 & 1.20 \\
Long re-opening & 25 & 3.5 & 4 & 0.15 & 1.35 \\
Long re-opening & 50 & 4.5 & 16 & 0.08 & 1.25 \\
Long re-opening & 1000 & 4.5 & 32 & 0.05 & 1.20 \\
Tail & final phase & 4.5 & 32 & 0.05 & 1.20 \\
\bottomrule
\end{tabular}
\end{table}

The cold policy starts from a uniform density. Warm policies start from the preceding rung's field after upsampling, one filter pass, and volume matching. A saturated sharp projection can prevent topology change during tail continuation. Therefore a passing tail certificate supports acceptance of the transferred density at the new grid; it does not demonstrate optimization from a uniform fine-grid start.

The scale comparison completed cantilever, bridge, torsion, L-bracket, and motor-mount problems. The bridge cold release needed an FP64 multigrid configuration after iterative-refinement failures, so that timing comparison includes a solver-configuration difference. The duct bridge and constant-physical-filter pilots failed to supply the planned successful comparisons and were excluded from the completed five-problem result.

\begin{figure}[!htb]
\centering\includegraphics[width=\linewidth]{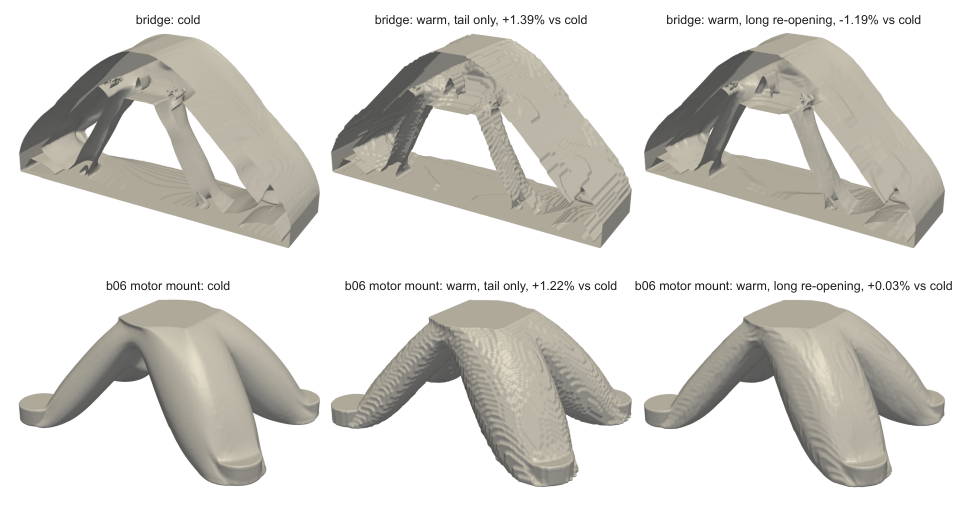}
\caption{Cold, tail, and long re-opening densities for selected release-policy problems. Surface similarity is descriptive; compliance is compared on the corresponding final grid. Binary overlap statistics use pooled masks for large fields. Tail primarily preserves the transferred topology. Surfaces are closed $\rho=0.5$ isosurfaces; the bridge's near half is cut away to expose the struts behind its full-depth side webs. Each design is a single connected solid.}
\end{figure}
\input{tables/supp_t9_x5_resolution}

In the complete agent sessions (Section~\ref{S:sec:suppsessions}) the certificate's regularization fields were reconstructed from each log (release job, solve call and released specification): release grids two (b01), four (ten sessions) or eight (b04, b11, b44) times finer than stated, filter radius 1.2 elements throughout, long re-opening with 61 iterations on the release grid in 13 sessions and the tail schedule with 12 iterations for the motor mount.

\section{Complete sessions, replay, and latency}
\label{S:sec:suppsessions}
The original-session column is first logged call to export/report. Across the 14 released tasks, its median is 179 s and its range 77--1877 s. Including CLI startup and surrounding provider execution, recorded wall time has median 199 s and range 93--1898 s. The padding-aware column is reconstructed, with median 132 s and range 86--334 s. Exact replay applies to the original logged configuration; the alternative coarse policy produces small compliance changes and can change exported surface hashes.
\input{tables/main_t8_x3_sessions}

The cantilever block was run three times. The first session was not released because of faults in multigrid padding, resolution consistency, and replay timing; the second exposed an additional event-delivery fault. After correction, both later runs released the same design; the third is reported.

\begin{figure}[!htb]
\centering\includegraphics{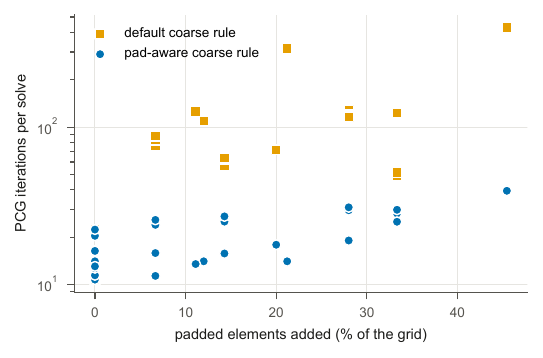}
\caption{Padding diagnostics for the coarse support policy. PCG denotes preconditioned conjugate gradient. The padding-aware policy changes the multigrid preconditioner; final state acceptance still uses a recomputed true residual. Changes in timings on unpadded grids include run-to-run noise.}
\end{figure}
\input{tables/t11_x6a_latency}
\begin{figure}[!htb]
\centering\includegraphics{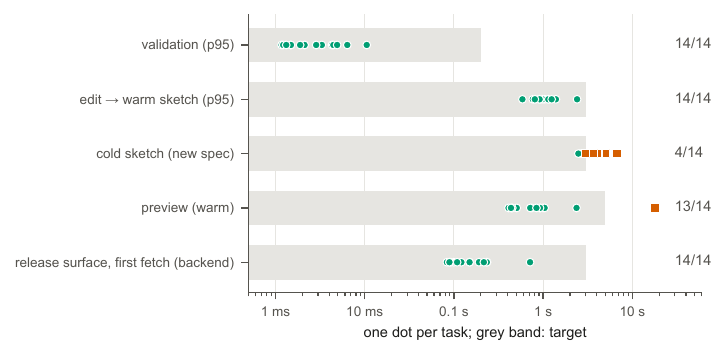}
\caption{Backend interaction latencies and the stated target bands; p95 denotes the 95th percentile. Fast interactions were repeated 20 times per task; each complete ladder was run once. Cold sketches often exceed the target. Warm release is tail continuation, rather than a cold fine-grid optimization. Frontend frame timing was not measured.}
\end{figure}

\section{Interaction-cost model and proposal cards}
The Keystroke-Level Model uses 0.28 s per keystroke, 1.1 s per pointing action, 0.1 s per button press or release, 0.4 s per hand movement between keyboard and mouse, and 1.35 s per mental operator. A click comprises two button events. Mental operators precede chunks such as typed values, choices, and code lines. The agent condition adds review of each proposed card, clarification-chip choices, and the same release/export clicks as the GUI condition. Solver and agent waiting are excluded. These are model assumptions, not estimated participant parameters.
\input{tables/t10_x6b_klm}
\begin{figure}[!htb]
\centering\includegraphics{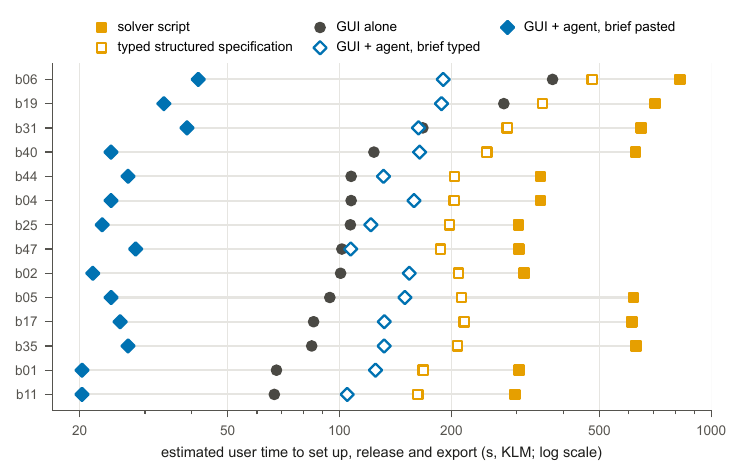}
\caption{Keystroke-Level Model (KLM) estimates of expert interaction time over the 14 released tasks. Existing pasted text removes translation actions; typing the brief can cost more than graphical-interface entry for a simple problem. The model excludes waiting, mistakes, debugging, learning, and subjective workload.}
\end{figure}
Proposal-card scoring replays each final specification as a proposal on an empty workspace. The scripted oracle accepts matching items, edits a wrong item corresponding to a requested one, and rejects extras. G+C and G+C+V each yield 423 accepted cards with no edits or rejections under the reference readings. G yields 436 cards with six edits and ten rejections. These counts are reference-based programmatic agreement, not a participant acceptance study.

\section{Rule findings in the benchmark runs and the rule-removal case}
\label{S:sec:rulecensus}
Every call of the validation tool in the saved session logs records its findings. Table~\ref{S:tab:rulecensus} counts them for the first-repeat runs of the tool-using conditions; the picture-check condition reuses the G+C specification and is not counted separately. Blocking errors fired only on the three designated refusal briefs, and each of those runs ended in a refusal. No run edited its specification after a rule error and validated again. Agents called validation about once per run, so rule feedback had little opportunity to drive repairs. The two second-provider unit slips passed every rule, because no rule checks the plausibility of the modulus.

\begin{table}[H]
\centering\small
\caption{Rule findings in the first-repeat benchmark runs. Calls: validation calls; Finding, Warning, Error: runs with at least one finding, warning or error; Repaired: runs that edited the specification after an error and validated again. Blocking errors concerned free rigid-body motion on b08 and b49 and a load inside a keep-out on b50 (b08 and b50 only for the second provider). The other findings were one warning about a load on fixed displacement components per condition and advisory coarse-multigrid support notes.}
\label{S:tab:rulecensus}
\begin{tabular}{lrrrrrr}
\toprule
Condition & Runs & Calls & Finding & Warning & Error & Repaired\\
\midrule
G, primary provider & 50 & 50 & 15 & 1 & 3 & 0\\
G+C, primary provider & 50 & 53 & 15 & 1 & 3 & 0\\
G+C, second provider & 50 & 49 & 14 & 1 & 2 & 0\\
\bottomrule
\end{tabular}
\end{table}

The rule-removal case (b50, valve bracket) was solved twice on a $60\times40\times20$ grid ($h=2$~mm, 48,000 elements): as specified, with the load pad inside the keep-out region, and with the pad moved to $x=40$~mm, outside it, as the corrected reference. Both runs converged in 76 iterations and passed every certificate predicate, with true relative residuals of $8.8\times10^{-6}$ and $3.9\times10^{-6}$ against the tolerance $10^{-5}$, exact volume and passive regions, no memory spill, and spill monitoring active. Their compliances were $1.449\times10^{8}$ and $0.3817$~N\,mm, a ratio of $3.79\times10^{8}$. The two rigid-body-deficient briefs failed in the solver: b08 at the first state solve (non-positive curvature) and b49 after 20,000 iterations with a true residual of $2.4\times10^{4}$ against the tolerance $10^{-5}$.

Two follow-up checks used the suite's reference specifications. In the first, each reference specification was checked unchanged and with one injected fault, on the grid of the rule-removal runs (the shortest side divided by 20, or the specification's own element size where that does not divide every side); a fault counts as caught when a blocking error appears that the unchanged specification lacks. Table~\ref{S:tab:faultmatrix} gives the counts for the 47 specifications that pass every rule; the three refusal briefs already fail. The three missed load-on-support faults moved one of several loads, and the load-on-support rule reports a warning when at most half of the load is lost. The four rule-class faults mirror the zero-load, load-on-support, rigid-body, and keep-out checks, so they test the rules as defined. Two defects outside the rule set were injected as well: the first load moved to straddle a keep-out face, half inside and half outside (5 references, b28, b30, b41, b43 and b46, in which the keep-out ends inside the domain along the axis with the most room beyond its upper face), and the modulus divided by 1000; no rule caught either.

The intent faults that were caught tripped other rules: the invented pad overlapped a keep-out region, which makes the problem infeasible (the passive-overlap rule), or the moved support put more than half of the applied load on fixed components, which the stricter load-on-support rule rejects.

In the second check, the share of the compliance carried by strain energy in keep-out elements, $E_{\min}\sum_{e\in\mathcal V}u_e^{\mathsf T}k_0u_e/C$, was computed after solving with the rules bypassed and the solver settings of the rule-removal runs. It was 0.99999997 for the b50 specification with the pad inside the keep-out, $2.9\times10^{-11}$ with the pad moved to $x=40$~mm, and $1.6\times10^{-10}$ to $2.4\times10^{-8}$ for the eight valid reference readings with keep-out regions (b05, b09, b18, b28, b30, b41, b43, b46); b09 stopped at the 150-iteration cap.

\begin{table}[H]
\centering\small
\caption{Rule outcomes for faults injected one at a time into the 47 reference specifications that pass every rule; rows with fewer specifications apply only where the reference has the needed region. A fault is caught when a blocking error appears that the unchanged specification lacks. The volume fraction was halved where $\times1.5$ would reach 0.9.}
\label{S:tab:faultmatrix}
\begin{tabular}{llrrl}
\toprule
Class & Injected fault & Specifications & Caught & Blocking rule\\
\midrule
Rule classes & Zero force on every load & 47 & 47 & R004\\
 & First load moved onto a fully fixed support & 46 & 43 & R005\\
 & Supports reduced to one, fixing one direction & 47 & 47 & R003\\
 & First load moved inside a keep-out region & 8 & 8 & R008\\
Outside rules & First load straddling a keep-out face & 5 & 0 & --\\
 & Modulus divided by 1000 & 47 & 0 & --\\
Intent & Force signs flipped & 47 & 0 & --\\
 & Forces $\times1000$, modulus kept (unit slip) & 47 & 0 & --\\
 & Solid pad of one fifth of each side at the center & 47 & 6 & R007\\
 & Volume fraction $\times1.5$ or halved & 47 & 0 & --\\
 & First support moved to the opposite face & 38 & 15 & R005\\
\bottomrule
\end{tabular}
\end{table}

\section{Prompts and coordinate convention}
\label{S:sec:suppprompts}
The following strings are reproduced verbatim from the benchmark runner. Braced fields are template substitutions. The coordinate frame text defines coordinates, the text-only prompt (T) requests JavaScript Object Notation (JSON) output or refusal, the grounding prompt (G) defines tool-grounded setup, the clarification text distinguishes guessing from questioning, the picture-check prompt requests image review, and the repair prompt applies actionable review feedback. For the follow-up conditions, the provenance text and the schema-feedback retry are verbatim, and the way the runner combines them with the prompts above is stated in words. The schema supplied to T is included as a separate machine-readable file in this package.
\input{supplement/prompts}

\section{All 50 design briefs}
\label{S:sec:briefs}
The following are the complete experimental inputs, reproduced without editorial correction. Expected response labels and source strings are experimental metadata. The intended and admissible specifications are provided in \path{anc/supplement/reference_readings.json}; the adjudication record is in Section~\ref{S:sec:adjudication}. A benchmark source indicates the originating problem family, not necessarily a verbatim quotation or an independently confirmed match in all engineering details.
\input{supplement/briefs}

%% file: tables/t5_x2_mcnemar.tex
\begin{table}[t]
\centering\small
\caption{Paired comparisons of right behavior (exact \href{https://doi.org/10.1007/BF02295996}{McNemar} comparison, 50 briefs)}
\label{S:tab:t5_x2_mcnemar}
\begin{adjustbox}{max width=\linewidth}
\begin{tabular}{lrrr}
\toprule
Comparison & Fixed & Broken & p \\
\midrule
R $\rightarrow$ T & 28 & 0 & 7.5e-09 \\
T $\rightarrow$ G & 8 & 1 & 0.039 \\
G $\rightarrow$ G+C & 6 & 0 & 0.031 \\
T $\rightarrow$ G+C & 13 & 0 & 0.00024 \\
G+C $\rightarrow$ G+C+V & 0 & 0 & 1 \\
T $\rightarrow$ T+S & 7 & 2 & 0.18 \\
T+S $\rightarrow$ G & 3 & 1 & 0.62 \\
G $\rightarrow$ GNR & 0 & 2 & 0.5 \\
G $\rightarrow$ G+P & 0 & 2 & 0.5 \\
G+C $\rightarrow$ G+C+P & 0 & 3 & 0.25 \\
\bottomrule
\end{tabular}
\end{adjustbox}
\par\smallskip\footnotesize Scores use the adjudicated reference readings.
\end{table}

%% file: tables/t6_x2_repeats.tex
\begin{table}[t]
\centering\small
\caption{Run-to-run variation (repeats 2 and 3 on subsets)}
\label{S:tab:t6_x2_repeats}
\begin{adjustbox}{max width=\linewidth}
\begin{tabular}{lrrrr}
\toprule
Condition & Repeat & Briefs & Right behavior & Delivered wrong \\
\midrule
T text only & 1 & 50 & 37 & 1 \\
T text only & 2 & 50 & 37 & 3 \\
T text only & 3 & 20 & 9 & 3 \\
G grounding tools & 1 & 50 & 44 & 6 \\
G grounding tools & 2 & 50 & 42 & 8 \\
G+C + clarification & 1 & 50 & 50 & 0 \\
G+C + clarification & 2 & 32 & 32 & 0 \\
G+C + clarification & 3 & 20 & 20 & 0 \\
G+C+V + picture check & 1 & 50 & 50 & 0 \\
G+C+V + picture check & 2 & 20 & 20 & 0 \\
G+C+V + picture check & 3 & 20 & 20 & 0 \\
G+P G + value provenance & 1 & 50 & 42 & 8 \\
G+P G + value provenance & 2 & 20 & 13 & 7 \\
\bottomrule
\end{tabular}
\end{adjustbox}
\par\smallskip\footnotesize Scores use the adjudicated reference readings.
\end{table}

%% file: tables/supp_t9_x5_resolution.tex
\begin{table}[H]
\centering\small
\caption{Part compliance along the release path, judged at the finest grid}
\label{S:tab:t9_x5_resolution}
\begin{adjustbox}{max width=\linewidth}
\begin{tabular}{lrrrrr}
\toprule
Problem & Arm & Elements & Part compliance (N\,mm) & vs cold fine & Compute (s) \\
\midrule
span bridge & release path & 0.06M & 2715 & -18.5\% & 24 \\
span bridge & release path & 0.51M & 2506 & -24.8\% & 32 \\
span bridge & release path & 4.10M & 2422 & -27.3\% & 124 \\
span bridge & release path & 32.77M & 2382 & -28.5\% & 232 \\
span bridge & cold fine run & 32.77M & 3333 & 0\% & 466 \\
b06 & release path & 0.14M & 4.012 & +2.6\% & 6 \\
b06 & release path & 1.15M & 3.97 & +1.5\% & 8 \\
b06 & release path & 9.22M & 3.933 & +0.6\% & 40 \\
b06 & release path & 73.73M & 3.915 & +0.1\% & 260 \\
b06 & cold fine run & 73.73M & 3.912 & 0\% & 466 \\
\bottomrule
\end{tabular}
\end{adjustbox}
\end{table}

%% file: tables/main_t8_x3_sessions.tex
\begin{table}[t]
\centering\small
\setlength{\tabcolsep}{4pt}
\caption{Complete agent sessions and reconstructed padding-aware timings.}
\label{S:tab:t8_x3_sessions}
\begin{tabular}{lrrrrrr}
\toprule
Task & \shortstack{Elements\\(million)} & Calls & Questions & \shortstack{Input\\tokens} & \shortstack{Original\\session (s)} & \shortstack{Aware\\estimate (s)} \\
\midrule
b01 & 16.0 & 23 & 1 & 475k & 125 & 92 \\
b02 & 12.3 & 22 & 1 & 422k & 77 & 89 \\
b04 & 65.5 & 21 & 0 & 478k & 140 & 154 \\
b05 & 25.6 & 26 & 1 & 564k & 236 & 140 \\
b06 & 73.7 & 36 & 4 & 549k & 176 & 154 \\
b11 & 49.2 & 25 & 1 & 567k & 273 & 232 \\
b17 & 14.4 & 27 & 1 & 823k & 351 & 119 \\
b19 & 35.8 & 59 & 1 & 2959k & 1877 & 334 \\
b25 & 16.0 & 21 & 0 & 611k & 182 & 102 \\
b31 & 18.4 & 28 & 2 & 584k & 128 & 124 \\
b35 & 15.4 & 21 & 0 & 419k & 116 & 86 \\
b40 & 13.8 & 32 & 0 & 1179k & 463 & 106 \\
b44 & 65.5 & 25 & 1 & 521k & 138 & 145 \\
b47 & 27.6 & 27 & 1 & 730k & 287 & 148 \\
b49 & -- & 7 & 0 & 92k & 4 & 4 \\
\bottomrule
\end{tabular}
\par\smallskip\footnotesize Fourteen tasks released a solver-certified stereolithography (STL) surface; b49 was refused. All 15 original sessions replayed exactly. Aware estimates replace solve times and retain original agent time; they do not imply bit-exact replay under the changed policy. Tokens include cached input.
\end{table}

%% file: tables/t11_x6a_latency.tex
\begin{table}[t]
\centering\small
\caption{Interactive latency through the backend (14 tasks, 20 repeats of the fast interactions)}
\label{S:tab:t11_x6a_latency}
\begin{adjustbox}{max width=\linewidth}
\begin{tabular}{lrrr}
\toprule
Interaction & Target & Median over tasks & Range \\
\midrule
Spec edit $\rightarrow$ validation (p50 per task) & < 200 ms & 2.15 ms & 1.01--10.2 ms \\
Spec edit $\rightarrow$ validation (p95 per task) & < 200 ms & 2.49 ms & 1.17--10.5 ms \\
Edit commit to the specification (p50) & -- & 0.925 ms & 0.88--1.14 ms \\
Edit $\rightarrow$ sketch, warm (p50 per task) & 1--3 s & 0.934 s & 0.577--2.13 s \\
Edit $\rightarrow$ sketch, warm (p95 per task) & 1--3 s & 0.951 s & 0.586--2.4 s \\
Edit $\rightarrow$ first sketch frame (p50) & -- & 0.0943 s & 0.0779--0.221 s \\
Sketch, cold (new spec) & 1--3 s & 3.32 s & 2.48--6.75 s \\
Preview, warm & $\approx$ 5 s & 0.794 s & 0.41--18 s \\
Refine, warm & 2--5 s compute & 1.52 s & 0.721--17.5 s \\
Release, warm, compute & $\approx$ 10--25 s & 6.65 s & 3.14--56.4 s \\
Release surface, first fetch (backend) & < 3 s & 0.136 s & 0.0834--0.714 s \\
STL export at release & -- & 0.593 s & 0.298--2.19 s \\
\bottomrule
\end{tabular}
\end{adjustbox}
\end{table}

%% file: tables/t10_x6b_klm.tex
\begin{table}[t]
\centering\small
\caption{Interaction cost by Keystroke-Level Model (14 released tasks)}
\label{S:tab:t10_x6b_klm}
\begin{adjustbox}{max width=\linewidth}
\begin{tabular}{lrrr}
\toprule
Condition & Median user time (s) & Range (s) & Median actions \\
\midrule
A1 solver script & 480 & 296--823 & 1530 \\
A2 typed structured specification & 208 & 162--479 & 680 \\
B GUI, no agent & 104 & 67--374 & 94 \\
C GUI + agent, brief typed & 141 & 105--190 & 418 \\
C GUI + agent, brief pasted & 25 & 20--42 & 8 \\
\bottomrule
\end{tabular}
\end{adjustbox}
\par\smallskip\footnotesize A model, not a measurement: error-free expert, solver and agent waiting excluded.
\end{table}

%% file: supplement/prompts.tex
\Needspace{8\baselineskip}
\subsection{Coordinate frame}
\begin{lstlisting}
Region coordinates are absolute in the domain frame, origin at the domain's min corner. A face patch gives
its centre and full size in the face's two in-plane axes in increasing axis order: (y, z) on x faces, (x, z) on y faces,
(x, y) on z faces. Loads are total forces on the region.
\end{lstlisting}
\Needspace{8\baselineskip}
\subsection{Text-only prompt (T)}
\begin{lstlisting}
You translate an engineering design brief into a topology-optimization spec.
Return ONLY one JSON object that validates against this JSON Schema (agentto.spec/v0). No prose, no code fences.
{frame}
If the brief cannot be solved as stated, return {{"refuse": "<one sentence why>"}} instead.

JSON Schema:
{schema}

Brief:
{brief}
\end{lstlisting}
\Needspace{8\baselineskip}
\subsection{Grounding prompt (G)}
\begin{lstlisting}
You set up a 3D topology-optimization problem from an engineering brief, using the agentto tools only.
Build the complete spec with the tools, check it with list_regions and validate_spec and fix any errors, and look at
render once to confirm the supports and loads are where the brief says. {clarify}
If the brief cannot be solved as stated (for example the part cannot be held still by the given supports), say so:
finish with REFUSE and one sentence why. Otherwise finish with one line: DONE, then any assumptions you made.

Brief:
{brief}
\end{lstlisting}
\Needspace{8\baselineskip}
\subsection{Clarification text}
\begin{lstlisting}
G: Do not ask questions: where the brief is silent, choose a reasonable value and state it.

GC: Where the brief is ambiguous or silent on something that changes the design, call ask_clarification with 2-4 concrete readings (the full item or parameter values each would give); the user answers. Do not ask about things the brief states.

GP: the G text, followed by the provenance text below.

GPC: the provenance text, followed by: Where an item you tagged assumed changes the design, you may call ask_clarification with 2-4 concrete readings (the full item or parameter values each would give) and its key in about; questions about items tagged stated or derived are rejected. After an answer, apply it and tag the item answered.

Provenance text: For every spec item (units, domain, material, volume_fraction, resolution, and each support, load and passive region) record where its value comes from with set_source: stated (the brief gives it), derived (computed from numbers the brief gives), or assumed (your own choice where the brief is silent or ambiguous). Call list_sources before you finish; every item must be tagged.
\end{lstlisting}
\Needspace{8\baselineskip}
\subsection{Picture-check prompt}
\begin{lstlisting}
You check a topology-optimization setup against its engineering brief. The pictures show the design
domain (box), supports (blue, with the fixed directions), loads (red arrows, labelled with force) and keep-in (green
solid) / keep-out (dashed) regions, from three views. Compare them with the brief: is every support, load and
keep-in/keep-out region where the brief says, with the right direction and size? Ignore colours and styling.
Return JSON: {{"match": true|false, "issues": ["<one concrete mismatch per item>", ...]}}.

Brief:
{brief}
\end{lstlisting}
\Needspace{8\baselineskip}
\subsection{Repair prompt}
\begin{lstlisting}
You set up a 3D topology-optimization problem from an engineering brief, using the agentto tools only. The
session already holds a spec built from the brief. A reviewer compared pictures of it with the brief and reported:
{issues}
Check each point against the brief with get_spec and list_regions. Fix the spec where the reviewer is right (keep
what is already correct), validate it, and where the brief is ambiguous call ask_clarification. Finish with one line:
DONE, then what you changed.

Brief:
{brief}
\end{lstlisting}
\Needspace{8\baselineskip}
\subsection{Schema-feedback retry (T+S)}
\begin{lstlisting}
<text-only prompt, as above>

Your previous reply was not a valid spec. The validator reported:
{errors}

Previous reply:
{previous}

Return ONLY the corrected JSON object (or the refusal object), nothing else.
\end{lstlisting}

%% file: supplement/briefs.tex
\Needspace{8\baselineskip}
\subsection{b01 cantilever block}

\noindent Expected response: solve.

\begin{quote}\small Bracket design, rev A

Envelope: a rectangular block 200 mm long (x), 100 mm tall (y) and 100 mm deep (z). Steel, E = 210 GPa,
nu = 0.3.

Mounting: the whole end face at x = 0 is welded to a wall (fully fixed).

Load: 1 kN pushing straight down (negative y) on a 10 mm x 10 mm pad in the middle of the free end face (x = 200).

Volume fraction 0.3. Mesh at 1 mm elements.\end{quote}

\noindent Source recorded by suite authors: restates the voxfem solver's 3D cantilever benchmark (after Sigmund) in physical units.

\noindent Adjudication: intended reading confirmed; cited source not traced to a specific published problem.

\Needspace{8\baselineskip}
\subsection{b02 mbb half}

\noindent Expected response: solve.

\begin{quote}\small Simply supported beam, symmetric half model

The full beam is 240 mm long (x), 40 mm deep (y, up) and 40 mm wide (z), aluminium 6061. It sits on two
supports along the bottom edges at its two ends, across the full width; the supports carry vertical and
sideways load only. A 2 kN downward line load acts across the full width at mid-span on the top surface.

Model only the left half (x from 0 to 120 mm) and use symmetry at mid-span. Volume fraction 0.3, 1 mm elements.\end{quote}

\noindent Source recorded by suite authors: 3D MBB beam (half model), as in \citet{Liu2014Top3d}, top3d.

\noindent Adjudication: intended reading confirmed; benchmark family confirmed.

\Needspace{8\baselineskip}
\subsection{b03 bridge deck}

\noindent Expected response: solve.

\begin{quote}\small Bridge segment

A deck segment spanning 400 mm (x), 100 mm high (y, up) and 100 mm wide (z), steel. It rests on two bearings
along the bottom edges at each end, running across the full width. The left bearing is pinned; the right one
lets the deck expand along the span. A uniform 5 kN load acts downward on the entire top surface.

Volume fraction 0.25, 2 mm elements.\end{quote}

\noindent Source recorded by suite authors: voxfem solver benchmark bridge with a distributed deck load.

\noindent Adjudication: intended reading confirmed; cited source not traced to a specific published problem.

\Needspace{8\baselineskip}
\subsection{b04 torsion shaft}

\noindent Expected response: solve.

\begin{quote}\small Shaft bracket in torsion

Envelope 160 mm long (x) with an 80 x 80 mm square cross-section (y up, z), steel. The x = 0 end is bolted
rigidly to a gearbox. At the far end (x = 160) a 40 N*m torque about +x (right-hand rule) is applied through two
16 x 16 mm drive lugs on the end face: one flush with the top edge (y = 80), one flush with the bottom edge
(y = 0), both centred at z = 40. Each lug carries an equal and opposite force along z.

Volume fraction 0.3, 2 mm elements.\end{quote}

\noindent Source recorded by suite authors: voxfem solver benchmark torsion in physical units.

\noindent Adjudication: intended reading confirmed; cited source not traced to a specific published problem.

\Needspace{8\baselineskip}
\subsection{b05 l bracket}

\noindent Expected response: solve.

\begin{quote}\small L-bracket

Envelope 100 x 100 x 40 mm (x, y, z), y up, steel. The upper-right corner region, x > 40 mm and y > 40 mm, is
occupied by another part and must stay empty through the full depth. The top face of the vertical arm
(y = 100, x from 0 to 40) is bolted to the ceiling. A 500 N downward load acts on a 10 x 10 mm pad on the
right end face of the horizontal arm (x = 100), centred at y = 20 and mid-depth.

Volume fraction 0.3, 1 mm elements.\end{quote}

\noindent Source recorded by suite authors: 3D L-bracket benchmark (e.g.\ \citet{Le2010Stress}), keep-out as a passive void.

\noindent Adjudication: intended reading confirmed; cited source not traced to a specific published problem.

\Needspace{8\baselineskip}
\subsection{b06 motor mount}

\noindent Expected response: solve.

\begin{quote}\small Motor mount

Plate envelope 160 x 60 x 120 mm (x, y, z), y up, aluminium 6061. It bolts to the machine bed through four M10
bolts on its bottom face (y = 0) at x = 20 and 140 mm, z = 20 and 100 mm; each bolt clamps through a 24 mm
washer. Keep a solid boss of 24 mm diameter and 10 mm height around each bolt.

The motor sits on a 40 x 40 mm, 5 mm thick solid pad on the top face, centred at x = 80, z = 60. Its
mounting pad carries 2 kN downward and 0.8 kN sideways (+x) from belt tension.

Volume fraction 0.2, 1 mm elements.\end{quote}

\noindent Source recorded by suite authors: engineering.

\noindent Adjudication: intended reading confirmed.

\Needspace{8\baselineskip}
\subsection{b07 sensor arm}

\noindent Expected response: clarify.

\begin{quote}\small Sensor arm

Envelope 150 x 30 x 30 mm, the arm running along x. It carries a 2 kg camera (about 20 N) at the far end,
mounted on a 20 x 20 mm plate in the middle of the end face at x = 150 mm. The arm attaches to the robot frame.

Steel, volume fraction 0.25, 1 mm elements.\end{quote}

\noindent Source recorded by suite authors: engineering.

\noindent Adjudication: intended reading confirmed.

\noindent Scripted clarification facts (topic: oracle answer): where and how the arm attaches: ``The x = 0 end face is bolted flat to the frame; treat it as fully fixed.''; direction of the camera load: ``Gravity acts along -y.''

\Needspace{8\baselineskip}
\subsection{b08 slider bracket}

\noindent Expected response: refuse.

\begin{quote}\small Slider bracket

Envelope 120 x 40 x 40 mm (x, y, z), y up, steel. It rides on a linear rail along its whole bottom face
(y = 0): the rail holds it vertically and sideways but lets it slide freely along x. A 300 N push along +x acts
on a 10 x 10 mm pad at the centre of the left end face (x = 0).

Volume fraction 0.3, 1 mm elements.\end{quote}

\noindent Source recorded by suite authors: engineering.

\noindent Adjudication: intended reading confirmed.

\noindent Scripted clarification facts (topic: oracle answer): any other support or an end stop: ``No, it is supported only by the rail.''

\Needspace{8\baselineskip}
\subsection{b09 duct bridge}

\noindent Expected response: solve.

\begin{quote}\small Pedestrian bridge model with cable duct

Span 300 mm (x), 60 mm high (y, up), 60 mm wide (z), steel. Both bottom end edges rest on bearings across the
full width: left pinned, right free to slide along the span. 3 kN acts downward on the full top surface.
A 16 mm diameter cable duct runs across the full width (along z) at x = 150 mm, y = 20 mm and must stay empty.
No member may be thinner than 3 mm.

Volume fraction 0.25, 1 mm elements.\end{quote}

\noindent Source recorded by suite authors: engineering.

\noindent Adjudication: intended reading confirmed.

\Needspace{8\baselineskip}
\subsection{b10 hook plate}

\noindent Expected response: clarify.

\begin{quote}\small Hook plate

Plate envelope 120 x 80 x 20 mm (x, y, z), y up, steel. It is bolted along its top face (y = 80) to a beam.
A 1.5 kN load hangs from the lower corner.

Volume fraction 0.3, 1 mm elements.\end{quote}

\noindent Source recorded by suite authors: engineering.

\noindent Adjudication: intended reading confirmed.

\noindent Scripted clarification facts (topic: oracle answer): which lower corner, and the attachment size: ``The lower right corner (x = 120). The hook hangs from a 10 mm wide lug on the bottom face at the right end (x from 110 to 120), across the full thickness.''

\Needspace{8\baselineskip}
\subsection{b11 cantilever line load}

\noindent Expected response: solve.

\begin{quote}\small Cantilever with an end line load

Beam envelope 120 mm long (x), 40 mm tall (y, up) and 20 mm wide (z), steel. The x = 0 end face is fully
fixed. A 1 kN downward load is spread evenly along the bottom edge of the free end (x = 120, y = 0), across
the full width.

Volume fraction 0.3, 1 mm elements.\end{quote}

\noindent Source recorded by suite authors: 3D cantilever with an end line load, the default problem of top3d (\citet{Liu2014Top3d}).

\noindent Adjudication: intended reading confirmed; benchmark family confirmed.

\Needspace{8\baselineskip}
\subsection{b12 wheel}

\noindent Expected response: solve.

\begin{quote}\small Wheel problem

Envelope 100 x 50 x 100 mm (x, y, z), y up, steel. The part stands on four 5 x 5 mm pads, one in each corner
of its bottom face (y = 0); each pad is fixed in all directions. A 2 kN downward load hangs from a 10 x 10 mm
pad at the centre of the bottom face.

Volume fraction 0.2, 1 mm elements.\end{quote}

\noindent Source recorded by suite authors: the `wheel' problem of top3d (\citet{Liu2014Top3d}): bottom corners fixed, load at the bottom centre.

\noindent Adjudication: intended reading confirmed; benchmark family confirmed.

\Needspace{8\baselineskip}
\subsection{b13 cantilever end traction}

\noindent Expected response: solve.

\begin{quote}\small Cantilever with a distributed end load

Envelope 160 x 80 x 80 mm (x, y, z), y up, steel. The x = 0 face is fully fixed. The last 4 mm of the free end
(x from 156 to 160) is a solid end plate that must be kept. A 4 kN downward shear load is spread uniformly over
the whole free end face (x = 160).

Volume fraction 0.12, 2 mm elements.\end{quote}

\noindent Source recorded by suite authors: 2:1:1 cantilever with a distributed end load, the large-scale benchmark of the PETSc TopOpt framework (\citet{Aage2015PETSc}).

\noindent Adjudication: intended reading confirmed; cited source not traced to a specific published problem.

\Needspace{8\baselineskip}
\subsection{b14 cube torsion}

\noindent Expected response: solve.

\begin{quote}\small Block in torsion

A 100 mm cube (x, y, z), z up, steel. The bottom face (z = 0) is bolted down over its full area. An 18 N*m
torque about +z (right-hand rule, counter-clockwise seen from above) is applied to the top face through four
10 x 10 mm pads, one at the midpoint of each top edge and flush with that edge. Each pad carries an equal force
tangential to the edge it sits on.

Volume fraction 0.15, 1 mm elements.\end{quote}

\noindent Source recorded by suite authors: 3D torsion benchmark family (Michell-type torsion in a box domain).

\noindent Adjudication: intended reading confirmed.

\Needspace{8\baselineskip}
\subsection{b15 corner plate}

\noindent Expected response: solve.

\begin{quote}\small Plate on four corner supports

Square plate envelope 200 x 20 x 200 mm (x, y, z), y up, aluminium 6061. It is pinned (fixed in all
directions) at four 6 x 6 mm pads in the corners of its bottom face. A 3 kN downward load acts on a
20 x 20 mm pad at the centre of the top face.

Volume fraction 0.3, 2 mm elements.\end{quote}

\noindent Source recorded by suite authors: point-loaded square plate on four corner supports (3D plate benchmark family).

\noindent Adjudication: intended reading confirmed.

\Needspace{8\baselineskip}
\subsection{b16 column}

\noindent Expected response: solve.

\begin{quote}\small Column under combined load

Column envelope 60 x 240 x 60 mm (x, y, z), y up, steel. It stands on four 10 x 10 mm footing pads in the
corners of its bottom face, each fixed in all directions. A 20 x 20 mm cap pad at the centre of the top face
carries 10 kN downward and 1 kN sideways along +x.

Volume fraction 0.2, 2 mm elements.\end{quote}

\noindent Source recorded by suite authors: tower / column under an axial and a lateral top load (3D benchmark family).

\noindent Adjudication: intended reading confirmed.

\Needspace{8\baselineskip}
\subsection{b17 arch bridge}

\noindent Expected response: solve.

\begin{quote}\small Arch bridge with a solid deck

Envelope 300 mm span (x), 100 mm high (y, up), 60 mm wide (z), steel. The top 6 mm (y from 94 to 100) is the
deck and must stay solid. The bridge is pinned (fixed in all directions) along both bottom end edges, across
the full width. A 6 kN load acts downward uniformly over the whole deck surface (y = 100).

Volume fraction 0.15, 2 mm elements.\end{quote}

\noindent Source recorded by suite authors: bridge benchmark with a passive solid deck and pinned abutments (arch-forming).

\noindent Adjudication: intended reading confirmed.

\Needspace{8\baselineskip}
\subsection{b18 cantilever hole}

\noindent Expected response: solve.

\begin{quote}\small Cantilever with a hole

Envelope 200 x 100 x 40 mm (x, y, z), y up, steel. The x = 0 face is fully fixed. A 2 kN downward load acts
on a 10 x 10 mm pad at the centre of the free end face (x = 200). A 40 mm diameter hole runs through the full
thickness (along z), centred at x = 100, y = 50; it must stay empty.

Volume fraction 0.3, 1 mm elements.\end{quote}

\noindent Source recorded by suite authors: cantilever with a circular keep-out (passive-void benchmark family).

\noindent Adjudication: intended reading confirmed.

\Needspace{8\baselineskip}
\subsection{b19 engine bracket}

\noindent Expected response: solve.

\begin{quote}\small Engine bracket (inch units)

Envelope 7.0 x 2.5 x 4.0 in (x, y, z), y up, Ti-6Al-4V (E = 16.5 Mpsi, nu = 0.34). Work in inches and pounds-force.
It bolts down through four bolts on its bottom face (y = 0) at x = 0.5 and 2.5 in, z = 0.5 and 3.5 in, each
clamping through a 0.5 in diameter washer. The clevis is a solid cylinder that must be kept: axis along z,
centred at x = 6.0 in, y = 1.75 in, radius 0.5 in, from z = 1.25 to 2.75 in. The design load is an 8000 lbf
vertical pull (+y) carried by the clevis.

Volume fraction 0.2, 0.05 in elements.\end{quote}

\noindent Source recorded by suite authors: adapted from the GE jet-engine bracket challenge~\citep{GE2013Bracket}: a box envelope with a distributed vertical load replaces the original geometry, pin load and fixed interfaces.

\noindent Adjudication: intended reading confirmed; benchmark family confirmed.

\Needspace{8\baselineskip}
\subsection{b20 quarter plate}

\noindent Expected response: solve.

\begin{quote}\small Plate, quarter model

The full plate is 240 x 24 x 240 mm (x, y, z), y up, steel. It is pinned (fixed in all directions) at four
8 x 8 mm pads in the corners of its bottom face, and carries 4 kN downward on a 20 x 20 mm pad at the centre of
its top face.

Model one quarter only (x and z from 0 to 120 mm) and use the two symmetry planes at x = 120 and z = 120.
Volume fraction 0.25, 1 mm elements.\end{quote}

\noindent Source recorded by suite authors: point-loaded plate on corner supports, quarter model with two symmetry planes.

\noindent Adjudication: intended reading confirmed.

\Needspace{8\baselineskip}
\subsection{b21 mbb z up}

\noindent Expected response: solve.

\begin{quote}\small Simply supported beam, z up

Beam envelope 320 mm long (x), 40 mm wide (y) and 40 mm tall (z), with z pointing up; steel. The left bottom
edge (x = 0, z = 0) is pinned across the full width. The right bottom edge (x = 320, z = 0) is a roller across
the full width: it carries vertical and sideways load but lets the beam move along x. A 3 kN downward line load
acts across the full width at mid-span on the top surface.

Volume fraction 0.3, 2 mm elements.\end{quote}

\noindent Source recorded by suite authors: 3D MBB beam, full model (\citet{Liu2014Top3d}), stated with z up.

\noindent Adjudication: intended reading confirmed; benchmark family confirmed.

\Needspace{8\baselineskip}
\subsection{b22 cantilever slab}

\noindent Expected response: solve.

\begin{quote}\small Cantilevered slab

Slab envelope 150 mm out from the wall (x), 20 mm thick (y, up) and 200 mm wide (z), steel. The wall face
(x = 0) is fully fixed. A 6 kN load acts downward, spread uniformly over the whole top surface.

Volume fraction 0.3, 2 mm elements.\end{quote}

\noindent Source recorded by suite authors: cantilevered plate under a uniform load (distributed-load cantilever family).

\noindent Adjudication: intended reading confirmed.

\Needspace{8\baselineskip}
\subsection{b23 clamped beam}

\noindent Expected response: solve.

\begin{quote}\small Clamped-clamped beam

Beam envelope 240 x 40 x 40 mm (x, y, z), y up, steel. Both end faces (x = 0 and x = 240) are fully clamped.
A 3 kN downward load acts on a 10 mm long strip across the full width of the top face at mid-span
(x from 115 to 125).

Volume fraction 0.25, 1 mm elements.\end{quote}

\noindent Source recorded by suite authors: clamped-clamped beam with a central load (3D benchmark family).

\noindent Adjudication: intended reading confirmed.

\Needspace{8\baselineskip}
\subsection{b24 shear block}

\noindent Expected response: solve.

\begin{quote}\small Block in shear

An 80 mm cube (x, y, z), y up, steel. The bottom face (y = 0) is fully fixed. A 4 kN horizontal load along +x
is spread uniformly over the whole top face.

Volume fraction 0.2, 1 mm elements.\end{quote}

\noindent Source recorded by suite authors: shear-loaded block (3D benchmark family).

\noindent Adjudication: intended reading confirmed.

\Needspace{8\baselineskip}
\subsection{b25 cantilever metres}

\noindent Expected response: solve.

\begin{quote}\small Large cantilever (metres and kilonewtons)

Envelope 2 m long (x), 1 m tall (y, up) and 1 m wide (z), structural steel, E = 210 GPa, nu = 0.3. Work in
metres and kilonewtons. The x = 0 face is fully fixed. A 100 kN downward load acts on a 0.2 x 0.2 m bearing
plate at the centre of the free end face (x = 2 m).

Volume fraction 0.15, 0.02 m elements.\end{quote}

\noindent Source recorded by suite authors: 2:1:1 cantilever (\citet{Aage2015PETSc} family), in metres and kN.

\noindent Adjudication: intended reading confirmed; cited source not traced to a specific published problem.

\Needspace{8\baselineskip}
\subsection{b26 canopy}

\noindent Expected response: solve.

\begin{quote}\small Canopy on four columns

Plate envelope 300 x 40 x 200 mm (x, y, z), y up, aluminium 6061. It rests on four columns whose 20 x 20 mm
heads meet its bottom face (y = 0) centred at x = 40 and 260 mm, z = 40 and 160 mm; each head is fixed in all
directions. An 8 kN load acts downward, spread uniformly over the whole top surface.

Volume fraction 0.25, 2 mm elements.\end{quote}

\noindent Source recorded by suite authors: plate on four interior column supports under a uniform load (multi-support slab family).

\noindent Adjudication: intended reading confirmed.

\Needspace{8\baselineskip}
\subsection{b27 eccentric cantilever}

\noindent Expected response: solve.

\begin{quote}\small Cantilever with an eccentric end load

Envelope 160 x 40 x 40 mm (x, y, z), y up, steel. The x = 0 face is fully fixed. A 1 kN downward load acts on
a 10 x 10 mm pad in the top back corner of the free end face (x = 160): y from 30 to 40 mm and z from 30 to
40 mm. The beam therefore carries bending and torsion.

Volume fraction 0.25, 1 mm elements.\end{quote}

\noindent Source recorded by suite authors: cantilever with an eccentric end load, bending plus torsion (3D benchmark family).

\noindent Adjudication: intended reading confirmed.

\Needspace{8\baselineskip}
\subsection{b28 stool}

\noindent Expected response: solve.

\begin{quote}\small Stool

Envelope 200 x 200 x 200 mm (x, y, z), y up, aluminium 6061. The seat, the top 10 mm (y from 190 to 200), must
stay solid. It stands on four 20 x 20 mm feet in the corners of its bottom face (y = 0), each fixed in all
directions. The space under the middle of the seat, x and z from 50 to 150 mm and y from 0 to 120 mm, must stay
empty for leg room. A 1.2 kN load acts downward, spread over the whole seat surface.

Volume fraction 0.1, 2 mm elements.\end{quote}

\noindent Source recorded by suite authors: stool / table problem: top load on a passive seat, four feet (3D benchmark family).

\noindent Adjudication: intended reading confirmed.

\Needspace{8\baselineskip}
\subsection{b29 michell cantilever}

\noindent Expected response: solve.

\begin{quote}\small Michell cantilever on a pin

Slab envelope 200 x 100 x 20 mm (x, y, z), y up, steel. It is held by a rigid 30 mm diameter pin through the
full thickness (along z) at x = 30, y = 50: everything inside the pin's circle is fixed in all directions. A
500 N downward load acts on a 10 mm tall strip of the right end face (x = 200), centred at y = 50, across the
full thickness.

Volume fraction 0.2, 1 mm elements.\end{quote}

\noindent Source recorded by suite authors: Michell cantilever on a circular support (\citet{Michell1904}), as a 3D slab.

\noindent Adjudication: intended reading confirmed; benchmark family confirmed.

\Needspace{8\baselineskip}
\subsection{b30 bridge clearance}

\noindent Expected response: solve.

\begin{quote}\small Bridge over a clearance

Envelope 320 mm span (x), 80 mm high (y, up), 80 mm wide (z), steel. The left bottom edge (x = 0, y = 0) is
pinned across the full width; the right bottom edge (x = 320, y = 0) is a roller across the full width that
lets the span expand along x. The top 4 mm (y from 76 to 80) is the deck and must stay solid. A 5 kN load acts
downward over the whole deck surface. Under mid-span, a clearance
x from 100 to 220 mm, y from 0 to 40 mm, through the full width, must stay empty.

Volume fraction 0.2, 2 mm elements.\end{quote}

\noindent Source recorded by suite authors: bridge with a passive-void clearance under the span (bridge benchmark family).

\noindent Adjudication: intended reading confirmed.

\Needspace{8\baselineskip}
\subsection{b31 pillow block}

\noindent Expected response: solve.

\begin{quote}\small Pillow-block bearing bracket

Envelope 120 x 60 x 40 mm (x, y, z), y up, grey cast iron (E = 110 GPa, nu = 0.26). Two M8 bolts hold its bottom
face (y = 0) down, at x = 15 and 105 mm, z = 20 mm, each through a 16 mm washer. The bearing seat is a solid
cylinder that must be kept: axis along z, centred at x = 60, y = 40, 30 mm diameter, through the full depth.
The shaft pushes 3 kN downward on the seat.

Volume fraction 0.3, 1 mm elements.\end{quote}

\noindent Source recorded by suite authors: engineering.

\noindent Adjudication: intended reading confirmed.

\Needspace{8\baselineskip}
\subsection{b32 drone arm}

\noindent Expected response: solve.

\begin{quote}\small Drone arm

Arm envelope 150 mm long (x), 20 mm wide (y) and 20 mm tall (z), with z pointing up; aluminium 7075. The root
end face (x = 0) is clamped in the frame. The motor mounts on a 20 x 20 mm, 5 mm thick solid pad on the top of
the far end (x from 130 to 150, the full width, z from 15 to 20), which must be kept. Full thrust is 15 N
upward (+z), acting on the top face of that pad.

Volume fraction 0.2, 0.5 mm elements.\end{quote}

\noindent Source recorded by suite authors: engineering.

\noindent Adjudication: intended reading confirmed.

\Needspace{8\baselineskip}
\subsection{b33 shelf bracket}

\noindent Expected response: solve.

\begin{quote}\small Shelf bracket

Envelope 150 mm out from the wall (x), 150 mm tall (y, up) and 20 mm thick (z), aluminium 6061. Two screws
fix it to the wall face (x = 0), at y = 20 and 130 mm, z = 10 mm, each through a 10 mm washer. The shelf rests
on the whole top face (y = 150) and puts 600 N downward on it. No member may be thinner than 3 mm.

Volume fraction 0.2, 1 mm elements.\end{quote}

\noindent Source recorded by suite authors: engineering.

\noindent Adjudication: intended reading confirmed.

\Needspace{8\baselineskip}
\subsection{b34 antenna bracket}

\noindent Expected response: solve.

\begin{quote}\small Antenna wall bracket (SI units)

Envelope 0.40 m out from the wall (x), 0.30 m tall (y, up) and 0.20 m wide (z); S355 steel, E = 210 GPa,
nu = 0.3. Work in metres and newtons. The wall face (x = 0) is welded over its full area. The mast clamp is a
solid vertical cylinder that must be kept: axis along y, centred at x = 0.33 m, z = 0.10 m, radius 0.04 m, from
y = 0.05 m to the top. The clamp carries the antenna's 300 N weight (downward) and a 1200 N wind load along +z.

Volume fraction 0.15, 5 mm elements.\end{quote}

\noindent Source recorded by suite authors: engineering.

\noindent Adjudication: intended reading confirmed.

\Needspace{8\baselineskip}
\subsection{b35 padeye}

\noindent Expected response: solve.

\begin{quote}\small Lifting padeye

Envelope 100 x 80 x 30 mm (x, y, z), y up, steel. The base (y = 0) is welded to the deck over its full area.
The shackle pin passes through a solid boss that must be kept: axis along z, centred at x = 50, y = 55, 30 mm
diameter, through the full thickness. The sling pulls the boss with 6 kN upward and 2.5 kN along +x.

Volume fraction 0.3, 1 mm elements.\end{quote}

\noindent Source recorded by suite authors: engineering.

\noindent Adjudication: intended reading confirmed.

\Needspace{8\baselineskip}
\subsection{b36 pedal arm}

\noindent Expected response: solve.

\begin{quote}\small Pedal arm

Arm envelope 200 mm long (x), 40 mm tall (y, up) and 20 mm thick (z), aluminium 6061. It is keyed to a 16 mm
shaft through its thickness at x = 15, y = 20 mm, so treat everything inside the shaft's circle as fixed in all
directions. The foot pushes 1 kN downward on a 30 mm long pad on the top face at the far end (x from 170 to 200),
across the full thickness.

Volume fraction 0.25, 1 mm elements.\end{quote}

\noindent Source recorded by suite authors: engineering.

\noindent Adjudication: intended reading confirmed.

\Needspace{8\baselineskip}
\subsection{b37 battery crossmember}

\noindent Expected response: solve.

\begin{quote}\small Battery-tray crossmember (kN and mm)

Envelope 600 mm long (x), 60 mm tall (y, up) and 80 mm deep (z), aluminium 6082-T6 (E = 70 GPa, nu = 0.33).
Work in millimetres and kilonewtons. Both end faces (x = 0 and x = 600) bolt to the side rails and are fully
fixed. The battery module stands on two 40 x 40 mm feet on the top face, centred at x = 200 and 400 mm,
z = 40 mm; each foot puts 1.5 kN downward.

Volume fraction 0.2, 2 mm elements.\end{quote}

\noindent Source recorded by suite authors: engineering.

\noindent Adjudication: intended reading confirmed.

\Needspace{8\baselineskip}
\subsection{b38 peek bracket}

\noindent Expected response: solve.

\begin{quote}\small Satellite equipment bracket

Envelope 80 x 60 x 40 mm (x, y, z), z normal to the panel; PEEK (E = 3.6 GPa, nu = 0.38). The base (z = 0) is
bonded to the panel over its full area. The unit mounts on a 30 x 30 mm, 4 mm thick solid interface pad on the
top face (z = 40), centred at x = 40, y = 30, which must be kept. Design for the launch load on the pad: 300 N
along +x and 300 N along -z together.

Volume fraction 0.25, 1 mm elements.\end{quote}

\noindent Source recorded by suite authors: engineering.

\noindent Adjudication: intended reading confirmed.

\Needspace{8\baselineskip}
\subsection{b39 tv bracket}

\noindent Expected response: solve.

\begin{quote}\small TV wall bracket

Envelope 300 mm out from the wall (x), 200 mm tall (y, up) and 40 mm wide (z), steel. Four lag screws fix it to
the wall face (x = 0) at y = 30 and 170 mm, z = 10 and 30 mm, each through a 12 mm washer. The TV hangs on a
mounting plate that covers the far end face (x = 300) from y = 50 to 150 mm across the full width; it puts
400 N downward and 50 N pulling away from the wall (+x) on that plate. No member may be thinner than 5 mm.

Volume fraction 0.15, 2 mm elements.\end{quote}

\noindent Source recorded by suite authors: engineering.

\noindent Adjudication: intended reading confirmed.

\Needspace{8\baselineskip}
\subsection{b40 crank arm}

\noindent Expected response: solve.

\begin{quote}\small Crank arm

Envelope 180 mm long (x), 40 mm tall (y, up) and 30 mm thick (z), aluminium 7075. The axle end is keyed to a
20 mm diameter spindle through the thickness at x = 20, y = 20 mm: everything inside that circle is fixed. The
pedal threads into a solid boss that must be kept: axis along z at x = 160, y = 20 mm, 16 mm diameter, through
the full thickness. The rider's 1.2 kN downward push arrives on the boss's outer face (z = 30), over the 16 mm
circle.

Volume fraction 0.3, 1 mm elements.\end{quote}

\noindent Source recorded by suite authors: engineering.

\noindent Adjudication: intended reading confirmed.

\Needspace{8\baselineskip}
\subsection{b41 hinge bracket}

\noindent Expected response: solve.

\begin{quote}\small Door hinge bracket

Envelope 140 x 80 x 30 mm (x, y, z), y up, steel. Two bolts hold its bottom face (y = 0) at x = 20 and 120 mm,
z = 15 mm, each through a 12 mm washer. The hinge pin runs through a solid boss that must be kept: axis along
z, centred at x = 70, y = 65 mm, 20 mm diameter, full thickness. The door hangs 800 N downward and pulls 300 N
along +x on the boss. A cable slot, x from 40 to 100 mm and y from 20 to 35 mm, through the full thickness,
must stay empty.

Volume fraction 0.3, 1 mm elements.\end{quote}

\noindent Source recorded by suite authors: engineering.

\noindent Adjudication: intended reading confirmed.

\Needspace{8\baselineskip}
\subsection{b42 snap hook}

\noindent Expected response: solve.

\begin{quote}\small Printed snap hook

Envelope 40 mm long (x), 10 mm tall (y, up) and 12 mm wide (z), printed in PA12 (E = 1.7 GPa, nu = 0.4). The
root end face (x = 0) is part of the housing and fully fixed. On assembly the tip is pushed 15 N upward (+y) on
a 4 mm long strip of its underside (y = 0, x from 36 to 40), across the full width. No member may be thinner
than 1.2 mm.

Volume fraction 0.35, 0.25 mm elements.\end{quote}

\noindent Source recorded by suite authors: engineering.

\noindent Adjudication: intended reading confirmed.

\Needspace{8\baselineskip}
\subsection{b43 press riser}

\noindent Expected response: solve.

\begin{quote}\small Press riser block

Envelope 300 x 100 x 150 mm (x, y, z), y up, steel. The bottom face (y = 0) is bolted to the press bed over its
full area. A coolant channel along the full length (x), z from 45 to 105 mm and y from 0 to 30 mm, must stay
empty. The tool presses on a 100 x 50 mm pad at the centre of the top face with 20 kN downward and 2 kN along +z.

Volume fraction 0.3. Mesh it with 50 elements across the shortest side.\end{quote}

\noindent Source recorded by suite authors: engineering.

\noindent Adjudication: intended reading corrected.

\Needspace{8\baselineskip}
\subsection{b44 gearbox bracket}

\noindent Expected response: clarify.

\begin{quote}\small Gearbox output bracket

Envelope 160 mm long (x) with an 80 x 80 mm square cross-section (y, z), steel. The x = 0 end is bolted
rigidly to a gearbox. At the far end a 40 N*m torque about the long axis is applied through two 16 x 16 mm
drive lugs on the end face, one at the top edge and one at the bottom edge, each centred in depth.

Volume fraction 0.3, 2 mm elements.\end{quote}

\noindent Source recorded by suite authors: engineering.

\noindent Adjudication: intended reading confirmed.

\noindent Scripted clarification facts (topic: oracle answer): which axis is up, and where the lugs sit: ``y is up. Both lugs sit inside the end face, flush with its top edge (y = 80) and bottom edge (y = 0), centred at z = 40, so their centres are 64 mm apart.''; the sense of the torque: ``Either sense is fine; take +x by the right-hand rule.''

\Needspace{8\baselineskip}
\subsection{b45 pump bracket}

\noindent Expected response: clarify.

\begin{quote}\small Pump bracket

Envelope 160 x 100 x 60 mm (x, y, z), y up, steel. The back face (z = 0) bolts to the wall over its full area.
The pump sits on the top face (y = 100) over a 60 x 40 mm footprint centred at x = 80, z = 35. Design it for the
pump.

Volume fraction 0.25, 2 mm elements.\end{quote}

\noindent Source recorded by suite authors: engineering.

\noindent Adjudication: intended reading confirmed.

\noindent Scripted clarification facts (topic: oracle answer): the pump's load: ``The pump weighs 25 kg: use 250 N downward (-y) spread over the footprint. Nothing else acts on it.''

\Needspace{8\baselineskip}
\subsection{b46 sensor housing}

\noindent Expected response: clarify.

\begin{quote}\small Sensor housing bracket

Envelope 80 x 50 x 40 mm (x, y, z), y up, aluminium 6061. The bottom face (y = 0) is bolted to the chassis over
its full area. The sensor head pushes 200 N along -x on a 10 x 10 mm pad at the centre of the x = 80 end face.
Leave room for the connector plug at the front.

Volume fraction 0.3, 1 mm elements.\end{quote}

\noindent Source recorded by suite authors: engineering.

\noindent Adjudication: intended reading confirmed.

\noindent Scripted clarification facts (topic: oracle answer): the connector clearance: ``The front is the z = 0 face. The plug needs an empty box from x = 20 to 50 mm, y = 10 to 30 mm, and from the front face 15 mm deep (z = 0 to 15).''

\Needspace{8\baselineskip}
\subsection{b47 camera mount}

\noindent Expected response: clarify.

\begin{quote}\small Inspection-camera mount

Envelope 120 x 60 x 60 mm (x, y, z), y up, aluminium 6061. The x = 0 face is bolted to the gantry over its full
area. The camera hangs 80 N downward on a 20 x 20 mm pad at the centre of the x = 120 end face. Make it as light
as practical.

1 mm elements.\end{quote}

\noindent Source recorded by suite authors: engineering.

\noindent Adjudication: intended reading confirmed.

\noindent Scripted clarification facts (topic: oracle answer): how much material: ``Use volume fraction 0.15.''

\Needspace{8\baselineskip}
\subsection{b48 cable anchor}

\noindent Expected response: clarify.

\begin{quote}\small Guy-cable anchor

Envelope 100 x 60 x 60 mm (x, y, z), y up, steel. The bottom face (y = 0) is bolted down over its full area. A
guy cable pulls 5 kN on a 16 mm diameter anchor boss face at the centre of the top face; the cable rises 3 for
every 4 it runs horizontally.

Volume fraction 0.25, 1 mm elements.\end{quote}

\noindent Source recorded by suite authors: engineering.

\noindent Adjudication: intended reading confirmed.

\noindent Scripted clarification facts (topic: oracle answer): which way the cable runs: ``The cable leaves towards +x: 4 kN along +x and 3 kN upward (+y).''

\Needspace{8\baselineskip}
\subsection{b49 hinged arm}

\noindent Expected response: refuse.

\begin{quote}\small Hinged support arm

Arm envelope 200 x 40 x 20 mm (x, y, z), y up, steel. The arm hangs on a single hinge pin that runs through its
thickness (along z) at x = 10, y = 20 mm, and turns freely on it; nothing else holds it. A 400 N downward load
acts on a 10 mm tall strip of the far end face (x = 200), centred at y = 20, across the full thickness.

Volume fraction 0.3, 1 mm elements.\end{quote}

\noindent Source recorded by suite authors: engineering.

\noindent Adjudication: intended reading confirmed.

\noindent Scripted clarification facts (topic: oracle answer): any other support or a stop: ``No. The arm must stay free to turn on the pin.''

\Needspace{8\baselineskip}
\subsection{b50 valve bracket}

\noindent Expected response: refuse.

\begin{quote}\small Valve actuator bracket

Envelope 120 x 80 x 40 mm (x, y, z), y up, aluminium 6061. The bottom face (y = 0) is bolted down over its full
area. The actuator's 600 N downward load enters through a 20 x 20 mm pad on the top face centred at x = 90,
z = 20. The region x > 70 mm, y > 50 mm must stay clear for the valve handle, through the full depth.

Volume fraction 0.3, 1 mm elements.\end{quote}

\noindent Source recorded by suite authors: engineering.

\noindent Adjudication: intended reading confirmed.

\noindent Scripted clarification facts (topic: oracle answer): moving the pad or the clearance: ``Neither can move: both are set by the actuator and the valve.''